\documentclass[]{pasj01}

\Received{}
\Accepted{}
 
\usepackage[switch,mathlines]{lineno} 
\usepackage{natbib} 
\usepackage{url}
\usepackage{chngcntr}
\usepackage{multirow}
\usepackage{txfonts}
\usepackage{array}
\usepackage{xcolor}
\renewcommand{\arraystretch}{1.2}

\begin{document} 

\title{ 
Unraveling the structure of the stratified ultra-fast outflows in PDS 456 with XRISM}

\author{Yerong \textsc{Xu}\altaffilmark{1}%
}

\email{yerong.xu@smu.ca}

\author{Luigi C. \textsc{Gallo}\altaffilmark{1}%
}
\author{Kouichi \textsc{Hagino}\altaffilmark{2}%
}
\author{James N. \textsc{Reeves}\altaffilmark{3,4}%
}
\author{Francesco \textsc{Tombesi}\altaffilmark{5}}

\author{Misaki \textsc{Mizumoto}\altaffilmark{6}}

\author{Alfredo \textsc{Luminari}\altaffilmark{7,8}}

\author{Adam G. \textsc{Gonzalez}\altaffilmark{1}}

\author{Ehud \textsc{Behar}\altaffilmark{9,10}}

\author{Rozenn \textsc{Boissay-Malaquin}\altaffilmark{11,12,13}}

\author{Valentina \textsc{Braito}\altaffilmark{3,4,14}}

\author{Pierpaolo  \textsc{Cond\'o}\altaffilmark{5}}

\author{Chris \textsc{Done}\altaffilmark{15}}

\author{Aiko \textsc{Miyamoto}\altaffilmark{16}}

\author{Ryuki \textsc{Mizukawa}\altaffilmark{17}}

\author{Hirokazu \textsc{Odaka}\altaffilmark{16}}

\author{Riki \textsc{Sato}\altaffilmark{2}}

\author{Atsushi \textsc{Tanimoto}\altaffilmark{18}}

\author{Makoto \textsc{Tashiro}\altaffilmark{17,19}}

\author{Tahir \textsc{Yaqoob}\altaffilmark{11,12,13}}

\author{Satoshi \textsc{Yamada}\altaffilmark{20}}


\altaffiltext{1}{Department of Astronomy \& Physics, Saint Mary's University, 923 Robie Street, Halifax, NS B3H 3C3, Canada}
\altaffiltext{2}{Department of Physics, University of Tokyo, Tokyo 113-0033, Japan}
\altaffiltext{3}{INAF, Osservatorio Astronomico di Brera, Via Bianchi 46 I-23807 Merate (LC), Italy}
\altaffiltext{4}{Department of Physics, Institute for Astrophysics and Computational Sciences, The Catholic University of America, Washington, DC 20064, USA}
\altaffiltext{5}{Physics Department, Tor Vergata University of Rome, Via della Ricerca Scientifica 1, 00133 Rome, Italy}
\altaffiltext{6}{Science Research Education Unit, University of Teacher Education Fukuoka, Fukuoka 811-4192, Japan}
\altaffiltext{7}{INAF, Istituto di Astrofisica e Planetologia Spaziali, Via del Fosso del Caveliere 100, I-00133 Roma, Italy}
\altaffiltext{8}{INAF, Osservatorio Astronomico di Roma, Via Frascati 33, 00078 Monteporzio, Italy}
\altaffiltext{9}{Department of Physics, Technion, Technion City, Haifa 3200003, Israel}
\altaffiltext{10}{Kavli Institute for Astrophysics and Space Research, Massachusetts Institute of Technology, MA 02139, USA}
\altaffiltext{11}{Center for Space Science and Technology, University of Maryland, Baltimore County (UMBC), 1000 Hilltop Circle, Baltimore, MD 21250, USA}
\altaffiltext{12}{NASA / Goddard Space Flight Center, Greenbelt, MD 20771, USA}
\altaffiltext{13}{Center for Research and Exploration in Space Science and Technology, NASA / GSFC (CRESST II), Greenbelt, MD 20771, USA}
\altaffiltext{14}{Dipartimento di Fisica, Universit`a di Trento, Via Sommarive 14, I-38123, Trento, Italy}
\altaffiltext{15}{Centre for Extragalactic Astronomy, Department of Physics, University of Durham, South Road, Durham DH1 3LE, UK}
\altaffiltext{16}{Department of Earth and Space Science, Osaka University, Osaka 560-0043, Japan}
\altaffiltext{17}{Department of Physics, Saitama University, Saitama 338-8570, Japan}
\altaffiltext{18}{Graduate School of Science and Engineering, Kagoshima University, Kagoshima, 890-8580, Japan}
\altaffiltext{19}{Institute of Space and Astronautical Science (ISAS), Japan Aerospace Exploration Agency (JAXA), Kanagawa 252-5210, Japan}
\altaffiltext{20}{Frontier Research Institute for Interdisciplinary Sciences, Tohoku University, Sendai 980-8578, Japan }


\maketitle

\begin{abstract}
Multiple clumpy wind components ($v_\mathrm{out}\sim0.2\mbox{--}0.3c$) in the luminous quasar PDS 456 have recently been resolved in the Fe-K band for the first time, thanks to the high-resolution X-ray spectrometer \textit{Resolve} onboard XRISM. In this paper, we investigate the structure of ultra-fast outflows (UFOs) using coordinated observations from XRISM, XMM-Newton, and NuSTAR, along with the self-consistently calculated photoionization model \texttt{PION}. Our results reveal a stratified ionization structure, characterized by a scaling relation between wind velocity and ionization parameter $v_\mathrm{out}\propto\xi^{(0.14\pm0.04)}$. To evaluate the screening effect in photoionization modeling, we tested all possible order permutations of six \texttt{PION} components. We find that highly ionized UFOs ($\log\xi>4.5$) are insensitive to their relative positions, whereas the soft X-ray UFO ($\log\xi\sim3$ and $v_\mathrm{out}\sim0.27c$) and the lowest-ionized hard X-ray UFO ($\log\xi\sim4.1$ and $v_\mathrm{out}\sim0.23c$) are statistically favored -- based on the evidence from both the C-statistic and Bayesian analysis -- to occupy the middle and innermost layers, respectively. This suggests a possible trend where slower UFOs are launched from regions closer to the supermassive black hole (SMBH). The soft X-ray UFO is found to be thermally unstable, regardless of its relative position. However, its location remains unclear. Our sequence analysis and its similarity to hard X-ray UFOs suggest that they may be co-spatial, while variability constraints support its location within the broad-line region at sub-parsec scales. Simulations with XRISM with an open gate valve show that high-resolution soft X-ray data can enhance the reliability of our results. Furthermore, simulations with the future X-ray mission NewAthena demonstrate its capability to resolve the absorber sequence and spatial distributions, enabling the determination of UFO structures and their roles in AGN feedback.
\end{abstract}

\section{Introduction}\label{sec:intro}

A large fraction of active galactic nuclei (AGNs) have been shown to host powerful outflows \citep[e.g,][]{2010Tombesi,2012Patrick,2013Gofford,2020Igo,2021Chartas,2023Matzeu}. The fastest and most powerful subclass is ultra-fast outflows (UFOs), usually detected by highly blueshifted absorption lines in hard X-rays, characterized by high ionization states ($\log[\xi/\mathrm{cm\,erg\,s^{-1}}]\sim3\mbox{--}6$, where $\xi\equiv L_\mathrm{ion}/n_\mathrm{e}R^2$, with $L_\mathrm{ion}$ representing the ionizing luminosity, $n_\mathrm{e}$ the electron number density, and $R$ the distance to the X-ray source) and mildly relativistic velocities \citep[$v \geq 10,000$\,km/s; see review in][]{2021Laha,2023Gallo}. Their extreme properties lead to a large amount of kinetic power, making them promising candidates to drive AGN feedback, which can affect the evolution of host galaxies and regulate the growth of supermassive black holes (SMBHs) \citep{2005DiMatteo,2010Hopkins,2015Tombesi}.

Over the past two decades, numerous UFOs have been detected in AGNs. However, their driving mechanism and precise structure remain unclear; these properties are crucial for determining how effectively UFOs transfer energy and momentum to their host galaxies. Radiatively-driven and magnetically-driven (MHD) mechanisms \citep[e.g.,][]{2000Proga,2010Sim,2016Hagino,2004Kato,2010Fukumura,2015Fukumura} have been invoked to accelerate winds to relativistic speeds. In highly accreting systems, UFOs are expected to be dominated by radiation pressure while MHD-driven UFOs are typically associated with lowly accreting systems, although both of them can co-exist in the same system. Observationally, a positive correlation between the wind velocity and X-ray luminosity has been reported in several high-/super-Eddington AGNs \citep{2017Matzeu,2018Pinto,2023Xu,2024Xu}, indicating radiative driving as the primary mechanism. However, MHD-driven winds cannot be ruled out since MHD models can also reproduce the same correlation under specific conditions \citep{2018Fukumura}. Alternatively, the wind launching mechanism may also be diagnosed through the profiles of their absorption lines \citep{2023Gallo}. MHD-driven winds are theoretically predicted to produce a characteristic extended blue tail (i.e., faster when closer to SMBHs), whereas radiatively-driven winds exhibit a pronounced red tail \citep[i.e., faster when farther from SMBHs due to an asymptotically increasing terminal velocity;][]{2022Fukumura}.

In addition to the driving mechanism, the structure of AGN outflows also plays a critical role in their energy dissipation and their impact on host galaxies. A shocked wind scenario has been proposed in which UFOs launched from very close regions of SMBHs collide with the interstellar medium (ISM), generating multiphase outflows \citep{2013Pounds}. Such collisions could produce mildly relativistic but relatively low-ionization UFOs ($\log[\xi/\mathrm{cm\,erg\,s^{-1}}]\sim1\mbox{--}3$; e.g., \citealt{2015Longinotti,2018Reeves,2019Serafinelli,2020Kosec,2021Xu,2022Xu,2024Xu}) and slow and low-ionization `warm absorbers' \citep[WAs;][]{2013Tombesi,2014Laha}, both commonly detected in the soft X-ray band. However, the intrinsic structure of UFOs, such as whether they consist of homogeneous or clumpy outflowing gas, remains unresolved due to the limited spectral resolution of CCDs in the hard X-ray band.


The X-ray calorimeter, \textit{Resolve} (R. Kelley et al. in prep.), onboard the X-Ray Imaging and Spectroscopy Mission \citep[XRISM;][]{2020Tashiro,2024Tashiro,2025Tashiro}, is led by JAXA in collaboration with NASA and ESA. With its unprecedented spectral resolution ($R=E/\Delta E>1000$ in the Fe-K band), \textit{Resolve} enables resolving CCD-based broad spectral lines in AGNs. The quasar PDS 456 is a prototype for UFOs in AGNs \citep{2003Reeves,2009Reeves,2015Nardini}. CCD resolution spectra disclosed a broad feature that was often attributed to high turbulence velocities \citep[e.g.,][]{2016Reeves}. With XRISM, for the first time, the UFO feature in PDS 456 is resolved into multiple narrow lines of UFOs in hard X-rays, revealing a highly inhomogeneous outflow with five discrete velocity components outflowing at $\simeq20\mbox{--}30$\% of the speed of light \citep[][Paper I hereafter]{2025XRISM}.

In this work, we examine the XRISM observations of PDS 456 to investigate the structure of UFOs using the advanced photoionization model \texttt{PION} \citep{2015Miller,2016Mehdipour}. Unlike pre-calculated photoionization models, such as \texttt{XSTAR} \citep{2001Kallman}, \texttt{Cloudy} \citep{1998Ferland}, \texttt{PHASE} \citep{2003Krongold}, and \texttt{XABS} \citep{2003Steenbrugge}, which assume a constant spectral energy distribution (SED), \texttt{PION} dynamically calculates the ionization balance of plasma in response to changes in the irradiating field during spectral fitting. While the differences between \texttt{PION} and other models are small (up to 30\%) for a single plasma component \citep{2016Mehdipour}, they may become significant when multiple absorbers lie along our line of sight (LOS). In such cases, `screening effects’, where absorbers closer to the X-ray source block radiation from reaching farther absorbers, can alter the ionization states of subsequent absorbers, indicating the importance of the relative positions of different outflow components. Therefore, this work aims to accurately characterize winds in PDS 456 and explore their structure by applying \texttt{PION} to the XRISM data.

PDS 456 is the most luminous \citep[$L_\mathrm{bol}\sim10^{47}\,\mathrm{erg}\,\mathrm{s}^{-1}$;][]{2024Gravity}, radio-quiet \citep{2019Bischetti} quasar in the nearby \citep[$z=0.184$;][]{1997Torres} Universe with an estimated black hole mass of $M_\mathrm{BH}\simeq5\times10^{8}\,M_\odot$ \citep{2023Gravity}. It has been extensively studied due to evidence of outflows spanning all scales and wavelengths. On sub-pc scales, UFOs have been detected in the hard X-ray \citep[$v\sim0.25\mbox{--}0.34c$, e.g.,][]{2015Nardini}, soft X-ray \citep[$v\sim0.17\mbox{--}0.27c$, e.g.,][]{2016Reeves}, and probably also far-UV bands \citep[$v\sim0.30c$,][]{2018Hamann}. At scales of $\sim$1\,pc, fast winds ($v\sim5000\,\mathrm{km}\,\mathrm{s}^{-1}$) have been observed through UV emission lines \citep{2005OBrien}. On larger scales ($1\mbox{--}10$\,kpc), slower outflows ($v\sim250\,\mathrm{km}\,\mathrm{s}^{-1}$) of molecular and ionized gas have been identified through MUSE (optical) and ALMA (radio) observations \citep{2019Bischetti,2024Travascio}. The presence of outflows across these diverse spatial scales makes PDS 456 an exceptional laboratory for studying AGN outflows and feedback.

The paper is structured as follows: Section \ref{sec:obs} briefly describes the XRISM, NuSTAR and XMM-Newton observations used in this work. In Section \ref{sec:results}, we detail our spectral fits and results. We discuss the implications of our results in Section \ref{sec:discussion}. Finally, Section \ref{sec:conclusion} restates our findings as distinct conclusions.

%



\section{Observations}\label{sec:obs}

XRISM observed PDS 456 as a Performance Verification (PV) target from 2024-03-11 to 2024-03-17, with a total exposure of 258\,ks, coordinated with five other telescopes: XMM-Newton, NuSTAR, Swift, NICER, and Seimei, spanning optical to hard X-ray bands. In this work, we focus on data from the X-ray calorimeter, \textit{Resolve}, and the CCD detector, \textit{Xtend} \citep{2025Noda} onboard XRISM, incorporating NuSTAR and XMM-Newton/OM data to constrain the broadband SED of PDS 456, which is essential for photoionization modeling. 

During the XRISM observation, an X-ray flare was captured, primarily during the first half of the exposure. NuSTAR, with an exposure time of 160\,ks, began two hours earlier than XRISM and fully covered the flare, while XMM-Newton, with an 80\,ks exposure, overlapped the second half of the XRISM observation. Due to the closed gate valve (GVC), the energy bandpass of \textit{Resolve} is restricted to $2\mbox{--}10\,$keV. The RGS instrument onboard XMM-Newton provides high-resolution spectroscopy between $0.5\mbox{--}2$\,keV. However, the RGS data were excluded in this work because of the limited number of X-ray photons caused by heavy absorption in the soft X-ray band. For our spectral modeling, we utilized data from \textit{Resolve}, \textit{Xtend}, NuSTAR/FPMA+B, and XMM-Newton/OM, while the XMM-Newton/EPIC-pn data were used only for cross-checking with \textit{Xtend}. Details of the data reduction for these observations and the treatment of the background were described in Paper I. For consistency, the \textit{Resolve} spectrum was binned to a resolution of 10\,eV, the \textit{Xtend} data were grouped to at least 20 counts per bin, and the NuSTAR/FPMA+B spectra were combined into a single spectrum with a minimum of 100 counts per bin. The combination of FPMA and FPMB was validated through their mutual consistency and shown to have no impact on the analysis. The optimal binning method \citep{2016Kaastra} was also tested, confirming that our binning choices do not introduce any systematic bias.

\section{Results}\label{sec:results}
The X-ray data analysis software SPEX \citep[v3.07.03,][]{1996Kaastra} and C-statistics \citep{1979Cash} were used for spectral analysis in this work. The energy ranges considered for spectral modeling were: \textit{Resolve} ($2\mbox{--}10$\,keV), \textit{Xtend} ($0.4\mbox{--}10$\,keV), NuSTAR/FPMA+B ($3.0\mbox{--}40$\,keV; background-dominated above 40\,keV), and XMM-Newton/OM with six UV/optical filters \citep[UVW2: $2120\mathrm{\AA}$, UVM2: $2310\mathrm{\AA}$, UVW1: $2910\mathrm{\AA}$, U: $3440\mathrm{\AA}$, B: $4500\mathrm{\AA}$, V: $5430\mathrm{\AA}$;][]{2001Mason}. To avoid the influence of the lower resolution but higher count rates of CCDs/FPMs on the detection of atomic features in \textit{Resolve}, the overlapping $5\mbox{--}10$\,keV band of NuSTAR and \textit{Xtend} was excluded from the analysis. Consistent results are obtained regardless of whether or not the NuSTAR and \textit{Xtend} data are included along with \textit{Resolve} over the 5-10 keV band (e.g. see Paper I, Extended Data Table 1). The uncertainty of parameters was estimated at $1\sigma$ (i.e., 68\% confidence level) throughout the paper. 
The flux difference between XRISM and NuSTAR spectra due to calibration was considered by a variable constant factor. The luminosity calculation in this paper was based on the following cosmological parameters: $H_\mathrm{0}=70.0\,\mathrm{km/s/Mpc}$, $\Omega_\mathrm{m}=0.3$, and $\Omega_\mathrm{\Lambda}=0.7$.

\subsection{Spectral modeling with a pre-calculated photoionization model}\label{subsec:modeling}

Following Paper I, the broadband X-ray continuum is explained by a redshifted Comptonization component modified by Galactic absorption, plus a Gaussian component for the Fe K$\alpha$ emission region. The optical and UV photometry data from XMM-Newton/OM are modeled by a disk-like blackbody component modified by Galactic extinction. We also included one soft X-ray UFO and five hard X-ray UFO components discovered in Paper I. Galactic interstellar absorption was described by the \texttt{hot} model with a column density of $N_\mathrm{H}^\mathrm{Gal}=3.1^{+0.3}_{-0.1}\times10^{21}\,\mathrm{cm}^{-2}$ and the default protosolar abundance, which is the elemental composition of the Sun when it was formed \citep{2009Lodders}. The dust reddening was modeled by \texttt{ebv} with a fixed $E(B-V)=0.45$ \citep{2011Schlafly}. The source redshift was taken into account by \texttt{reds}. The models \texttt{comt} and \texttt{dbb} accounted for the emission from the hot Comptonization component \citep[i.e. hot corona;][]{1993Haardt} and the accretion disk, respectively, where the disk temperature $kT_\mathrm{disk}$ was tied with that of seed photons $kT_\mathrm{seed}$ in \texttt{comt}. A Gaussian profile (\texttt{gaus}) is used as a phenomenological model for broad emission lines associated with the hard X-ray UFOs, while the exploration of its physics is beyond the scope of this work (A. Luminari in prep.). UFOs were initially modeled by \texttt{XABS}, a pre-calculated photoionization absorption model with the ionization balance calculated by \texttt{XABSINPUT} package assuming a SED taken from Paper I. Then they were replaced by \texttt{PION} in Section \ref{subsec:PION}. The covering factors were assumed at unity (i.e. fully covering, $C_\mathrm{F}=1$). The turbulence velocities $v_\mathrm{turb}$ of the hard X-ray UFOs were linked since their variations were not significant in the modeling with $\Delta\mathrm{C-stat}<10$ for four additional parameters. The remaining free parameters of \texttt{XABS} are column density ($N_\mathrm{H}$), ionization parameter ($\xi$), and outflow velocity ($v_\mathrm{out}$). The baseline model for our spectral fitting is: \texttt{hot*ebv*reds*6UFOs*(dbb+comt+gaus)}, where normalizations of additional components were left free.

\begin{figure*}
 \begin{center}
  \includegraphics[width=\textwidth]{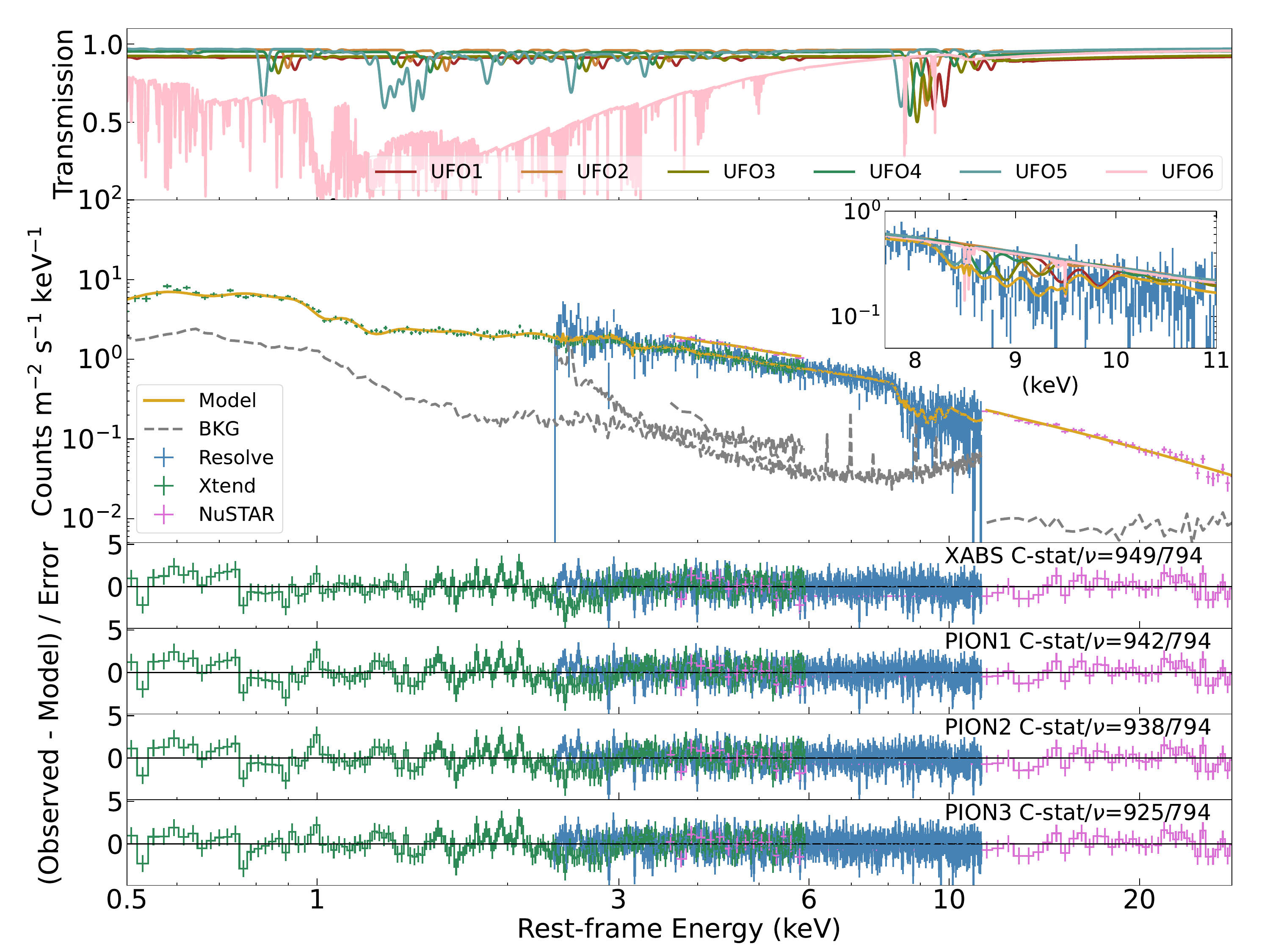} 
 \end{center}
\caption{Time-averaged \textit{Resolve} (\textit{blue}), \textit{Xtend} (\textit{green}), and NuSTAR (\textit{magenta}) spectra of PDS 456 with the best-fit model. The top panel shows the transmission of each absorption component. The middle panel presents the source spectra, background (\textit{grey}), and best-fit model (\textit{yellow}) with a zoom-in panel of the Fe-K region showing the contributions of UFOs. The lower four panels show the spectral residuals fitted with \texttt{XABS} and \texttt{PION} in different sequential combinations (see details in Section \ref{subsec:PION}), where PION3 is the combination of absorber components yielding the best fit. The differences between them mainly lie in the $<3$\,keV band.}

\label{fig:spectrum+ratio}
\end{figure*}

\begin{figure*}
 \begin{center}
  \includegraphics[width=\textwidth]{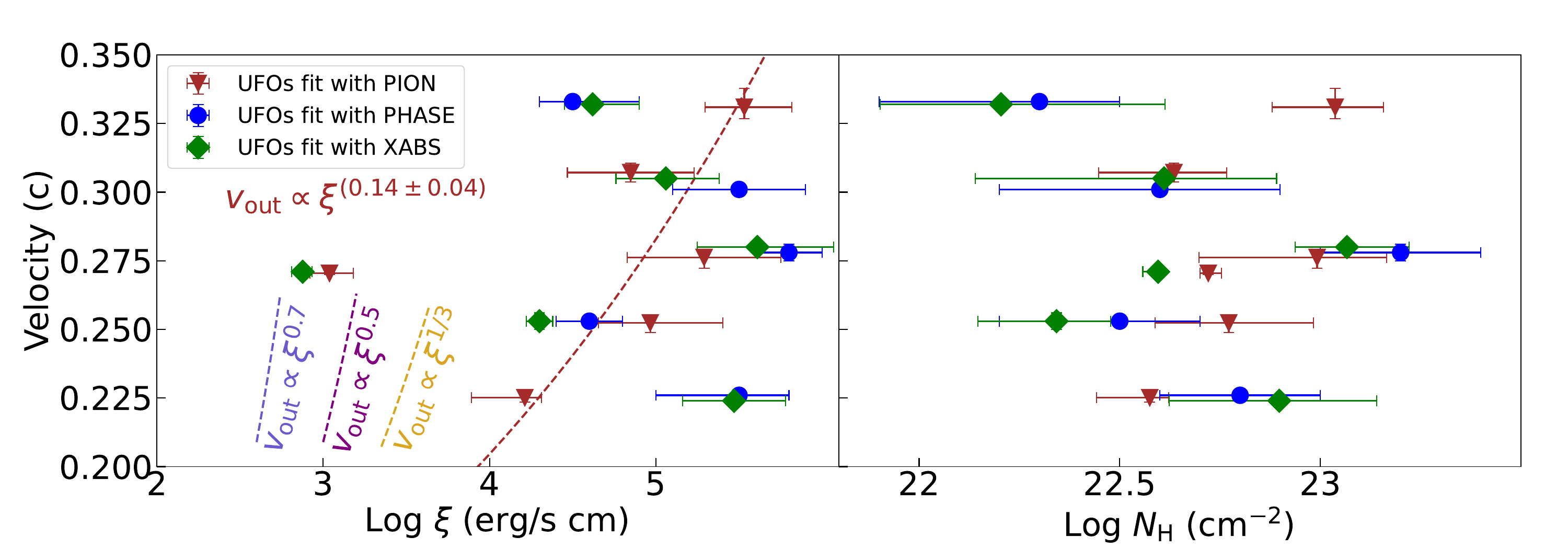}
 \end{center}
\caption{Wind velocities ($v_\mathrm{out}$) versus ionization parameters ($\xi$, left) and column densities ($N_\mathrm{H}$, right) of six UFOs derived from different photoionization codes. Results for \texttt{PION} (i.e., PION3, see Section \ref{subsubsec:pion3}) and \texttt{XABS} are taken from Table \ref{tab:fits}, while those for \texttt{PHASE} are from Table 3 in Paper I. \texttt{XABS} and \texttt{PHASE} yield consistent results within their uncertainties, whereas \texttt{PION} exhibits notable differences in the slowest and fastest hard X-ray UFOs. A power-law fit to the velocities and ionization parameters of five hard X-ray UFOs gives $v_\mathrm{out}\propto\xi^{(0.14\pm0.04)}$. For comparison, theoretical relationships predicted by MHD-driven ($v_\mathrm{out}\propto\xi^{0.7}$), momentum-conserving ($v_\mathrm{out}\propto\xi^{0.5}$), and energy-conserving ($v_\mathrm{out}\propto\xi^{1/3}$) outflows are presented (see details in Section \ref{subsec:mechanism}).
}\label{fig:parameter_comparison}
\end{figure*}

The best fit yielded an accretion disk with a temperature of $kT_\mathrm{disk}=10^{+15}_{-3}$\,eV, a hot corona ($kT_\mathrm{e}=26^{+30}_{-12}$\,keV) with an optical thickness of $\tau=0.97^{+0.88}_{-0.25}$, and a broad ($\mathrm{FWHM}=3.3^{+0.7}_{-0.6}$\,keV) Gaussian line at $7.5^{+0.4}_{-1.0}$\,keV in the rest frame. These continuum parameters remain constant within their uncertainties over changes in photoionization codes (i.e. \texttt{XABS} and \texttt{PION} models). The fitted spectra and the corresponding residuals are presented in the second and third panels of Figure \ref{fig:spectrum+ratio}. The best-fit results of the photoionization components are listed in Table \ref{tab:fits}, where the observed column densities $N_\mathrm{H,obs}$ are corrected by a factor of $(1-\beta)/(1+\beta)$, where $\beta=v_\mathrm{out}/c$, considering special relativistic effects \citep{2020Luminari}. Our \texttt{XABS} results are randomly distributed in parameter space, consistent with \texttt{PHASE} results presented in Table 3 of Paper I (see green diamonds and blue points in Fig.\ref{fig:parameter_comparison}), where both of them are pre-calculated photoionization codes.

\begin{table*}[]
  \tbl{Best fits of six UFOs in the time-averaged XRISM spectra of PDS 456, modeled with XABS and PION in different sequences (C-statistic analysis).}{%
  \begin{tabular}{cccccccc}
      \hline
      Parameters & UFO1 & UFO2 & UFO3 & UFO4 & UFO5 & UFO6 & C-stat/$\nu$\footnotemark[$\star$]  \\ 
      \hline
      \multicolumn{8}{c}{XABS}\\
      \hline
      $\log\xi$\footnotemark[a] & $4.62^{+0.28}_{-0.17}$ & $5.06^{+0.32}_{-0.30}$ & $5.61^{+0.46}_{-0.36}$ & $4.30^{+0.08}_{-0.08}$ & $5.47^{+0.31}_{-0.31}$ & $2.88^{+0.06}_{-0.07}$ & \multirow{5}{*}{$949/794$}\\
      $\log N_\mathrm{H,obs}$\footnotemark[b] & $22.20^{+0.41}_{-0.30}$ & $22.61^{+0.28}_{-0.47}$ & $23.07^{+0.16}_{-0.13}$ & $22.34^{+0.13}_{-0.19}$ & $22.90^{+0.24}_{-0.27}$ & $22.60^{+0.01}_{-0.04}$ &  \\
      $\log N_\mathrm{H,corr}$\footnotemark[c] & $22.50^{+0.41}_{-0.30}$ & $22.88^{+0.28}_{-0.47}$ & $23.32^{+0.16}_{-0.13}$ & $22.57^{+0.13}_{-0.19}$ & $23.10^{+0.24}_{-0.27}$ & $22.84^{+0.01}_{-0.04}$ &  \\
      $v_\mathrm{turb}$\footnotemark[d] & $2300^{+300}_{-300}$ & $2300$\footnotemark[t] & $2300$\footnotemark[t] & $2300$\footnotemark[t] & $2300$\footnotemark[t] & $<100$  \\
      $v_\mathrm{out}$\footnotemark[e] & $-0.330^{+0.004}_{-0.003}$ & $-0.305^{+0.002}_{-0.002}$ & $-0.280^{+0.004}_{-0.004}$ & $-0.253^{+0.002}_{-0.002}$ & $-0.224^{+0.002}_{-0.002}$ & $-0.271^{+0.002}_{-0.002}$ \\
      \hline
      \multicolumn{8}{c}{PION1\footnotemark[$\dag$]}\\
      \hline
      $\log\xi$\footnotemark[a] & $5.49^{+0.11}_{-0.08}$ & $4.86^{+0.10}_{-0.24}$ & $5.00^{+0.31}_{-0.32}$ & $4.89^{+0.37}_{-0.21}$ & $4.10^{+0.07}_{-0.07}$ & $2.89^{+0.14}_{-0.20}$ & \multirow{5}{*}{$942/794$}\\
      $\log N_\mathrm{H,obs}$\footnotemark[b] & $23.10^{+0.08}_{-0.07}$ & $22.74^{+0.13}_{-0.20}$ & $22.86^{+0.17}_{-0.13}$ & $22.74^{+0.20}_{-0.16}$ & $22.54^{+0.24}_{-0.23}$ & $22.70^{+0.01}_{-0.01}$   \\
      $\log N_\mathrm{H,corr}$\footnotemark[c] &  $23.40^{+0.08}_{-0.07}$ & $23.01^{+0.13}_{-0.20}$ & $23.11^{+0.17}_{-0.13}$ & $22.96^{+0.20}_{-0.16}$ & $22.73^{+0.24}_{-0.23}$ & $22.94^{+0.01}_{-0.01}$   \\
      $v_\mathrm{turb}$\footnotemark[d] & $2900^{+400}_{-300}$ & $2900$\footnotemark[t] & $2900$\footnotemark[t] & $2900$\footnotemark[t] & $2900$\footnotemark[t] & $<100$  \\
      $v_\mathrm{out}$\footnotemark[e] & $-0.330^{+0.003}_{-0.003}$ & $-0.304^{+0.002}_{-0.002}$ & $-0.278^{+0.002}_{-0.002}$ & $-0.253^{+0.002}_{-0.003}$ & $-0.225^{+0.002}_{-0.002}$ & $-0.270^{+0.001}_{-0.001}$ \\
      \hline
      \multicolumn{8}{c}{PION2\footnotemark[$\ddag$]}\\
      \hline
      $\log\xi$\footnotemark[a] & $5.49^{+0.14}_{-0.41}$ & $4.80^{+0.27}_{-0.31}$ & $4.99^{+0.30}_{-0.41}$ & $4.75^{+0.31}_{-0.24}$ & $4.07^{+0.18}_{-0.15}$ & $2.93^{+0.17}_{-0.12}$ & \multirow{5}{*}{$938/794$}\\
      $\log N_\mathrm{H,obs}$\footnotemark[b] & $23.05^{+0.11}_{-0.10}$ & $22.75^{+0.15}_{-0.25}$ & $22.89^{+0.19}_{-0.17}$ & $22.73^{+0.14}_{-0.11}$ & $22.57^{+0.05}_{-0.06}$ & $22.68^{+0.02}_{-0.01}$   \\
      $\log N_\mathrm{H,corr}$\footnotemark[c] & $23.35^{+0.11}_{-0.10}$ & $23.02^{+0.15}_{-0.25}$ & $23.14^{+0.19}_{-0.17}$ & $22.96^{+0.14}_{-0.11}$ & $22.77^{+0.05}_{-0.06}$ & $22.92^{+0.02}_{-0.01}$   \\
      $v_\mathrm{turb}$\footnotemark[d] & $2700^{+300}_{-200}$ & $2700$\footnotemark[t] & $2700$\footnotemark[t] & $2700$\footnotemark[t] & $2700$\footnotemark[t] & $<100$  \\
      $v_\mathrm{out}$\footnotemark[e] & $-0.330^{+0.003}_{-0.003}$ & $-0.307^{+0.002}_{-0.002}$ & $-0.277^{+0.002}_{-0.002}$ & $-0.253^{+0.002}_{-0.003}$ & $-0.226^{+0.002}_{-0.002}$ & $-0.270^{+0.001}_{-0.001}$ \\
      \hline
      \multicolumn{8}{c}{PION3\footnotemark[$\dag\dag$]}\\
      \hline
      $\log\xi$\footnotemark[a] & $5.53^{+0.28}_{-0.24}$ & $4.89^{+0.38}_{-0.38}$ & $5.29^{+0.46}_{-0.46}$ & $4.97^{+0.43}_{-0.31}$ & $4.21^{+0.11}_{-0.32}$ & $3.04^{+0.14}_{-0.12}$ & \multirow{5}{*}{$925/794$}\\
      $\log N_\mathrm{H,obs}$\footnotemark[b] & $23.04^{+0.12}_{-0.16}$ & $22.64^{+0.13}_{-0.19}$ & $22.99^{+0.17}_{-0.29}$ & $22.77^{+0.21}_{-0.18}$ & $22.57^{+0.05}_{-0.13}$ & $22.72^{+0.02}_{-0.03}$ &  \\
      $\log N_\mathrm{H,corr}$\footnotemark[c] & $23.34^{+0.12}_{-0.16}$ & $22.92^{+0.13}_{-0.19}$ & $23.24^{+0.17}_{-0.29}$ & $22.99^{+0.21}_{-0.18}$ & $22.77^{+0.05}_{-0.13}$ & $22.96^{+0.02}_{-0.03}$ &  \\
      $v_\mathrm{turb}$\footnotemark[d] & $2700^{+300}_{-200}$ & $2700$\footnotemark[t] & $2700$\footnotemark[t] & $2700$\footnotemark[t] & $2700$\footnotemark[t] & $<100$  \\
      $v_\mathrm{out}$\footnotemark[e] & $-0.331^{+0.004}_{-0.007}$ & $-0.307^{+0.003}_{-0.003}$ & $-0.276^{+0.004}_{-0.002}$ & $-0.252^{+0.003}_{-0.003}$ & $-0.225^{+0.002}_{-0.003}$ & $-0.270^{+0.001}_{-0.001}$ \\
      \hline
    \end{tabular}}\label{tab:fits}
\begin{tabnote}
\footnotemark[$\star$] The C-stat statistics and degrees of freedom.\\
\footnotemark[$\dag$]  The PION components are ordered from inner to outer layers as UFO 1-2-3-4-5-6, with faster components in the inner layers and the soft X-ray UFO in the outermost layer. \\ 
\footnotemark[$\ddag$]  The PION components are ordered from inner to outer layers as UFO 5-4-3-2-1-6, with slower components in the inner layers, except for the outermost soft X-ray UFO.  \\ 
\footnotemark[$\dag\dag$]  The PION components are ordered from inner to outer as UFO 5-4-6-3-2-1, with slower components in the inner layers, except for the soft X-ray UFO in the third layer. \\ 
\footnotemark[a]  Ionization parameter in units of erg cm $\mathrm{s}^{-1}$. \\ 
\footnotemark[b]  Observed column density in units of $\mathrm{cm}^{-2}$. \\ 
\footnotemark[c]  Intrinsic column density in units of $\mathrm{cm}^{-2}$ (i.e. corrected for the relativistic effects). \\ 
\footnotemark[d]  Turbulence velocity in units of km s$^{-1}$.\\ 
\footnotemark[e]  Outflow velocity in units of the speed of light $c$. Negative values refer to outflowing gas. \\ 
\footnotemark[t]  The parameter is tied. \\ 
\end{tabnote}
\end{table*}

\subsection{Self-consistently calculated photoionization modeling}\label{subsec:PION}

In this section, we replaced the \texttt{XABS} components in the baseline model with the self-consistently calculated photoionization code \texttt{PION}, where we only applied it to absorption components assuming that absorbers fully cover the X-ray source along our LOS ($C_\mathrm{F}=1$ and opening angle $\Omega=0$). In SPEX, the order of any multiplicative model components is adjustable and can affect the irradiating luminosity on the outer components. Unlike \texttt{XABS}, which assumes a pre-calculated constant ionization balance, \texttt{PION} dynamically computes the ionization balance based on the irradiating field during spectral fitting, significantly increasing the computational time. The ionization parameters of \texttt{PION} may vary depending on the sequence of components if an inner \texttt{PION} component significantly alters the SED. To account for this, we systematically explore the possible sequential combinations of \texttt{PION} in this section. For clarity, UFOs were labeled UFO1, UFO2,..., UFO5 in decreasing order of velocity (with UFO1 being the fastest), while the soft X-ray UFO was referred to as UFO6, given its low ionization state despite having a velocity comparable to that of the other UFOs.

\subsubsection{PION1: inner-when-faster hard X-ray UFOs}\label{subsubsec:pion1}

\begin{figure*}[!t]
 \begin{center}
  \includegraphics[width=0.49\textwidth]{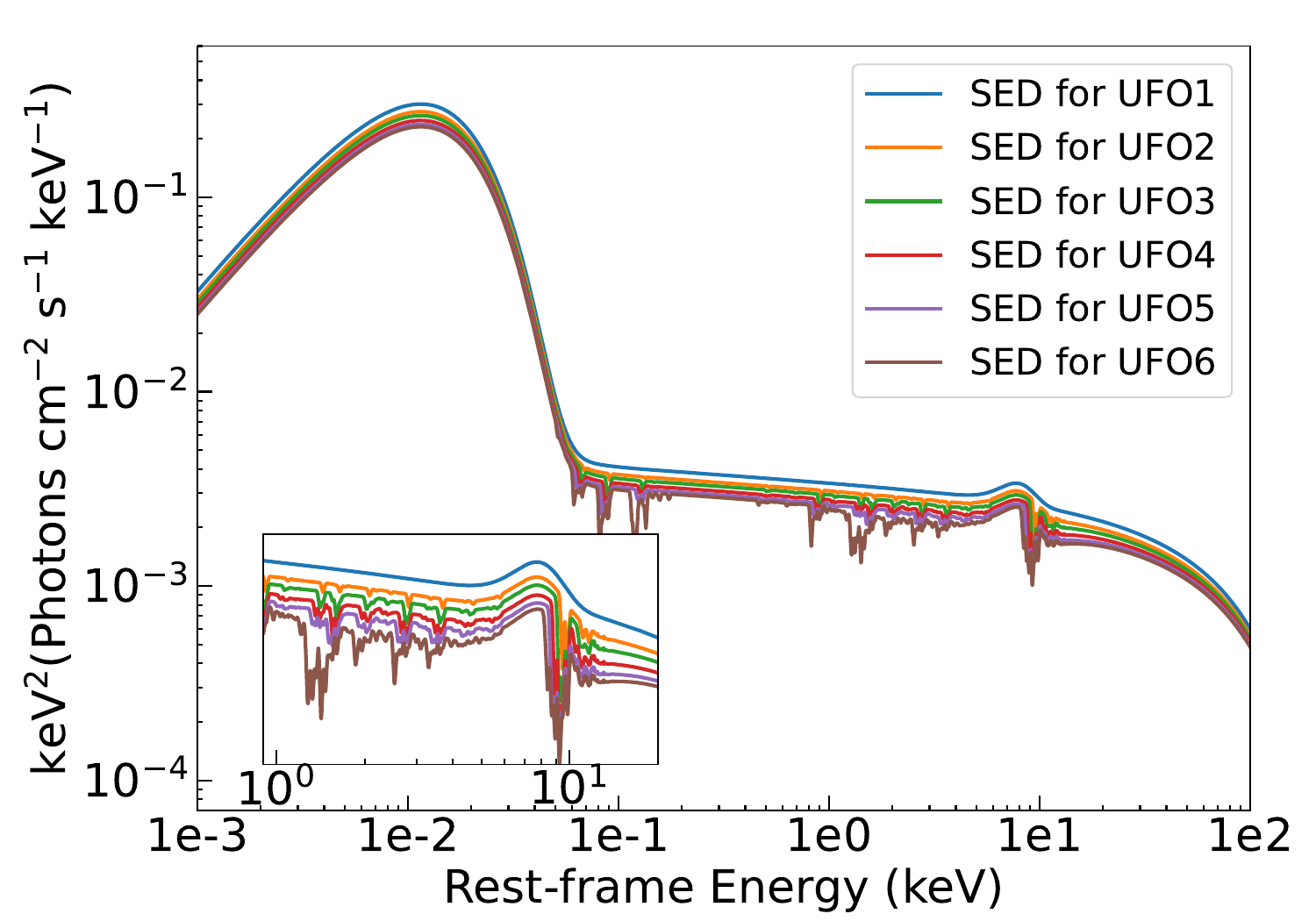} 
  \includegraphics[width=0.49\textwidth]{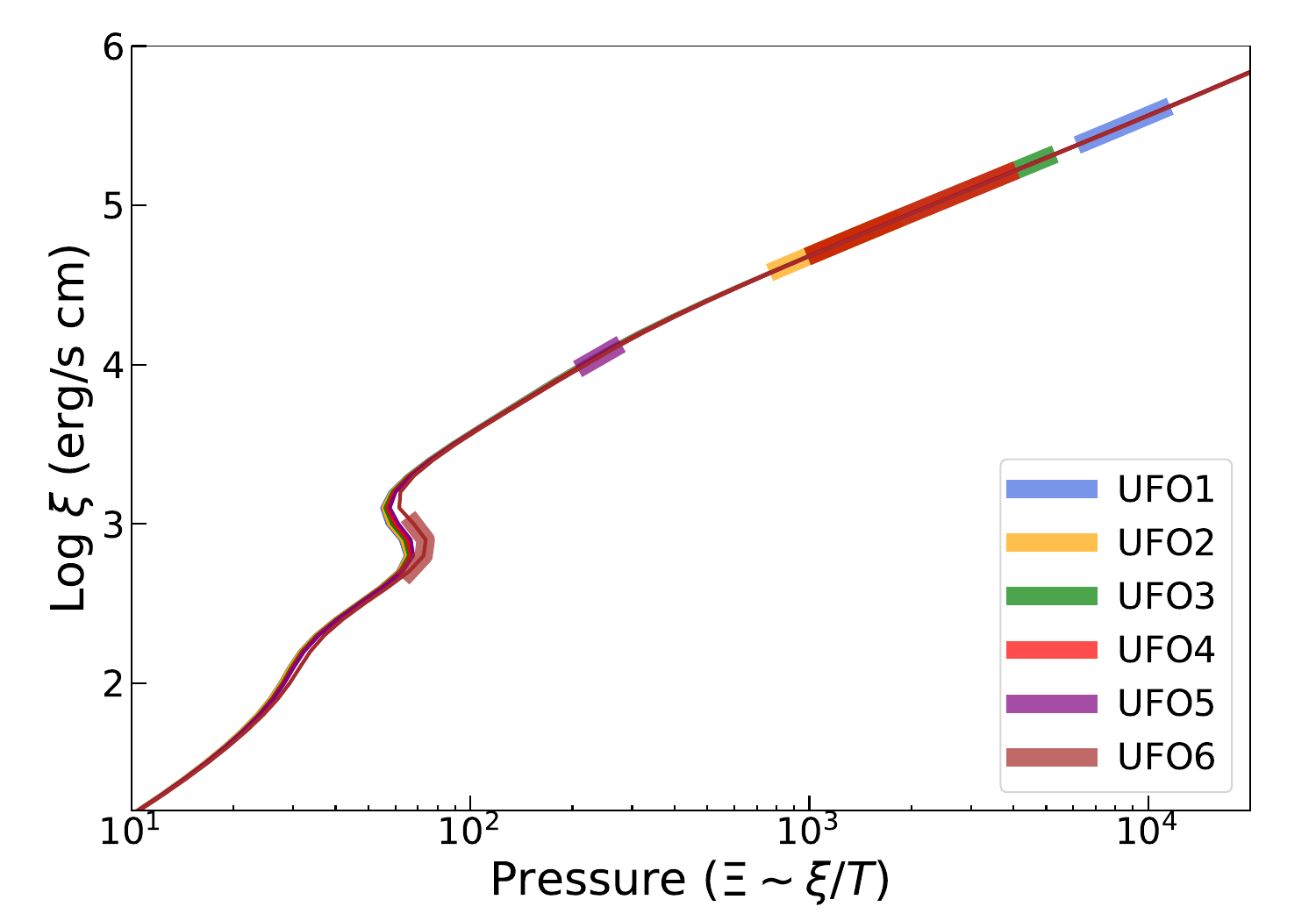} 
 \end{center}
\caption{The irradiating SEDs (left) and corresponding S-curves (i.e. stability curves, right) for each UFO layer in the order of PION1. The inset box in the left panel shows a zoom-in of the region where the SEDs differ most significantly. UFO solutions are overlaid on the S-curves, showing that the hard X-ray UFOs (UFO1–5) are thermally stable, while the soft X-ray UFO (UFO6) is situated in a thermally unstable region.
}
\label{fig:PION1_SED-S-curve}
\end{figure*}

We initially assumed that faster UFOs are closer to the SMBH, aligning with simple MHD-driven models and relations between UFOs and WAs on larger scales \citep{2010Fukumura,2013Tombesi,2024Gianolli,2024Yamada}. Under this assumption, the order from the innermost to the outermost layer is UFO 1-2-3-4-5-6, with UFO6 placed at the outermost layer due to its considerably lower ionization state compared to the other UFOs. This sequential combination is referred to as `PION1'.

The best-fit parameters of PION1 and its residuals are presented in Table \ref{tab:fits} and Figure \ref{fig:spectrum+ratio}, respectively. The statistical difference between PION1 and \texttt{XABS} fits is $\Delta \mathrm{C-stat}=7$, indicating that \texttt{PION} can achieve a better fit than pre-calculated codes. The improvements mainly lie in the region below 3\,keV (e.g. $\sim2.5$\,keV residuals in Figure \ref{fig:spectrum+ratio}). The ionization parameters of UFO1 and UFO5 differ significantly ($>2\mbox{--}3\sigma$ difference, details discussed in Section \ref{subsec:systematics}), while those of the other UFOs remain consistent within uncertainties, as illustrated in Figure \ref{fig:parameter_comparison} (noting that it displays results from PION3, but the \texttt{PION} parameters remain stable in all scenarios). 
Replacing \texttt{XABS} with \texttt{PION} caused the previously scattered properties of the hard X-ray UFOs to be more ordered (also confirmed in the later Bayesian analysis; see Section \ref{subsubsec:Bayesian}), revealing a stratified structure where faster UFOs possess higher ionization states. Coupled with their sequence, this suggests that UFOs slow down and their ionization states decrease as they move away from the X-ray source if they belong to the same outflowing stream. Alternatively, if the UFOs along our LOS originated from different streams, it means that the outer stream is slower and lower ionized.

The irradiating SED of each UFO layer is shown on the left panel of Figure \ref{fig:PION1_SED-S-curve}. The shapes of illuminating SEDs for UFO1-5 are similar, but look very different for UFO6 (i.e. UFO5 absorbs more line photons than UFO1-4). It demonstrates that the screening effect is negligible for highly ionized UFOs since most elements are over-ionized, and the impacts of relatively low ionized UFOs are non-negligible. The right panel of Figure \ref{fig:PION1_SED-S-curve} presents corresponding S-curves (i.e. stability curves), which is the plasma ionization parameter as a function of the pressure ratio $\Xi$. The pressure ratio is defined as the ratio between the radiation pressure ($F_\mathrm{ion}/c$) and thermal pressure
($n_\mathrm{H}k_\mathrm{B}T$), $\Xi=\frac{F_\mathrm{ion}}{n_\mathrm{H}ck_\mathrm{B}T}\propto\xi/T$ \citep{1981Krolik}. The plasma temperature $T$ corresponding to a given ionization parameter $\xi$ is self-consistently calculated in \texttt{PION}, given an input SED. The S-curve illustrates the heating-cooling balance of a plasma irradiated by a given SED, where branches with a positive gradient correspond to a thermally stable plasma and those with a negative gradient indicate a thermally unstable plasma. The UFO solutions are overlaid on the S-curves, indicating five thermally stable UFOs (UFO1-5) and one unstable UFO (UFO6) in PDS 456.


\subsubsection{PION2: inner-when-slower hard X-ray UFOs}\label{subsubsec:pion2}

To investigate the influence of the \texttt{PION} sequence on the spectroscopy, we thoroughly analyzed all 120 possible sequential combinations of the five hard X-ray UFOs, assuming the location of soft X-ray UFO (UFO6) is at the outermost layer. For each combination, the spectral fitting was initialized with $N_\mathrm{H}=5\times10^{22}\,\mathrm{cm}^{-2}$ and $\log\xi=5$ for each hard X-ray \texttt{PION} component. The initial values for turbulence velocity, outflow velocity, and UFO6 were set to their best-fit values from PION1.

\begin{figure*}
 \begin{center}
  \includegraphics[width=\textwidth]{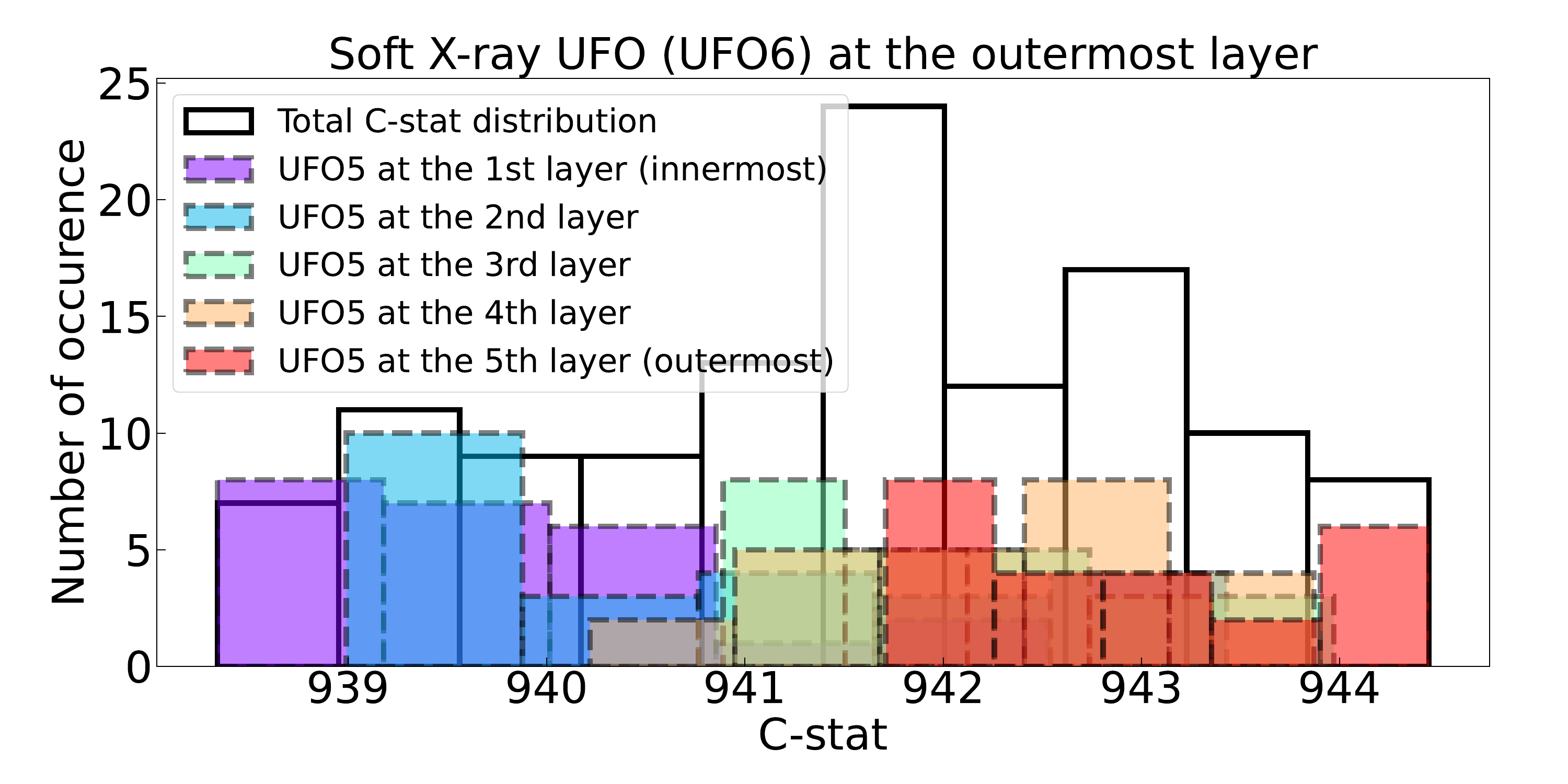} 
 \end{center}
\caption{
C-stat distribution of spectral fits with UFO5 placed at different layers (coded by colors), while the soft X-ray UFO (UFO6) is fixed at the outermost layer. It suggests that UFO5 prefers to situate at the innermost layer, with a clear disfavor for the outer layers.
}\label{fig:histogram-soft-outermost}
\end{figure*}

The C-stat distributions of fits with individual UFO components placed at different layers are presented in Figure \ref{fig:histogram-soft-outermost} (UFO5) and \ref{app:fig:histogram-soft-outermost} (UFO1-4). The maximal statistical difference among sequential combinations is $\Delta\mathrm{C-stat}=6$. Unlike UFO1-4, which are insensitive to their positions, UFO5 (the slowest hard X-ray UFO) displays a stratified C-stat distribution for various layers (Figure \ref{fig:histogram-soft-outermost}). The scenario where UFO5 lies at the innermost layer is tentatively preferred over those with UFO5 at the outermost layer by $\Delta$C-stat$=4$ without changes in degrees of freedom.
Based on this indication, we propose a stratified structure contrary to PION1, where slower UFOs are closer to the SMBH (i.e., UFO 5-4-3-2-1-6), referred to as `PION2', although the exact order of components UFO1-4 cannot be solely determined by the statistical quality of the fit.

The best-fit parameters of PION2 and residuals are presented in Table \ref{tab:fits} and Figure \ref{fig:spectrum+ratio}. The parameters of UFOs in PION2 remain consistent with those in PION1, with tiny improvements in soft X-rays.
In the PION2 scenario, hard X-ray UFOs appear to be accelerated and become more ionized as they move away from the SMBH. However, this indication only relies on the preference of UFO5 at the innermost layer with a moderate statistical improvement of $\Delta\mathrm{C-stat}=4$. The remaining hard X-ray UFOs are insensitive to their locations, limiting the justification of this scenario. This is consistent with our expectations because their high ionization states ($\log\xi>4.5$) have relatively small influence on the incident SED upon each layer.

\subsubsection{PION3: inner-when-slower UFOs}\label{subsubsec:pion3}

Then we incorporated the soft X-ray UFO (UFO6) into our analysis of sequential combinations. By re-performing the same analysis described in Section \ref{subsubsec:pion2}, we examined all possible order permutations of the six UFOs, resulting in a total of 720 combinations.

\begin{figure*}
 \begin{center}
  \includegraphics[width=0.49\textwidth]{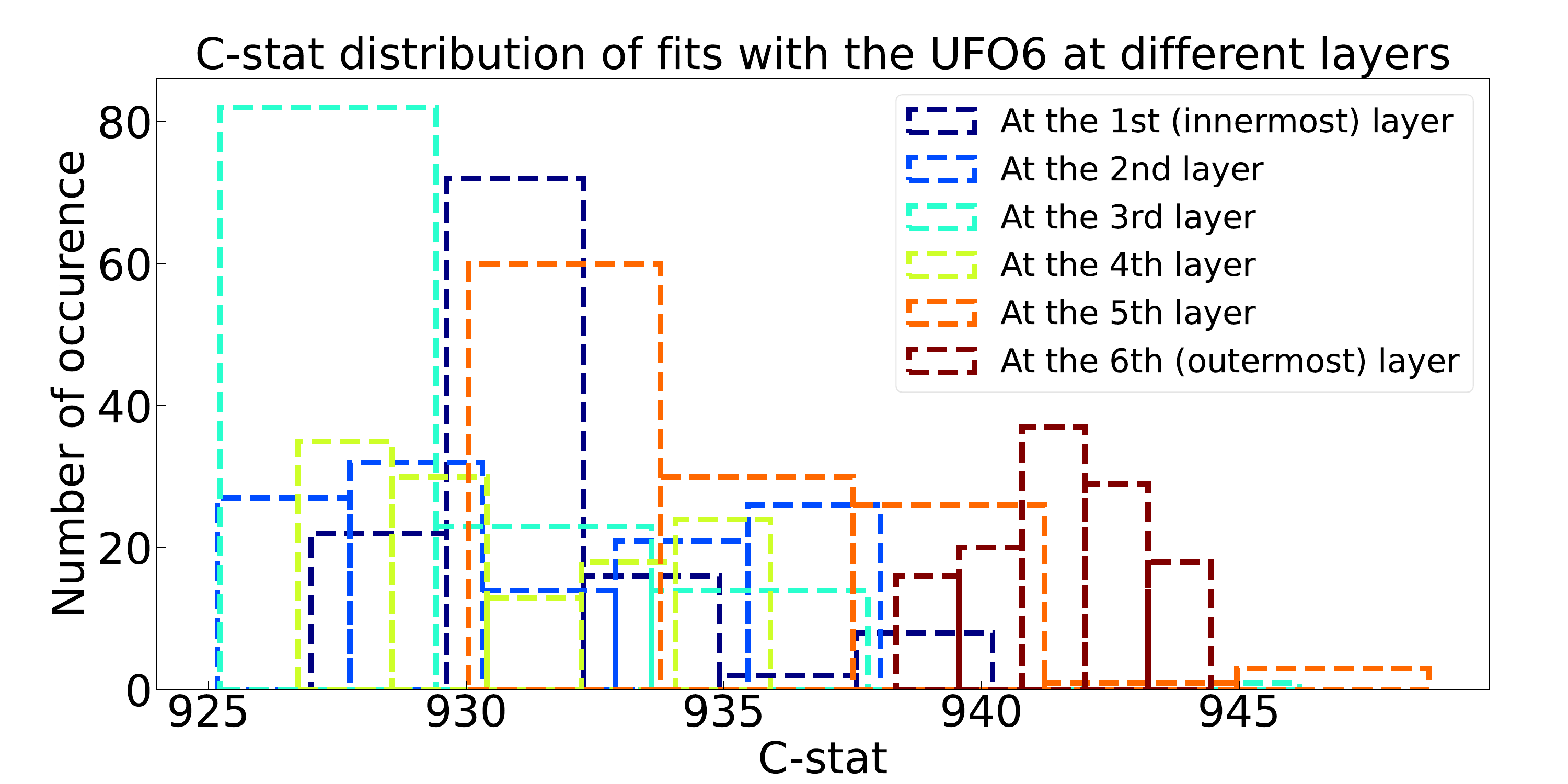} 
  \includegraphics[width=0.49\textwidth]{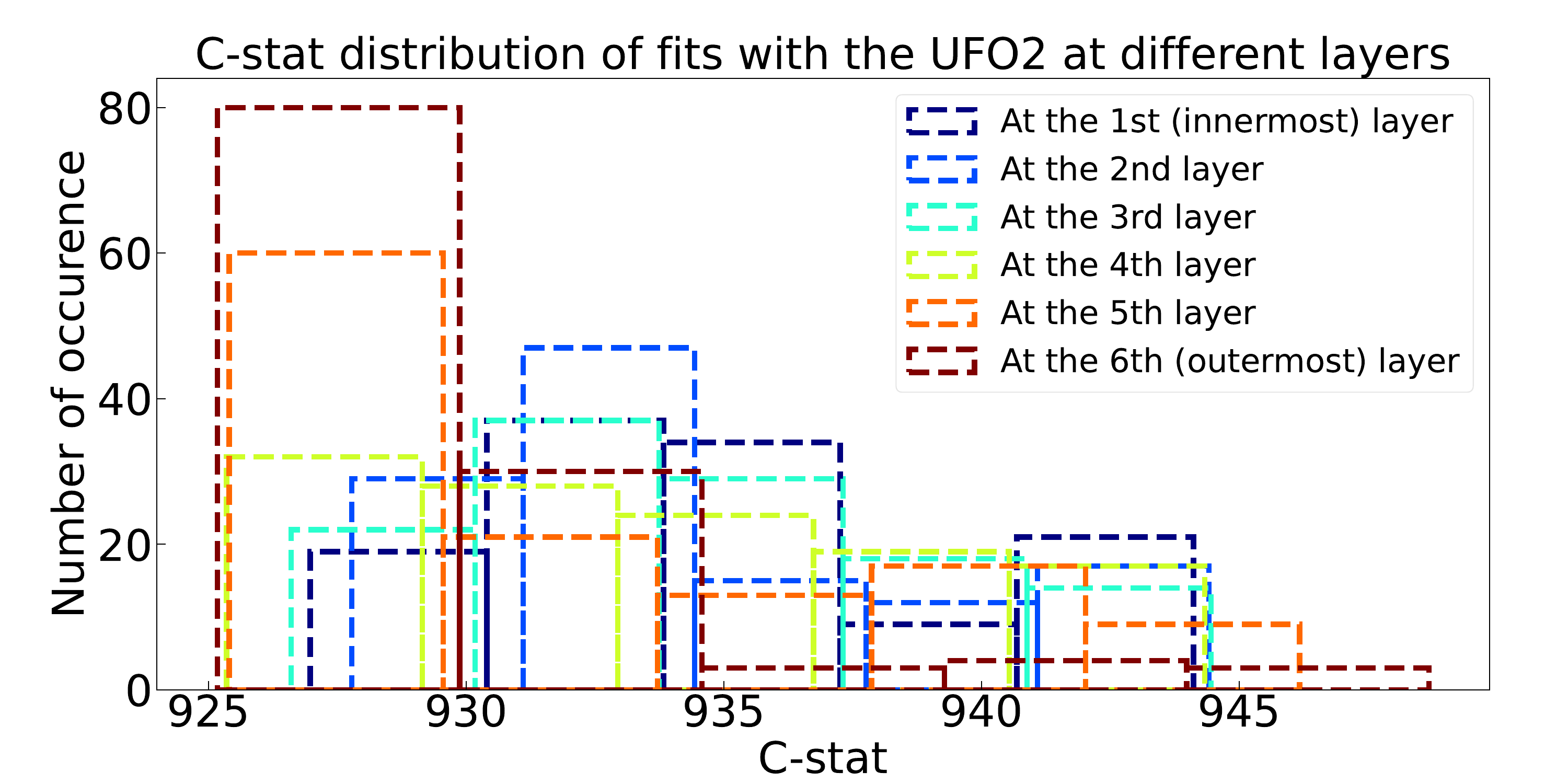} 
 \end{center}
\caption{Similar to Figure \ref{fig:histogram-soft-outermost} but the location of UFO6 is modifiable. C-stat distribution of fits with UFO6 (left) and UFO2 (right) placed at different layers. They indicate that UFO6 is most likely to reside in the third layer, while UFO2 shows a higher probability of being located in the outer layers (fifth or sixth layer). 
}\label{fig:histogram-all}
\end{figure*}

The C-stat distributions of fits with individual UFOs at various layers are shown in Figure \ref{fig:histogram-all} (UFO6 and UFO2) and \ref{app:fig:histogram-all} (other UFOs). Including UFO6 increases the maximal statistical difference between different possibilities to $\Delta\mathrm{C-stat}=22$, highlighting the significance of UFO orders. Unlike PION2, no specific sequential combination emerges as statistically superior to the others. For instance, in the case of UFO6 (see the left panel of Figure \ref{fig:histogram-all}), the best-fit statistics occur simultaneously when it is placed at the second and third layers, with a $\Delta\mathrm{C-stat}=2$ improvement over the second-best position. Given this statistical ambiguity, we determined UFO location preferences based on their probability distributions rather than fit statistics alone, identifying the layer with the highest occurrence within the best C-stat bin. Therefore, we found that UFO6 is most likely to reside in the third layer, UFO2 tends to be placed in the outer layers (the fifth or sixth layer), and UFO5 maintains its preference for the innermost layer, while the other UFOs remain insensitive to their positions. Any sequential combination satisfying these three conditions results in at most a $\Delta\mathrm{C-stat}=2$ difference. As a representative example, we showed details of the fit with the order of UFO 5-4-6-3-2-1, referred to as `PION3'.

\begin{figure}
 \begin{center}
  \includegraphics[width=0.49\textwidth]{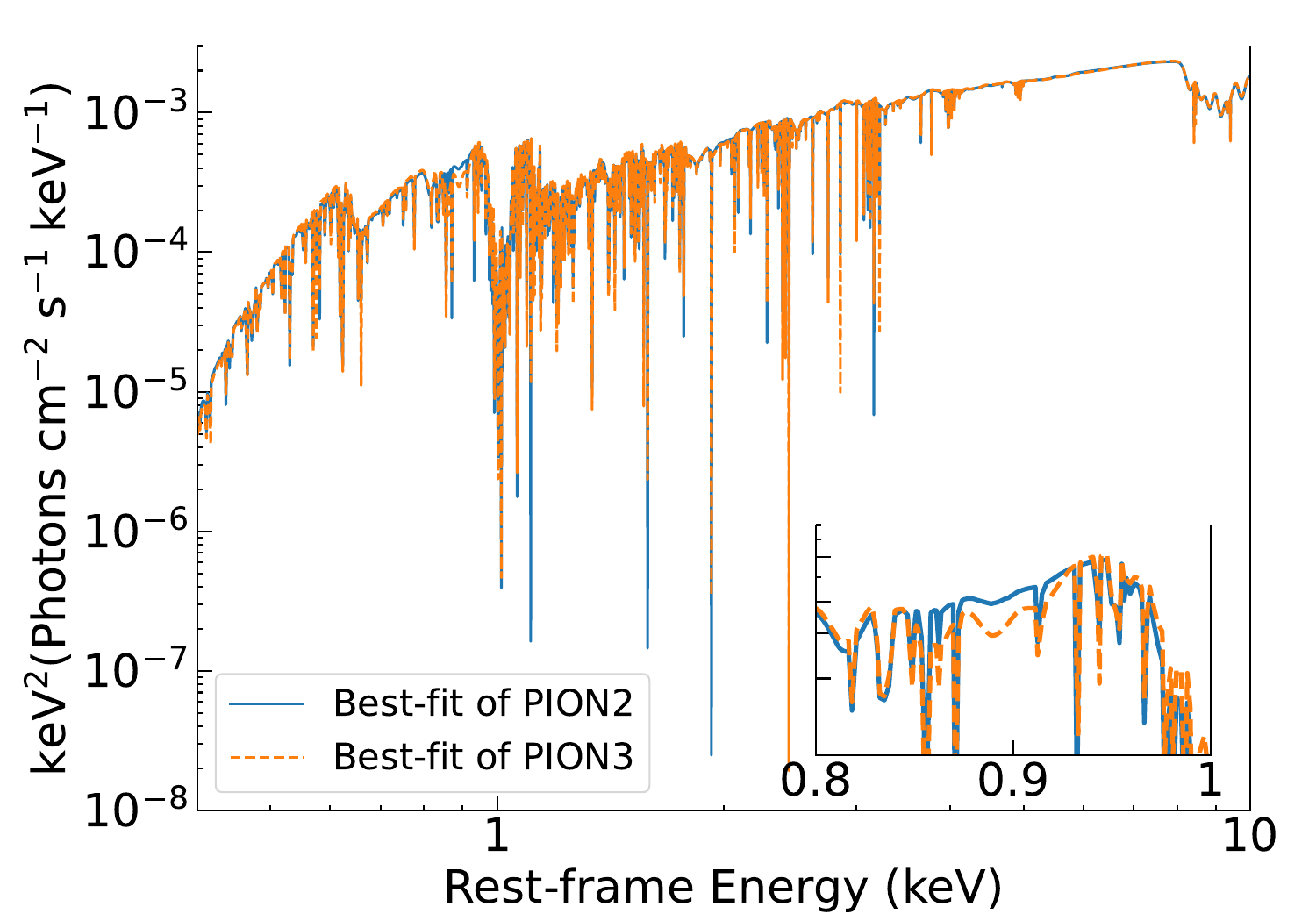} 
 \end{center}
\caption{Comparison between the best-fit models of PION2 (solid) and PION3 (dashed), with the zoom-in plot of the visible difference located at the $0.8\mbox{--}0.9$\,keV band in the rest frame.
}\label{fig:PION2-PION3-comparison}
\end{figure}

\begin{figure*}
 \begin{center}
  \includegraphics[width=0.49\textwidth]{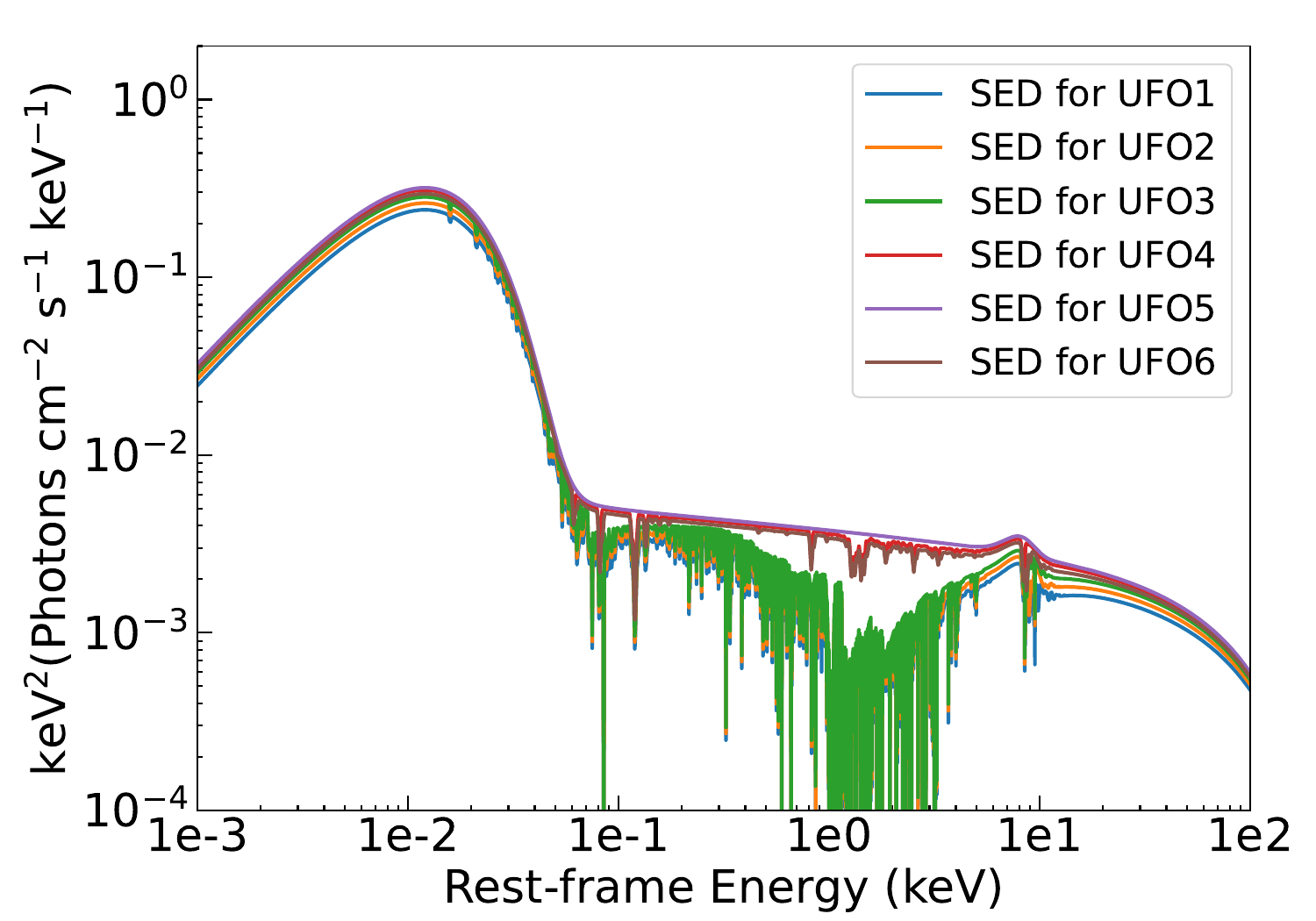} 
  \includegraphics[width=0.49\textwidth]{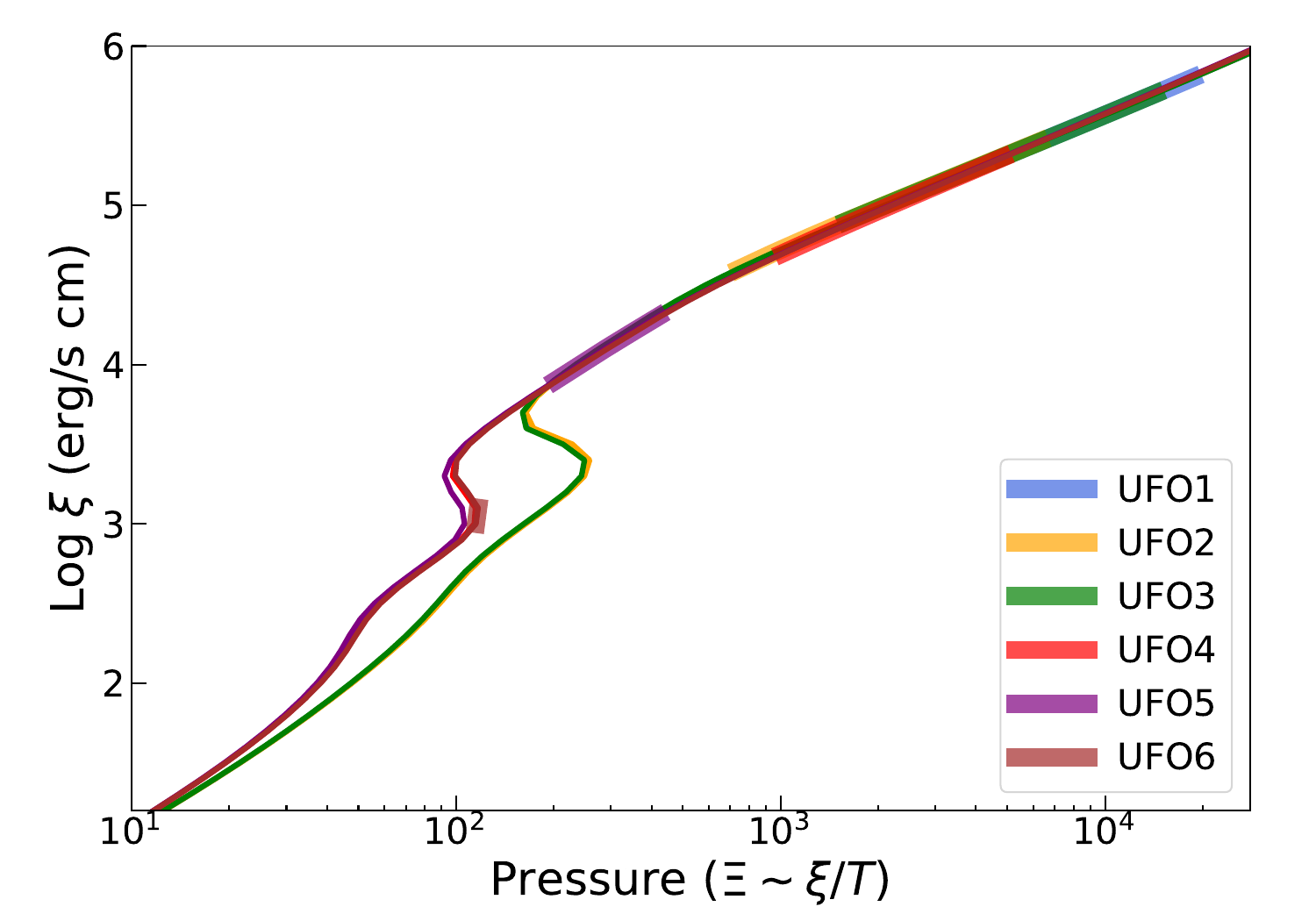} 
 \end{center}
\caption{Similar to Figure \ref{fig:PION1_SED-S-curve} but UFOs are in the `PION3' order. Although UFO6 at the third layer affects the irradiating SED and stability curves of the outer UFOs, it remains that UFO1-5 are thermally stable and UFO6 is thermally unstable.
}\label{fig:PION3_SED-S-curve}
\end{figure*}

The top panel of Figure \ref{fig:spectrum+ratio} shows the transmission of different \texttt{PION} components in the PION3 order, along with the best-fit model and its residuals in the second and sixth panels. The corresponding best-fit parameters are listed in Table \ref{tab:fits} and illustrated in Figure \ref{fig:parameter_comparison}. Compared with PION2, the UFO properties remain comparable within uncertainties, but the modified order yields a statistical improvement of $\Delta\mathrm{C-stat}=13$. Most of this improvement arises from better modeling of the soft X-ray band of the \textit{Xtend} data, particularly for the absorption residuals within the $0.8\mbox{--}0.9$\,keV range (see lower three panels of Figure \ref{fig:spectrum+ratio} and Figure \ref{fig:PION2-PION3-comparison}), which are attributed to O VIII ions. This enhancement is driven by the placement of UFO6, a moderately ionized UFO, which substantially contributes to the absorption of soft X-ray photons. As its transmission illustrated, UFO6 involves numerous ion transitions that are especially sensitive to photons within narrow energy bands, emphasizing the importance of its placement in the outflow structure.

The irradiating SEDs of each UFO components and corresponding stability curves are presented in Figure \ref{fig:PION3_SED-S-curve}. Although the soft X-ray UFO placed at the third layer significantly affects the irradiating SEDs and S-curves of the outer UFOs, the PION3 sequence does not change the conclusion that UFO6 remains in the thermally unstable region and the other UFOs are thermally stable. Therefore, regardless of the order combination, the soft X-ray UFO (UFO6) is thermally unstable in PDS 456, since the other UFOs are highly ionized and will not significantly affect the irradiating SED on UFO6, leaving it on the unstable branch. In this regime, small perturbations in temperature are amplified rather than damped. Physically, this implies that the outflowing gas cannot stably exist at those specific ionization parameters, potentially leading to the fragmentation of the outflow into multiple phases and influencing both the dynamics and detectability of the wind. This is consistent with archival observations that the soft X-ray UFO is more variable on timescales of less than one month \citep[e.g.,][]{2020Reeves}.

Considering the tentative position preferences of UFO6, UFO5, and UFO2, along with their properties, particularly the comparable velocities of UFO6 and UFO3, we propose that the UFOs in PDS 456 might represent velocity-stratified outflows (i.e. PION3). In this scenario, faster outflows are located in the outer layers, suggesting that UFOs undergo acceleration as they propagate outward.

\subsection{Simulations of future missions}\label{subsec:simulations}

Although the coordinated X-ray observations suggested a preferred order for the multiple \texttt{PION} components, the lack of resolved soft X-ray lines limits the ability to statistically determine the exact position of each component. Therefore, we explored whether a more definitive sequence of multiple photoionization components could be identified in the future. We considered XRISM/\textit{Resolve} in the gate valve open (GVO) configuration (if GV can be open in the AO-2 cycle) and the future X-ray mission, NewAthena \citep{2025Cruise}, with high spectral resolution covering the soft X-ray band.

\begin{figure*}
 \begin{center}
  \includegraphics[width=0.49\textwidth]{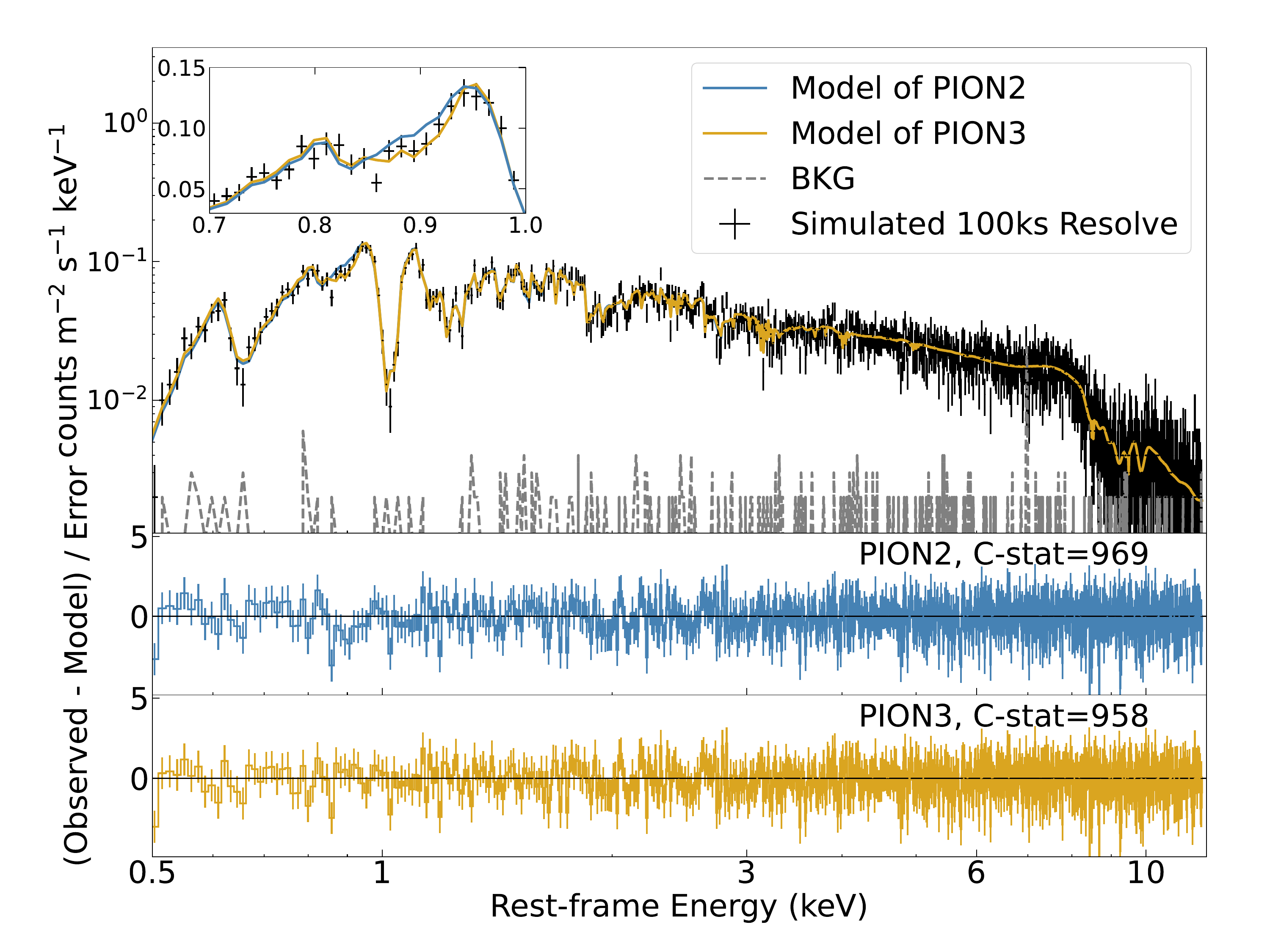}
  \includegraphics[width=0.49\textwidth]{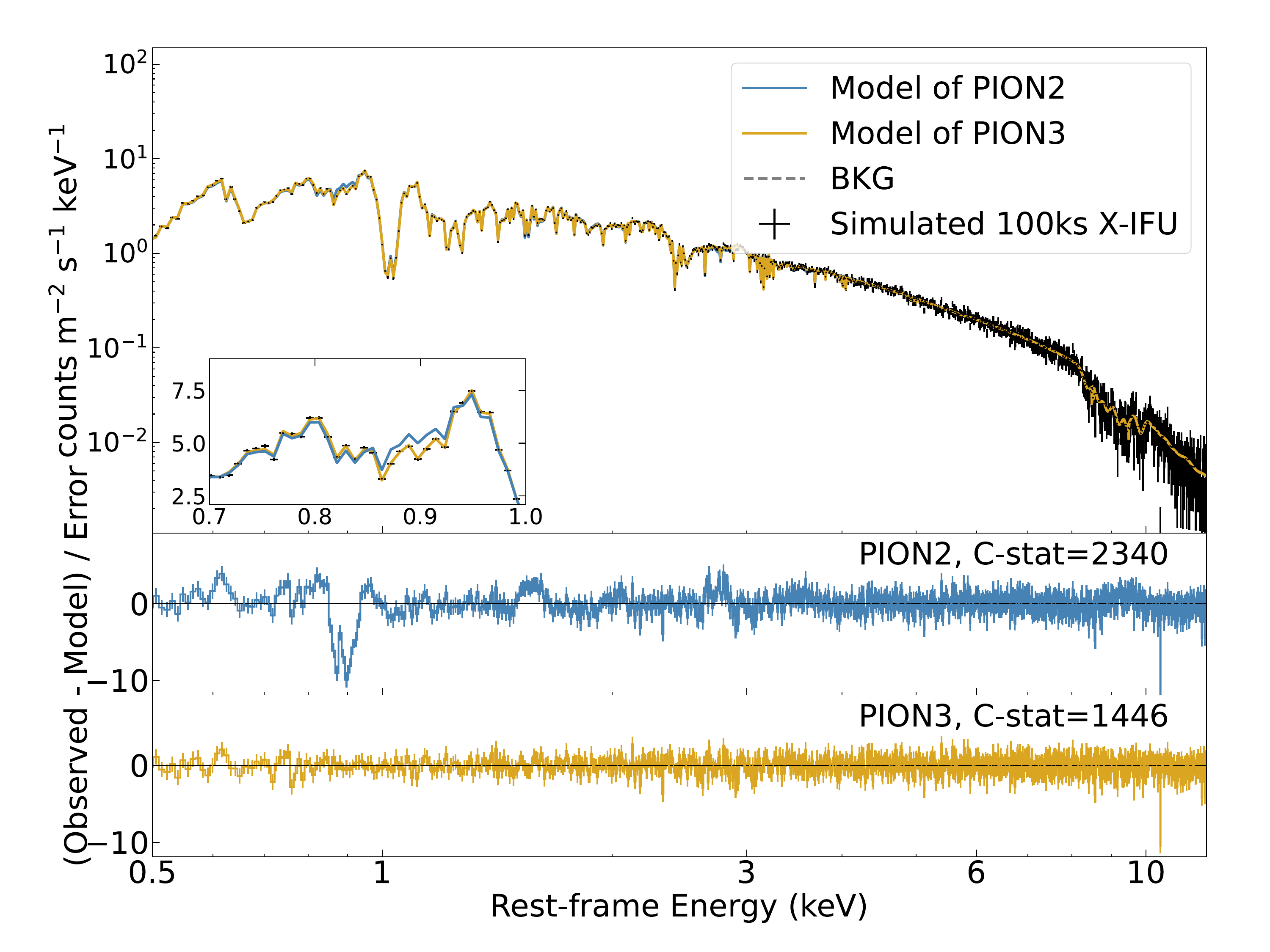} 
 \end{center}
\caption{Simulated 100\,ks-exposure XRISM/\textit{Resolve} with an open Gate Valve (GVO, left) and NewAthena/\textit{X-IFU} (right) spectrum (top) based on the best fit of PION3. The inset panel highlights the $0.7\mbox{--}1$\,keV band. Spectra are binned for clarity. The corresponding residuals against the PION2 (middle) and PION3 (bottom) models illustrate that XRISM/\textit{Resolve} (GVO) can achieve a statistical difference between PION2 and PION3 of $\Delta\mathrm{C-stat}\sim11$, comparable to XRISM/\textit{Xtend} but with resolved features. NewAthena further distinguishes between sequential combinations of absorbers with high statistical significance ($\Delta\mathrm{C-stat}\sim900$). NuSTAR and XMM-Newton/OM data were included in the fit but not shown for clarity.
}
\label{fig:simulations}
\end{figure*}

We first simulated a 100\,ks-exposure XRISM/\textit{Resolve} (GVO) spectrum based on the best-fit parameters of PION3 using the \texttt{simulate} task in SPEX. The simulated spectra and the corresponding residuals, compared against the PION2 and PION3 models, are shown in the left panel of Figure \ref{fig:simulations}. Due to a higher spectral resolution but a lower effective area of \textit{Resolve} than \textit{Xtend}, the GVO spectrum can achieve a statistical difference between PION2 and PION3 sequences of $\Delta\mathrm{C-stat}\sim11$, comparable to that from \textit{Xtend}.

A similar detailed analysis of sequential combinations was carried out using the simulated spectrum, incorporating NuSTAR and XMM-Newton/OM data to constrain the SED. Figure \ref{app:fig:histogram-xrism} presents the C-stat distributions for each UFO at different layers with a maximal statistical difference of $\Delta\mathrm{C-stat}\sim35$ among various sequential combinations. The results are similar to those obtained from the \textit{Xtend} spectrum, where UFO2, UFO5, and UFO6 show tentative preferences for the outer, innermost, and third layers, respectively, while the remaining UFOs are insensitive to their positions. However, with resolved soft X-ray lines enabled by the open gate valve, the soft X-ray absorber (UFO6) statistically disfavors the placement in the innermost or outermost layer ($\Delta\mathrm{C-stat}>10$). It suggests that XRISM with GVO can improve the reliability of our results.


Then we simulated a 100\,ks NewAthena/\textit{X-IFU} \citep{2023Barret} spectrum and conducted the same analysis of sequential combinations with results shown in the right panel of Figure \ref{fig:simulations} and Figure \ref{app:fig:histogram-athena}, separately. While the best-fit parameters of PION2 and PION3 are comparable, different sequences result in a statistical difference of $\Delta\mathrm{C-stat}\sim900$. The statistical difference among various sequential combinations can reach up to $\Delta\mathrm{C-stat}\sim1750$. With a 100\,ks NewAthena exposure, the position of UFO6 can be statistically constrained to the third layer ($\Delta\mathrm{C-stat}\sim100$ better than the other layers). For the other UFOs, although NewAthena may not fully resolve degeneracies in their positions due to similar ionization states, our simulation indicates that UFO4-5 are more likely to occupy the inner two layers, while UFO1-3 tend to be located in the outer three layers ($\Delta\mathrm{C-stat}\sim100\mbox{--}200$). Therefore, with its high spectral resolution and large effective area, NewAthena can significantly improve our ability to determine the order of multiple UFO components, providing insights into the structure of winds in AGNs.


\section{Discussions}\label{sec:discussion}
Using photoionization modeling of XRISM observations, coordinated with XMM-Newton and NuSTAR spectra of PDS 456, we derived UFO parameters with the self-consistent photoionization model \texttt{PION}. These results differ from those obtained using pre-calculated photoionization codes. The advanced capabilities of \texttt{PION} also allowed us to investigate the order of multiple photoionization components. We identified a tentative order preference, indicating that slower UFOs are located closer to the central SMBH.

\subsection{Systematics}\label{subsec:systematics}
In this subsection, we examine several potential systematic effects before discussing our results, including the origin of differences between \texttt{XABS} and \texttt{PION} results, the existence of residuals driving the sequence, the non-solar element abundance, the fully covering assumption, and the validation of order preferences.

\subsubsection{Differences between \texttt{XABS} and \texttt{PION}}\label{subsubsec:xabs-vs-pion}
\begin{figure*}
 \begin{center}
  \includegraphics[trim={0 50 0 50}, width=\textwidth]{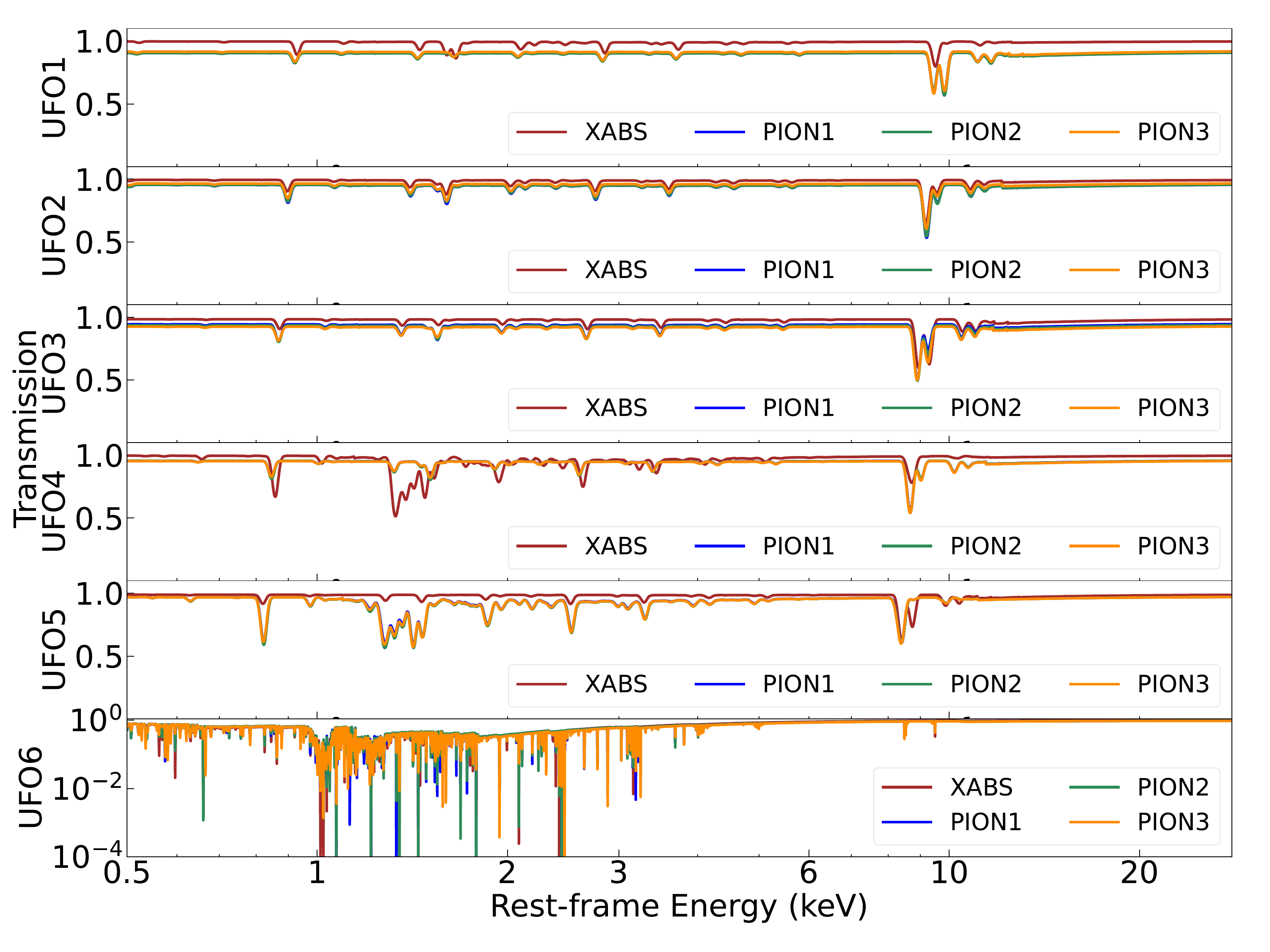}  
 \end{center}
\caption{The transmission of UFO components (from top to bottom: UFO1 to UFO6) in different order combinations. The main differences between \texttt{XABS} and \texttt{PION} lie in the modeling of UFO1, UFO4, and UFO5 in the Fe-L and Fe-K bands, as well as UFO6 in the $<3$\,keV range.
} \label{fig:transmission}
\end{figure*}

Figure \ref{fig:transmission} illustrates the transmission of UFO components in different order combinations. The main differences between \texttt{XABS} and \texttt{PION} arise from the modeling of UFO1, UFO4, and UFO5 in the Fe-L and Fe-K bands, as well as UFO6 in the $<3$\,keV range.
In the \texttt{XABS} case, UFO4 accounts for the Fe-L region, whereas in the \texttt{PION} cases, this feature is instead modeled by UFO5. Regardless of the sequence, \texttt{PION} always reinforces UFO1 and UFO5 to have the highest and lowest ionization states among Fe-K UFOs, while the best-fit solution of \texttt{XABS} requires a lower ionized UFO1 and a higher ionized UFO5.

These discrepancies should originate from differences in the underlying algorithms and the atomic data employed by the two models. \texttt{XABS} components are irradiated by the same intrinsic SED, whereas each \texttt{PION} component can be obscured by foreground absorbers or influenced by those behind it. Therefore, spectral fitting with multiple \texttt{PION} components is treated as an entire self-consistent procedure.
Additionally, the atomic database used in \texttt{PION} is more up to date than that used in \texttt{XABS} (via private communication with Dr. Jelle de Plaa)\footnote{The Fe-L atomic data in SPEX was updated in 2020 for version 3.06.00, based on \citet{2019Gu}; see the SPEX changelog.}. These differences in atomic data can accumulate and lead to noticeable discrepancies between the \texttt{PION} and \texttt{XABS} model results.


\subsubsection{Existence of residuals driving the sequence}\label{subsubsec:xabs-vs-pion}

\begin{figure}
 \begin{center}
  \includegraphics[trim={0 50 0 50}, width=0.49\textwidth]{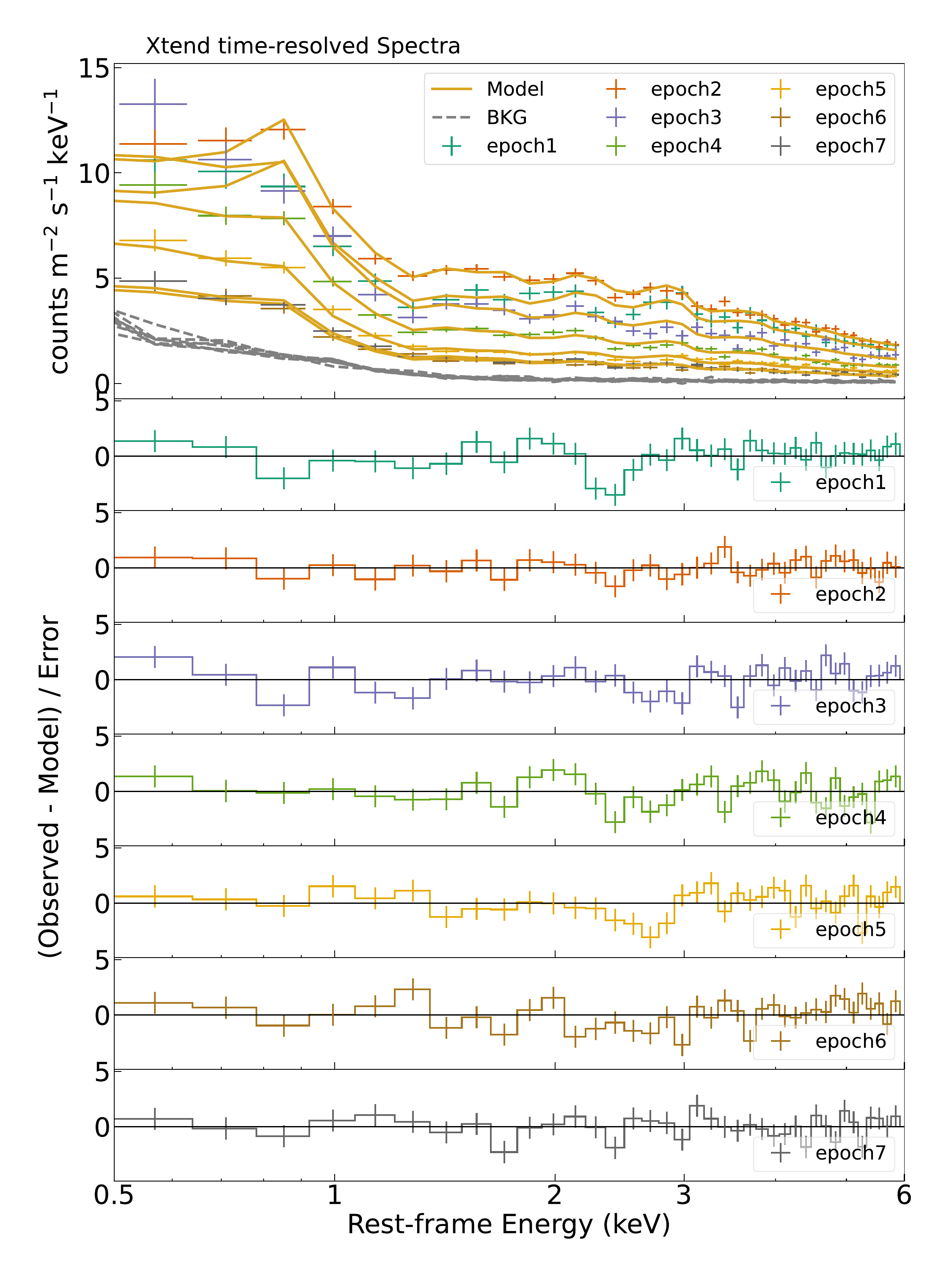}  
 \end{center}
\caption{Time-resolved \textit{Xtend} spectra (top panel) and corresponding residuals against the best-fit model (lower seven panels), obtained from R. Sato et al. (in prep.). Residuals within $0.8\mbox{--}0.9$\,keV driving the preferred order combination were significant in Epochs 1 and 3 but disappeared over the duration of XMM-Newton (Epochs 4-5). }
\label{fig:time-resolved-spectra}
\end{figure}

Since the residuals influencing the sequence are primarily located in the soft X-ray band of the \textit{Xtend} spectrum, we attempted to cross-check our results using the well-calibrated XMM-Newton/\textit{EPIC} spectrum. However, the XMM observation only covers the low-flux state and thus limits our ability to cross-check the result over the soft X-ray band. Time-resolved spectroscopy of the \textit{Xtend} observations revealed that such a comparison is not feasible. R. Sato et al. (in prep.) divided the XRISM observations into seven epochs, where Epochs 1–3 correspond to the X-ray flare and Epochs 4–5 coordinate with the XMM-Newton exposure. Figure \ref{fig:time-resolved-spectra} presents the time-resolved \textit{Xtend} spectra and the corresponding residuals against the best-fit model. It shows that the residuals between $0.8\mbox{--}0.9$\,keV, which drives the sequential preference, were only significant during the flare but disappeared throughout the XMM-Newton observation. Therefore, applying the same order analysis to XMM-Newton/\textit{EPIC} data does not make it possible to cross-check our results. This variability, in turn, suggests that those residuals are intrinsic rather than from calibration issues.

\subsubsection{Variable abundance}\label{subsubsec:abundance}

A potential systematic effect in our results is whether non-solar abundances, instead of sequence re-ordering, could account for those soft X-ray residuals ($0.8\mbox{--}0.9$\,keV) that drive the sequence. To investigate this, we allowed the oxygen abundance ($A_\mathrm{O}$) of \texttt{PION} components under the PION2 scenario to vary, linking the abundances across all six components, since the abundance should not vary significantly in the galaxy. This yielded a negligible improvement in the fit, with $A_\mathrm{O}=1.22^{+0.24}_{-0.17}$ and $\Delta\mathrm{C-stat}=3$ for one additional degree of freedom. The residuals around $0.8\mbox{--}0.9$\,keV remain unchanged. 


\subsubsection{Fully covering assumption}\label{subsubsec:covering-factor}
Our results are based on the assumption that each absorber fully covers the X-ray source, with $C_\mathrm{F}=1$, as would be expected for consecutive shells. However, Paper I reported evidence for a clumpy outflow, which would consist of relatively small clumps that are not necessarily aligned along the line of sight and are likely unaffected by screening effects. To test this scenario, we allowed the covering factors of the six UFOs in PION3 to vary independently.

We found that only UFO6 yielded a well-constrained covering factor, with $C_\mathrm{F}=0.88\pm0.02$ and a few improvements in residuals between $1.5\mbox{--}3$\,keV, in agreement with the value reported in Paper I and R. Sato et al. (in prep.). For the remaining UFOs, we could only constrain their lower limits that $C_\mathrm{F}>0.14$ for UFO1 and $C_\mathrm{F}>0.5$ for UFO2-5, with a total statistical improvement of $\Delta\mathrm{C-stat}/\nu=15/6$. These unconstrained upper limits prevent us from definitively distinguishing between scenarios of consecutive shells and clumpy outflows. Nevertheless, the results of the Fe-K UFOs remain consistent with the fully covering assumption, demonstrating that our assumption does not introduce significant systematic biases into our analysis.

\subsubsection{Validation of model order preferences}\label{subsubsec:Bayesian}
During the model order exploration, our results are based on C-stat, which may have difficulties identifying the global best solution in the parameter space. Therefore, to validate the model preferences obtained in Section \ref{subsec:PION}, we adopted the Bayesian approach to explore the entire parameter space. For the sake of computational resources, we only performed Bayesian inference on the PION1, PION2, and PION3 models as representatives. For each model, we evaluated the column density $N_\mathrm{H}$, ionization parameter $\log\xi$, and outflow velocity $v_\mathrm{out}$ of each UFO component, totaling 18 parameters. 

We performed Bayesian analysis using the Python package \texttt{nautilus} \citep{2023Lange}, a neural-network-based algorithm.  Similar to the traditional Markov chain Monte Carlo (MCMC) method, Bayesian inference can produce posterior samples for each parameter to evaluate their values, uncertainties, and marginal probability distributions. Moreover, it can produce Bayesian evidence for model comparison. We used \texttt{PYSPEX}, the Python interface to SPEX, to connect SPEX with \texttt{nautilus}. Uniform priors were adopted: $N_\mathrm{H}=[0.8\mbox{--}300]\times10^{22}\,\mathrm{cm}^{-2}$ and $\log\xi=[3.5\mbox{--}6.0]$ for all components, except for UFO6, where $\log\xi=[2\mbox{--}4]$ was used.
The prior on outflow velocity was centered on the best-fit values from Table \ref{tab:fits}, with a range of $\pm5000$\,km/s to avoid degeneracies between UFOs. The other setups are at default.

\begin{figure}
 \begin{center}
  \includegraphics[trim={0 20 0 0}, width=0.49\textwidth]{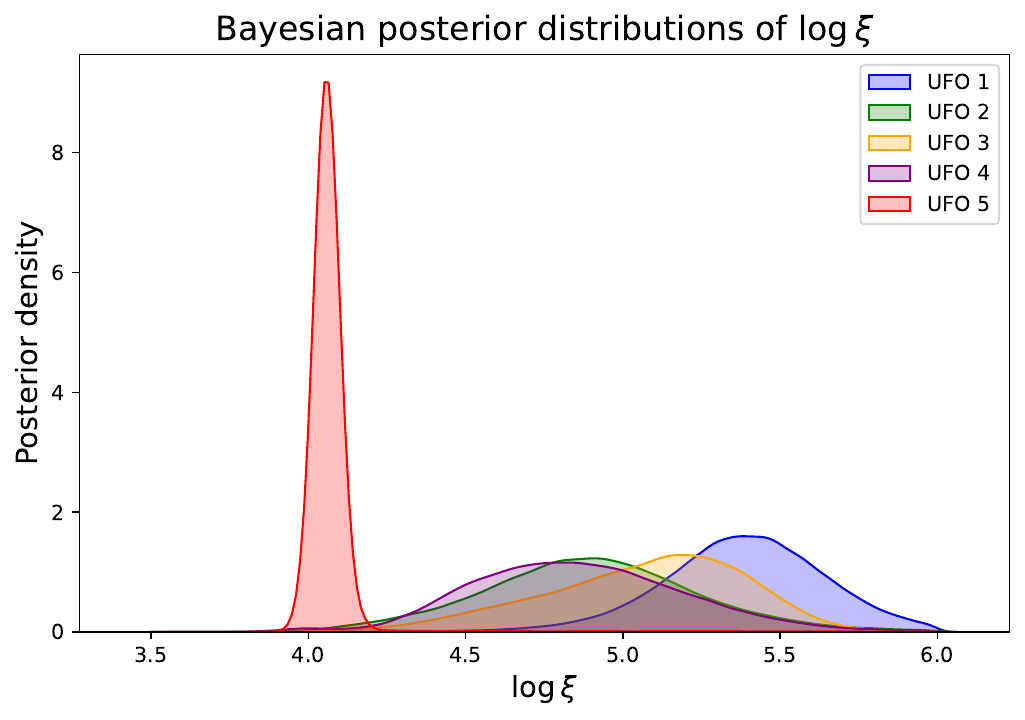}  
 \end{center}
\caption{The Bayesian posterior probability distribution of the ionization parameter $\log\xi$ for each UFO component in PION2, revealing the distinct difference between UFO1 and UFO5. }\label{fig:logxi-probability}
\end{figure}

Figures \ref{app:fig:nautilus-PION1}--\ref{app:fig:nautilus-PION3} display the normalized probability distributions of the PION1-3 parameters, respectively, with Bayesian evidence marked. The estimated values with their uncertainties are reported at the top of each panel. The two-dimensional distributions of the probability of pairing the free parameters with each other are also presented. The UFO parameters are consistent with those derived from the C-stat analysis (Table \ref{tab:fits}). For example, Figure \ref{fig:logxi-probability} shows the Bayesian posterior distributions of the ionization parameter for each hard X-ray UFO in PION2, revealing a clear distinction between UFO1 and UFO5. Although the best-fit value of UFO3 is slightly higher than that of UFO2, they are consistent within $1\sigma$ uncertainty and are thus considered effectively identical. Overall, the results show a general trend in which slower components tend to be less ionized. Only the outflow velocity of UFO6 exhibits multiple degenerate solutions. However, the UFO6 velocity is derived from the soft X-ray CCD-resolution spectra from \textit{Xtend}, and the degenerate values lie within the instrumental resolution. Therefore, this degeneracy has no significant impact on our conclusions.

The Bayesian evidence for PION1, PION2 and PION3 are $\log Z\sim-516$, $-513$ and $-508$, respectively. The Bayes factor is defined as $B_\mathrm{1,2}=Z_1/Z_2=p(y|M_1)/p(y|M_2)$, with $y$ corresponding to the data and $M$ corresponding to the model. According to the Jeffreys scale \citep{jeffreys1961theory}, $B_{1,2}>30$ ($\log B_\mathrm{1,2}>1.5$) represents `very strong evidence' against $M_2$. Therefore, PION3 is preferred over PION2 ($\log B_{3,2} = 5$), and PION2 over PION1 ($\log B_{2,1} = 3$), confirming the conclusions of the C-stat analysis.

\subsection{Soft X-ray absorber}
\label{subsec:soft-X-ray-absorber}

The soft X-ray UFO was previously detected by XMM-Newton/\textit{RGS} in 2013 and 2019, showing similar properties \citep[$v_\mathrm{out}\sim-0.26c$, $\log\xi=3\mbox{--}4$;][]{2016Reeves,2020Reeves}, but with a column density over an order of magnitude lower ($\log (N_\mathrm{H}/\mathrm{cm}^{-2})=21\mbox{--}22$) compared to our 2024 observations. In 2019, the soft X-ray UFO exhibited substantial variability, with its column density decreasing from $N_\mathrm{H}\sim10^{23}\,\mathrm{cm}^{-2}$ (due to bound-free continuum opacity) to $N_\mathrm{H}\sim10^{21}\,\mathrm{cm}^{-2}$ within 20 days, while the hard X-ray component remained remarkably stable. Similar behaviors were observed in 2024, where hard X-ray emission and absorption were stable (Paper I), while the covering factor of the soft X-ray absorber varied from $\sim0.88$ to $\sim0.95$ over the $\Delta t\sim500\,\mathrm{ks}$ exposure ($\sim6\,\mathrm{days}$) based on the time-resolved spectroscopy if the covering factor of UFO6 $C_\mathrm{F}$ is allowed to vary (R. Sato et al. in prep.). These comparable variability timescales suggest that UFO6 is located at the same place over these years and is continuously replenished. 

The location of the soft X-ray absorber (UFO6) was thus estimated to be $R_\mathrm{UFO6}\geq1000\,R_\mathrm{g}$ through the variability, assuming angular momentum conservation ($R_\mathrm{UFO}v_\mathrm{UFO}^{(\phi)}=R_\mathrm{launch}v_\mathrm{launch}^{(\phi)}$; R. Sato et al. in prep.), where $v_\mathrm{launch}^{(\phi)}=\sqrt{GM_\mathrm{BH}/R_\mathrm{launch}}$, $v_\mathrm{UFO}^{(\phi)}\sim \frac{\Delta C_\mathrm{F}}{\Delta t}d_\mathrm{corona}<5\times10^{-3}c$, and the diameter of the hot corona is $d_\mathrm{corona}<16\,R_\mathrm{g}$ (Paper I). The launching radius of UFO6 can be constrained to be $R_\mathrm{launch}\geq R_\mathrm{escape}\sim27\,R_\mathrm{g}$ assuming it exceeds the escape radius, $R_\mathrm{escape}=2GM_\mathrm{BH}/v_\mathrm{out}^2$ (here the averaged velocity $v_\mathrm{out}\sim0.27c$ is taken). In this case, the placement of UFO6 is compatible with the broad line region \citep[BLR, $R_\mathrm{BLR}\sim3000\mbox{--}10000\,R_\mathrm{g}$;][]{2023Gravity}, much farther out than the hard X-ray UFOs, which are located at $200\mbox{--}600\,R_\mathrm{g}$ based on their stability during X-ray flares (Paper I). This suggests that UFO6 occupies the outermost layer among the six UFOs, contradicting the PION3 scenario. The corresponding number density is estimated at $n_\mathrm{e}^\mathrm{UFO6}\leq6\times10^{8}\,\mathrm{cm}^{-3}$ according to the definition of ionization parameter ($n_\mathrm{e}=L_\mathrm{ion}/\xi R^{2}$, $R$ is the UFO distance $R_\mathrm{UFO}$ and $L_\mathrm{ion}\sim1.6\times10^{46}\,\mathrm{erg/s}$; Paper I). The clump size of UFO6 is thus $d_\mathrm{clump}\sim \Delta R\sim N_\mathrm{H}/n_\mathrm{H}\geq2\,R_\mathrm{g}$ if we assume that the shell thickness is similar to the clump size, as in Paper I.  



Alternatively, statistically favored PION3 scenario suggests that UFO6 is co-spatial with hard X-ray UFOs at $R_\mathrm{UFO6}\sim200\mbox{--}600\,R_\mathrm{g}$. In this case, the required number density of UFO6 is exceptionally high $n_\mathrm{e}\sim10^{10}\mbox{--}10^{11}\,\mathrm{cm}^{-3}$, exceeding that of hard X-ray UFOs ($n_\mathrm{e}\sim10^{7.5}\mbox{--}10^{9.5}\,\mathrm{cm}^{-3}$). This high density implies an absorber size of 
$d_\mathrm{clump}\sim\Delta R\sim 0.01\mbox{--}0.16\,R_\mathrm{g}$, significantly smaller than the size of hot corona $\sim16\,R_\mathrm{g}$ (Paper I), making it challenging to achieve the observed covering factor ($C_\mathrm{F}=0.88\pm0.02$). One possibility is that UFO6 exhibits, for example, a geometry consisting of a fine spray of many clumps.
However, the rapid variability exhibited by UFO6, in contrast to the stable UFO1–5 throughout the observation, suggests that such a high density is not implausible. Moreover, except for its ionization state, UFO6 shares comparable column densities and velocities with hard X-ray UFOs, indicating that UFO6 may belong to the same outflowing stream as UFO1-5. A possible explanation is the shocked wind scenario where rapid Compton cooling occurs as hot fast-moving UFOs shock against the cold ISM, producing highly inhomogeneous low-ionization clumps  \citep{2011King,2013Pounds}. The Compton cooling time of UFO6 is only $t_C<2\,\mathrm{ks}$ \citep[see Eq.7 in][]{2011King}, making this process feasible. As the plasma cools, its density increases with the inverse plasma temperature $T^{-1}$, and the temperature difference between UFO1-5 and UFO6, nearly one order of magnitude, could account for UFO6's higher density. A potential drawback of this scenario is that post-shocked winds should be slower than pre-shocked outflows, which contradicts observations. However, the slow, cool, and dense post-shock materials could be entrained by a bulk of fast-moving UFOs, retrieving high velocity \citep{2019Serafinelli}. This complex scenario presents a plausible mechanism for reconciling the observed properties of UFO6 with theoretical expectations.

The stability curves in Figures \ref{fig:PION1_SED-S-curve} and \ref{fig:PION3_SED-S-curve} show that hard X-ray UFOs are thermally stable, whereas the soft X-ray UFO remains thermally unstable, regardless of the sequential combination. The thermal equilibrium timescale of the gas can be approximately expressed as \citep{2021Burke}:
\begin{equation}\label{eq:thermal}
    t_\mathrm{th}=1680\left(\frac{\alpha}{0.01}\right)^{-1}\times\left(\frac{M_\mathrm{BH}}{10^{8}M_\odot}\right)\left(\frac{R}{200R_\mathrm{g}}\right)^{3/2}\mathrm{days}
\end{equation}
where $\alpha$ is the viscosity parameter and $R$ is the radial position. Observations suggest a typical range for the viscosity parameter, $\alpha=0.1\mbox{--}0.4$ \citep{2007King}. For gas at distances of either $200\mbox{--}600\,R_\mathrm{g}$ or $>1000\,R_\mathrm{g}$ to a black hole with a mass of $M_\mathrm{BH}=5\times10^{8}\,M_\odot$, the thermal equilibrium timescales are $t_\mathrm{th}\sim0.6\mbox{--}12\,\mathrm{yrs}$ or $>6\mbox{--}25\,\mathrm{yrs}$, respectively. Both timescales are much longer than our observation, implying that the thermal instability of UFO6 is not a concern for our photoionization modeling.

Therefore, the location of UFO6 remains uncertain. Based on variability estimates, UFO6 appears far from hard X-ray UFOs and is likely situated in the BLR. However, it could also be cospatial with UFO1–5, as suggested by the order permutations and its similar properties to UFO1–5. Nevertheless, we caution that UFO6 is inferred from bound-free absorption rather than resolved lines, limited by the spectral resolution in the soft X-ray CCD spectrum. Moreover, its variable nature suggests that derived UFO6 parameters could be a mixture of absorbers at different states. As a result, UFO6 may represent an average solution of multiple absorbers or a variable absorber itself. The lack of high-resolution soft X-ray spectra limits our ability to investigate the sequence of multiple UFOs. However, the order permutation methodology has been validated through XRISM (GVO) and NewAthena simulations (see Section \ref{subsec:simulations}), demonstrating its potential for future studies.

\subsection{Driving mechanism of outflows}
\label{subsec:mechanism}

Regardless of the location of UFO6, the best-fit parameters of UFO1-5 remain constant in \texttt{PION} scenarios. Figure \ref{fig:parameter_comparison} exhibits the relationship between the ionization parameter and outflow velocity of these hard X-ray UFOs. We fitted them with a power-law function using the \texttt{scipy.odr.ODR} package in Python:
\begin{equation}\label{eq:fit-logxi-v}
    v_\mathrm{out}\propto\xi^{(0.14\pm0.04)}.
\end{equation} 
The same fitting was applied to other models. Spearman's ranks and Pearson correlation coefficients with corresponding $p$-values are summarized in Table \ref{tab:xi-v-fits}. Although none of the fits meet the conventional threshold for statistical significance ($p<0.05$), compared with \texttt{XABS} and \texttt{PHASE} ($|\rho_\mathrm{sp/pear}|<0.5$ and $p>0.6$), the \texttt{PION} results display a marginally significant correlation ($\rho_\mathrm{sp/pear}>0.7$ and $p\sim0.1$) between $v_\mathrm{out}$ and $\xi$, indicating a stratified outflow. 


\begin{table}[!t]
  \tbl{Results of the power-law fit between velocities $v_\mathrm{out}$ and ionization parameters $\xi$ among hard X-ray UFOs in different fits.}{
  \begin{tabular}{cccc}
      \hline
      Fit Name& Slope & Spearman's rank   & Pearson correlation   \\
      & & $\rho_\mathrm{sp}$ ($p$-value) & $\rho_\mathrm{pear}$ ($p$-value) \\
      \hline
      XABS & $0.18\pm0.13$ & 0.00 (0.997)  & -0.05 (0.91) \\
      PHASE & $0.70\pm0.99$  & -0.25 (0.68) & -0.28 (0.65) \\
      PION & $0.14\pm0.04$ & 0.70 (0.13)  & 0.81 (0.09) \\
      \hline\\
    \end{tabular}}\label{tab:xi-v-fits}
\vspace{-0.5cm}
\end{table}


This possible correlation indicates that five hard X-ray UFOs originate from the same outflowing stream, enabling us to explore the driving mechanism of outflows in PDS 456. In the radiatively driven scenario, momentum-conserving outflows predict that the momentum rate ($\dot{P}_\mathrm{out}=\dot{M}_\mathrm{out}v_\mathrm{out}\propto n_\mathrm{H}r^2v_\mathrm{out}^2$, where $\dot{M}_\mathrm{out}$ is the outflow mass loss rate) approximates the momentum flux of the radiation field $\dot{P}_\mathrm{rad}=L_\mathrm{bol}/c$ \citep{2015Gofford}. This produces a scaling relation of $v_\mathrm{out}\propto\xi^{0.5}$ \citep{2013Tombesi}. The energy-conserving outflows require a conserved kinetic power $\dot{E}_\mathrm{kin}=0.5\dot{M}_\mathrm{out}v_\mathrm{out}^2\propto v_\mathrm{out}^3/\xi$, leading to $v_\mathrm{out}\propto\xi^{1/3}$ \citep{2015King}. For the MHD-driven outflows, based on a self-similar prescription in the radial direction with the Keplerian velocity profile ($v_\mathrm{out}\propto r^{-1/2}$), \citet{2010Fukumura} suggested $v_\mathrm{out}\propto\xi^{1/(4-2p)}$, where $p$ is the density slope $n_\mathrm{H}\propto r^{-p}$. Observations \citep{2009Behar,2013Tombesi,2018Fukumura} suggested that $p$ ranges between 1 and 1.3, yielding a scaling relation between $v_\mathrm{out}\propto\xi^{0.5}$ and $v_\mathrm{out}\propto\xi^{0.7}$. These scaling relations were plotted in the left panel of Figure \ref{fig:parameter_comparison}, showing that UFOs in PDS 456 are inconsistent with any predictions, implying a complex wind driving mechanism. The observed positive correlation between UFO velocities and source luminosity supports the radiation-pressure-driven scenario, compatible with the super-Eddington accretion in PDS 456 \citep{2017Matzeu}. However, given the relativistic velocities of these UFOs, which cannot be solely accelerated by radiation pressure \citep{2021Luminari}, and their energetics, exceeding the momentum flux of the radiation (Paper I), MHD-driving may also be required, though with modifications to the frequently-used simple self-similar solutions. Therefore, we can only infer that the ionization structure of winds in PDS 456 is stratified, while the accelerating mechanism remain unknown, probably contributed by both radiation and the magnetic field.

\subsection{Structure of outflows}
\label{subsec:structure}
The presence of five resolved hard X-ray UFOs has revealed highly inhomogeneous and clumpy winds in PDS 456 (Paper I). Based on the assumption that these outflows fully cover the hard X-ray source, our spectral modeling hints that these winds are not randomly distributed but are instead stratified. Statistically, the slowest UFO (UFO5) tends to occupy the innermost layer. Furthermore, multiple UFOs generally indicate a tentative trend of accelerating outflows as they propagate outward. This exploration of the intrinsic UFO structure is, for the first time, enabled by the high-resolution XRISM spectrum and the self-consistently calculated \texttt{PION} model.

The indicated sequence of stratified and accelerated winds is consistent with the radiatively driven mechanism \citep{2023Gallo}, where the outflow is asymptotically accelerated to a terminal velocity as it moves outward \citep[e.g.,][]{2010Sim}. In contrast, the magnetically driven scenario predicts slower UFOs at larger radii because magnetic dynamos typically exhibit a poloidal configuration \citep{2015Fukumura,2022Fukumura}.

Although the position of the soft X-ray absorber remains uncertain, the scenario where it is embedded within hard X-ray UFOs may resolve the long-standing challenge for X-ray UFOs: whereby their high ionization states lead to low opacities, rendering radiative acceleration inefficient and making it difficult to achieve relativistic speeds \citep{2020Luminari,2021Luminari}. However, dense clumps with lower ionization parameters, including the soft X-ray absorber and UV broad absorption lines (BALs) with similar velocities of $\sim0.3c$ \citep{2018Hamann}, embedded within hard X-ray UFOs might play a crucial role in boosting the opacities for radiation-pressure acceleration \citep[e.g.,][]{2015Hagino}.

\section{Conclusion}\label{sec:conclusion}

We investigated the structure of UFOs in a luminous quasar PDS 456 by examing XRISM observations coordinated with XMM-Newton and NuSTAR spectra. By applying the advanced photoionization model \texttt{PION}, we derived UFO parameters and explored the sequential combinations of multiple absorbers. The main results of our analysis are as follows:
\begin{itemize}
    \item We derived evidence for a possibly stratified ionization structure of hard X-ray UFOs in PDS 456 using \texttt{PION}, which differs from results obtained with pre-calculated photoionization codes. 
    \item The screening effect of ionized winds may play a crucial role in the spectroscopy of AGN with multiple absorption components, although its effects are less pronounced for the highly ionized absorbers. It can be utilized to determine relative positions among multiple absorbers, particularly through their sensitivity to ion transitions in the soft X-ray band. We identified a tentative trend in PDS 456, where slower UFOs appear closer to the SMBH. If later confirmed, this may support a scenario of outflow acceleration during propagation, consistent with radiatively driven mechanisms.
    \item The hard X-ray UFOs (UFO1–5) are thermally stable, whereas the soft X-ray UFO (UFO6) is thermally unstable regardless of their sequential combinations. The location of UFO6 remains uncertain, as it may reside in the BLR ($R_\mathrm{UFO6}>1000\,R_\mathrm{g}$) suggested by the variability estimates, or cospatial with hard X-ray UFOs ($R_\mathrm{UFO6}\sim200\mbox{--}600\,R_\mathrm{g}$) supported by the order permutation analysis and its similar properties to UFO1–5. 
    \item  The XRISM simulation with GVO demonstrates improved reliability of our results by leveraging high-resolution soft X-ray spectra. The NewAthena simulation highlights its potential to robustly constrain the structure of AGN ionized winds. Its high spectral resolution and large effective area will enable definitive conclusions for outflow structures, providing deeper insights into AGN feedback mechanisms.
\end{itemize}

\section*{Funding}
{\fontsize{8}{11}\selectfont
LCG acknowledges financial support from Canadian Space Agency grant 18XARMSTMA. KH is supported by JSPS KAKENHI Grant Numbers JP21K13963, 24K00638. JR acknowledges support from NASA XRISM grant 80NSSC23K0645. FT acknowledges funding from the European Union - Next Generation EU, PRIN/MUR 2022 (2022K9N5B4). MM acknowledges support from Yamada Science Foundation and JSPS KAKENHI Grant Number JP21K13958. EB acknowledges support from NASA grants 80NSSC20K0733, 80NSSC24K1148, and 80NSSC24K1774. RB acknowledges the support from NASA under award number 80GSFC21M0002. CD acknowledges support from STFC through grant ST/T000244/1. AT is supported by the Kagoshima University postdoctoral research program (KU-DREAM). TY acknowledges support by NASA under award number 80GSFC24M0006. SY acknowledges support by the RIKEN SPDR Program and JSPS KAKENHI Grant 23K13154. }

\begin{ack}
The authors thank Junjie Mao for fruitful discussions and comments. The manuscript was subject to an internal review process, with Dr. Matteo Guainazzi as the internal reviewer. The authors thank the anonymous referee for the helpful comments and suggestions.
\end{ack}

\begin{appendix}

\renewcommand{\thefigure}{A\arabic{figure}}

\setcounter{figure}{0}

\section{C-stat histograms}\label{app:histogram}
\begin{figure*}
 \begin{center}
  \includegraphics[width=0.49\textwidth]{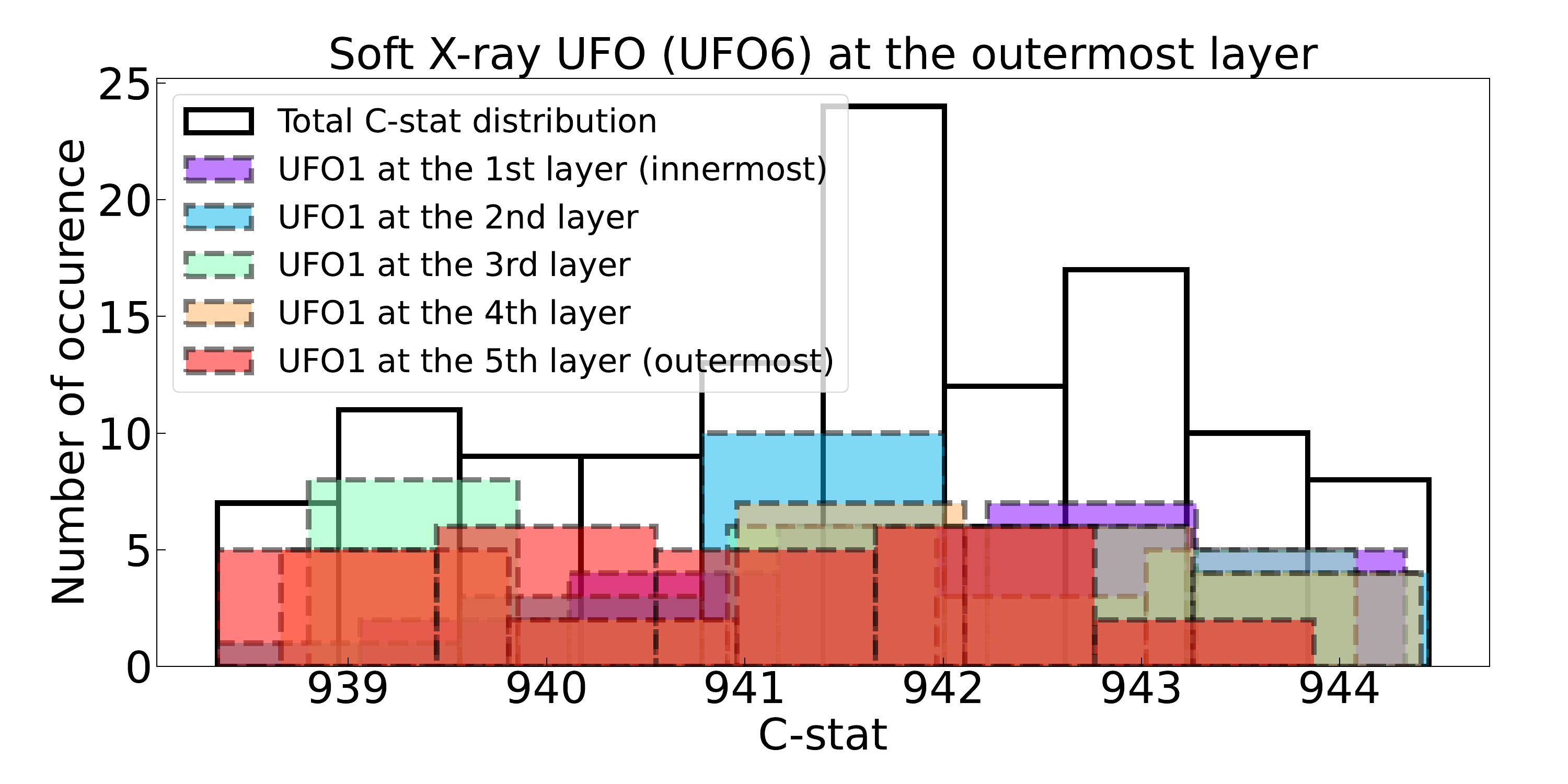} 
  \includegraphics[width=0.49\textwidth]{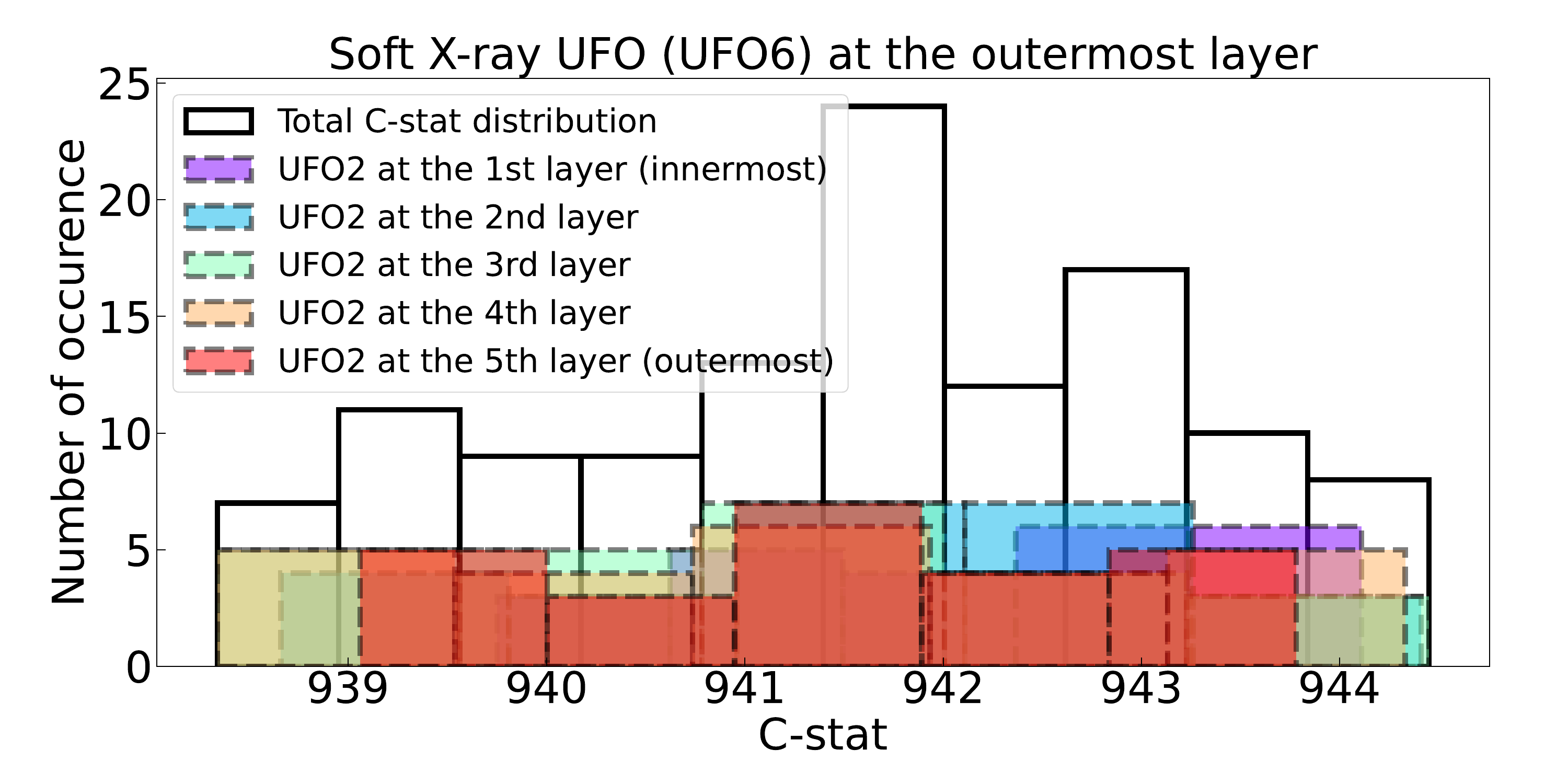} 
  \includegraphics[width=0.49\textwidth]{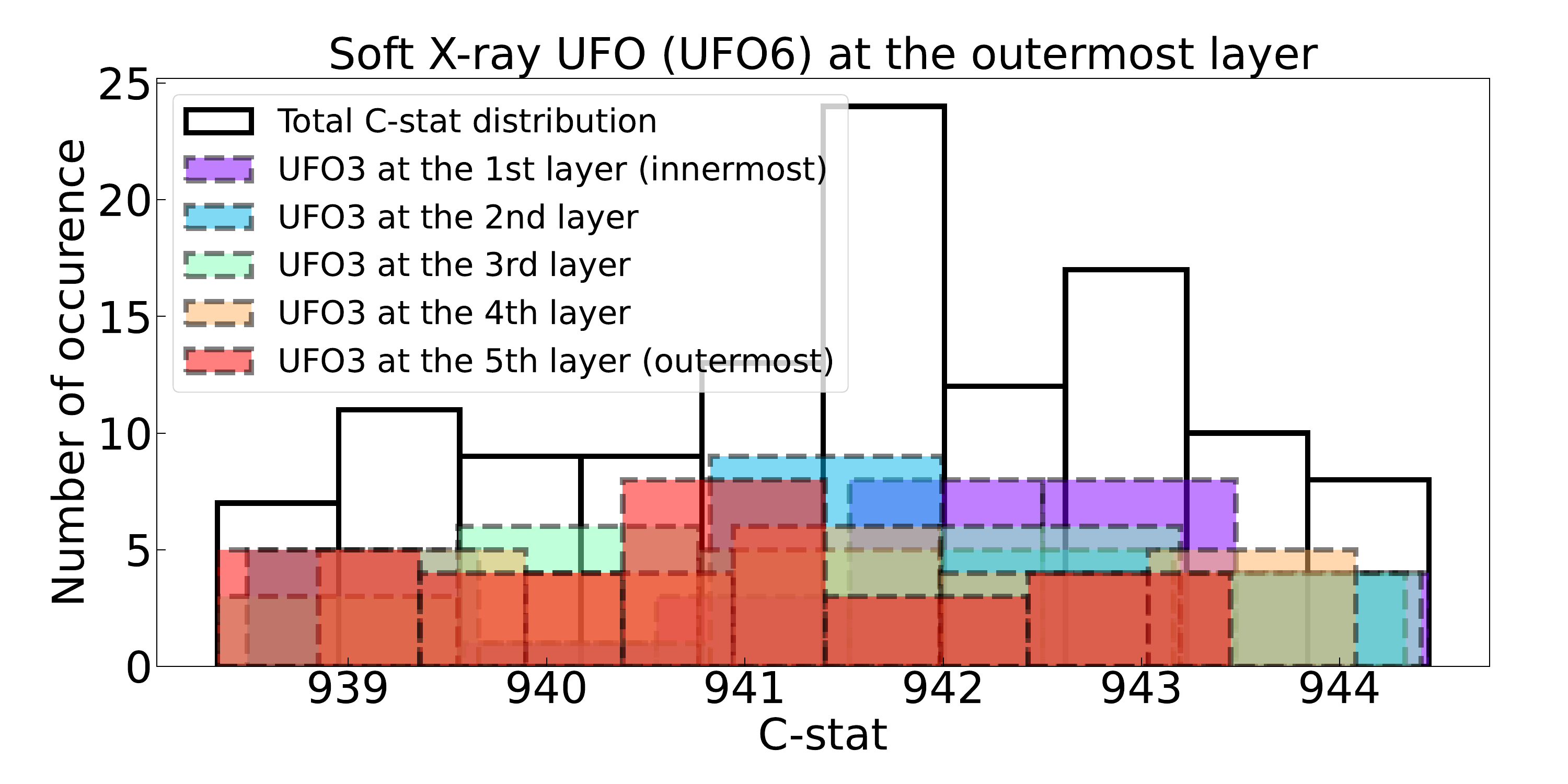} 
  \includegraphics[width=0.49\textwidth]{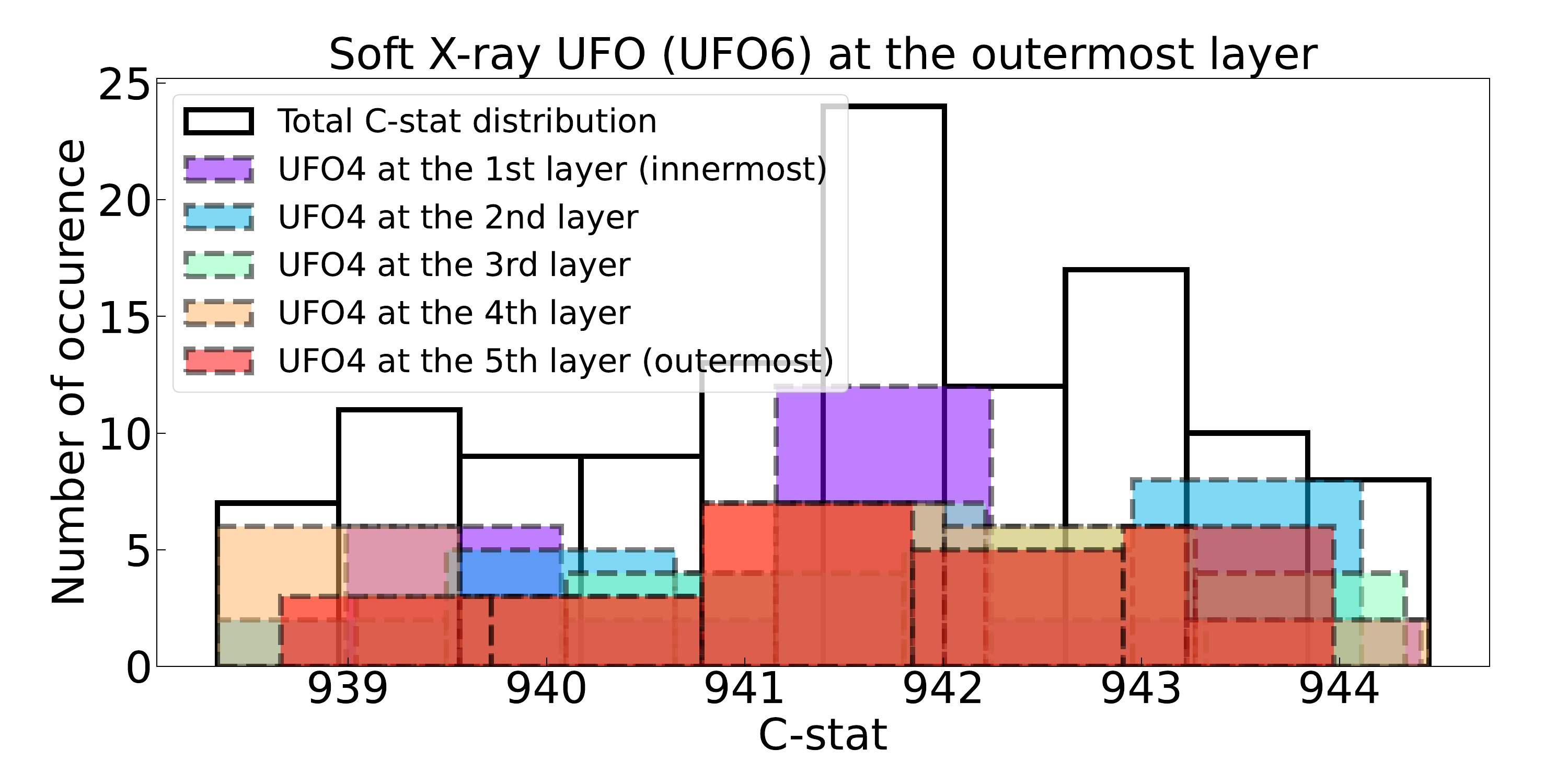} 
 \end{center}
\caption{Similar to Figure \ref{fig:histogram-soft-outermost}. C-stat distribution of spectral fits with UFO1 (upper left), UFO2 (upper right), UFO3 (lower left), and UFO4 (lower right) placed at different layers, while the soft X-ray UFO (UFO6) is fixed at the outermost layer. No preferred location for UFO1–4 was identified.
}\label{app:fig:histogram-soft-outermost}
\end{figure*}

\begin{figure*}
 \begin{center}
  \includegraphics[width=0.49\textwidth]{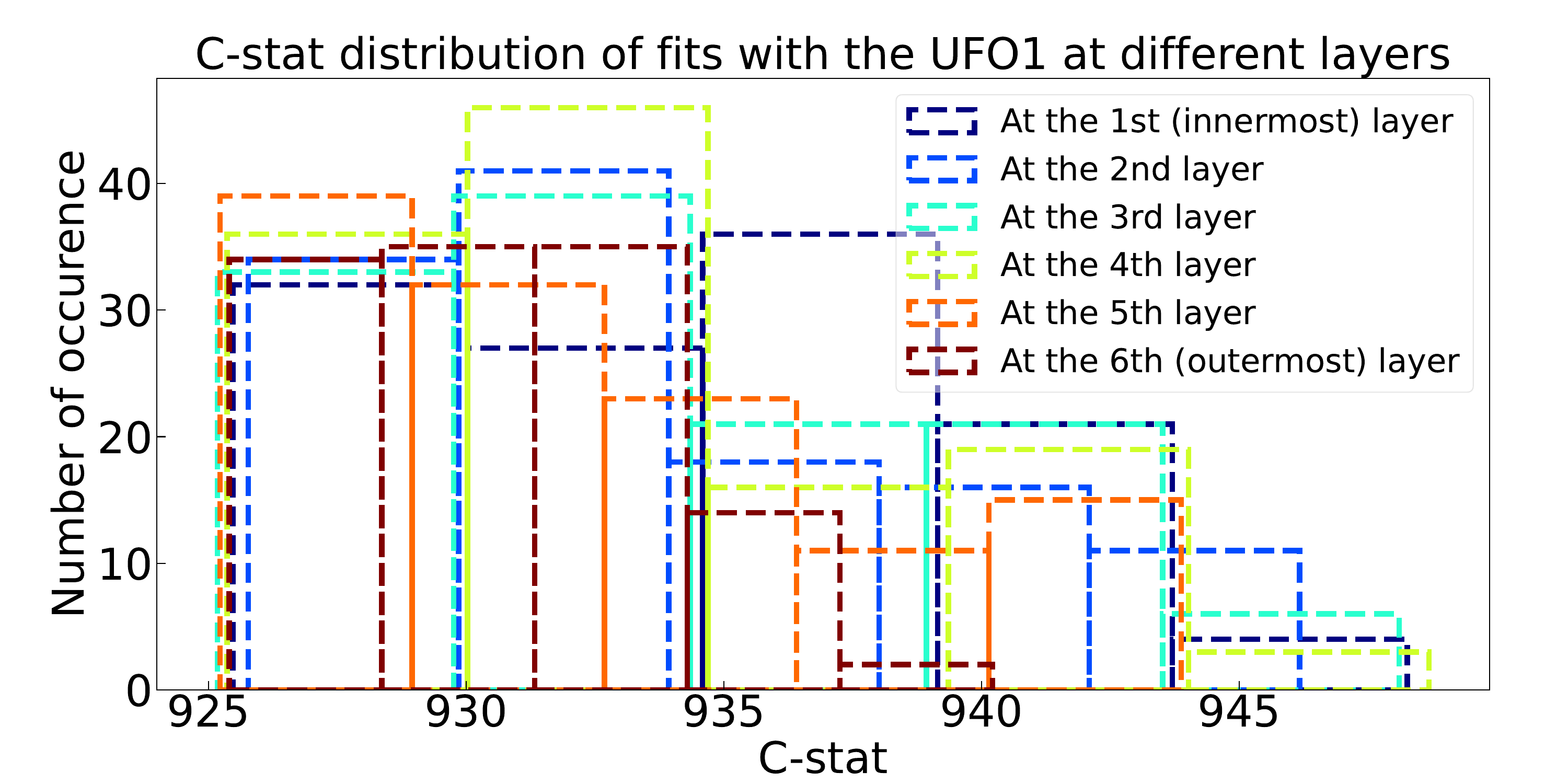} 
  \includegraphics[width=0.49\textwidth]{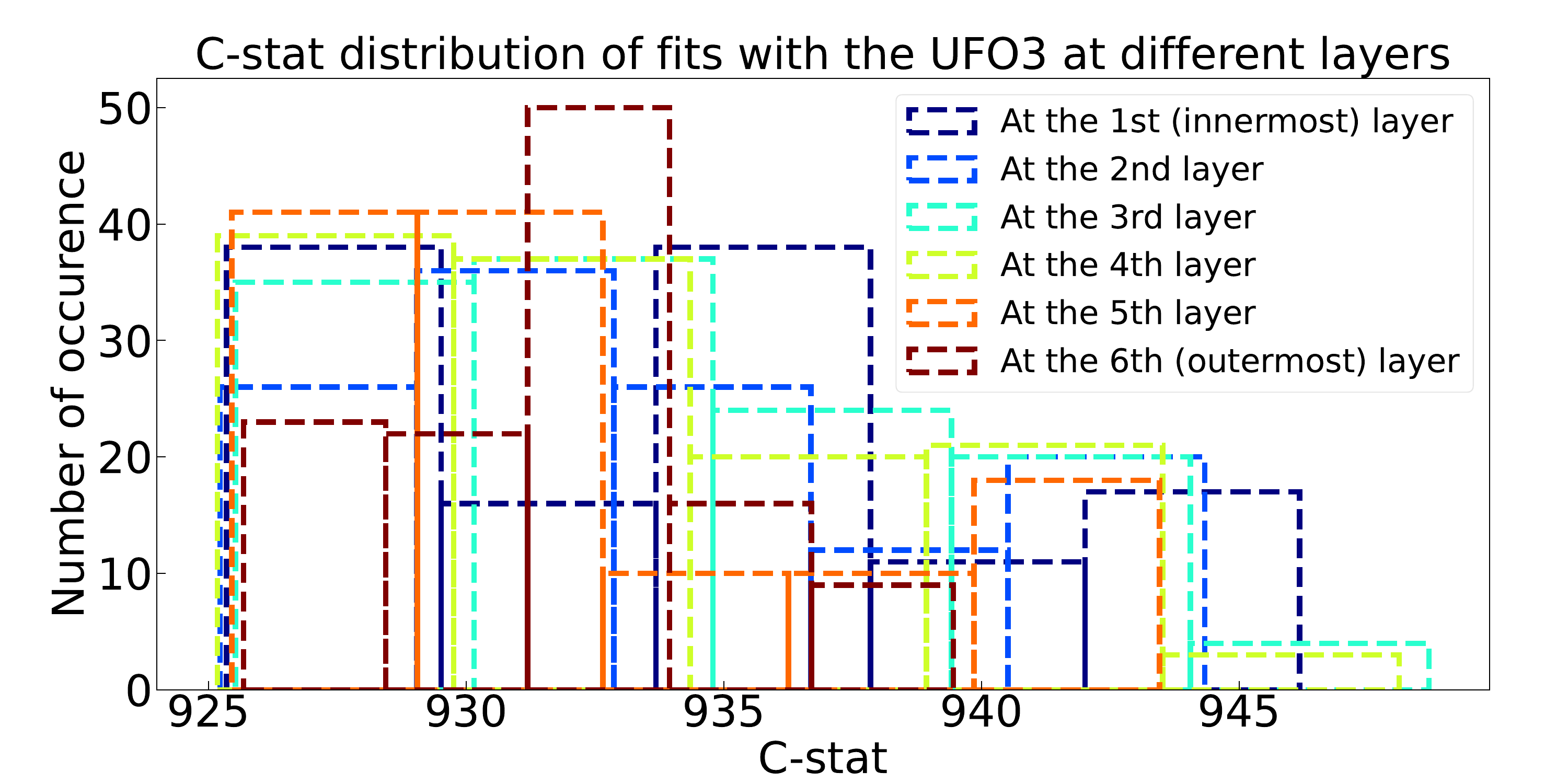} 
  \includegraphics[width=0.49\textwidth]{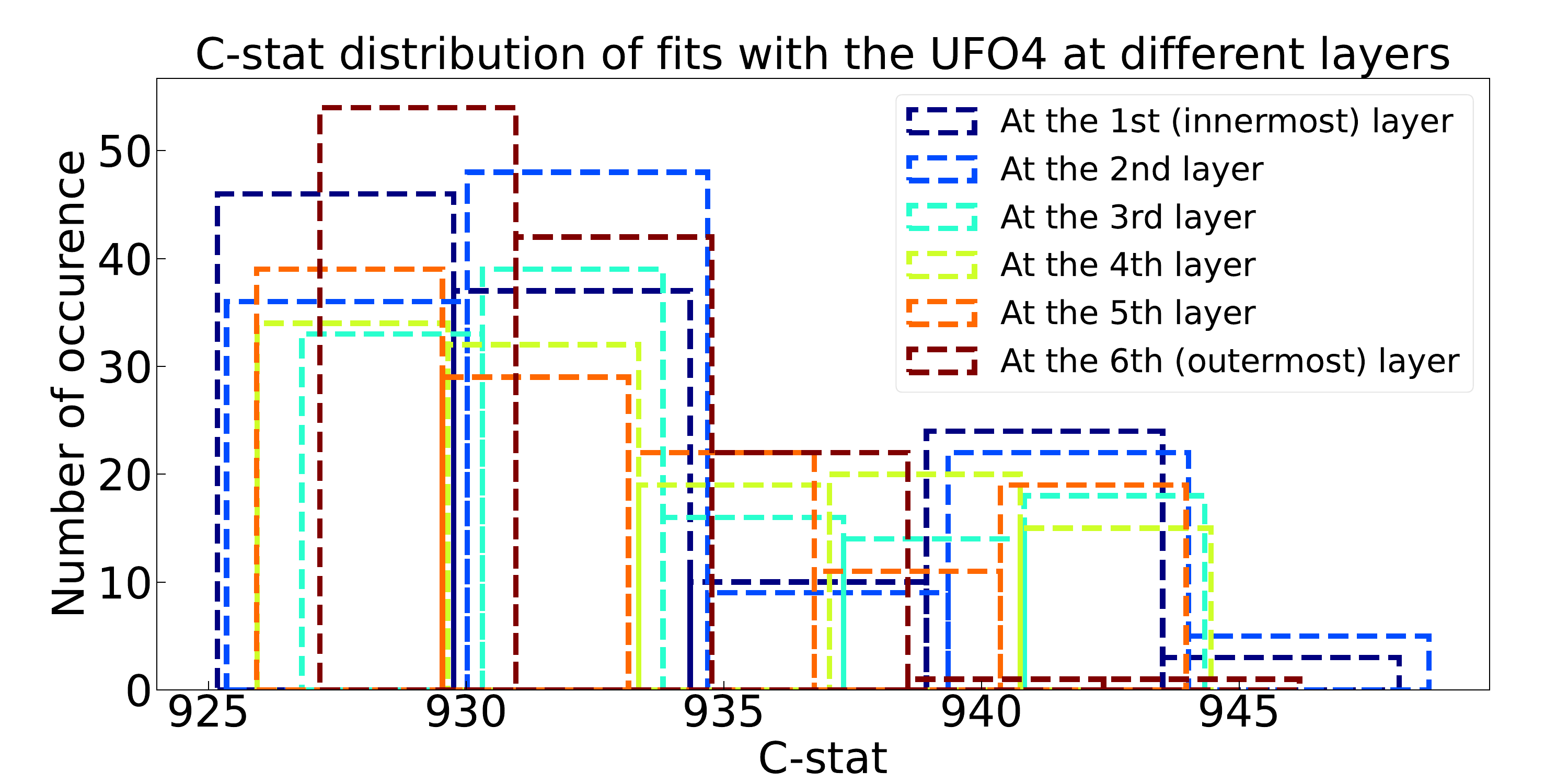} 
  \includegraphics[width=0.49\textwidth]{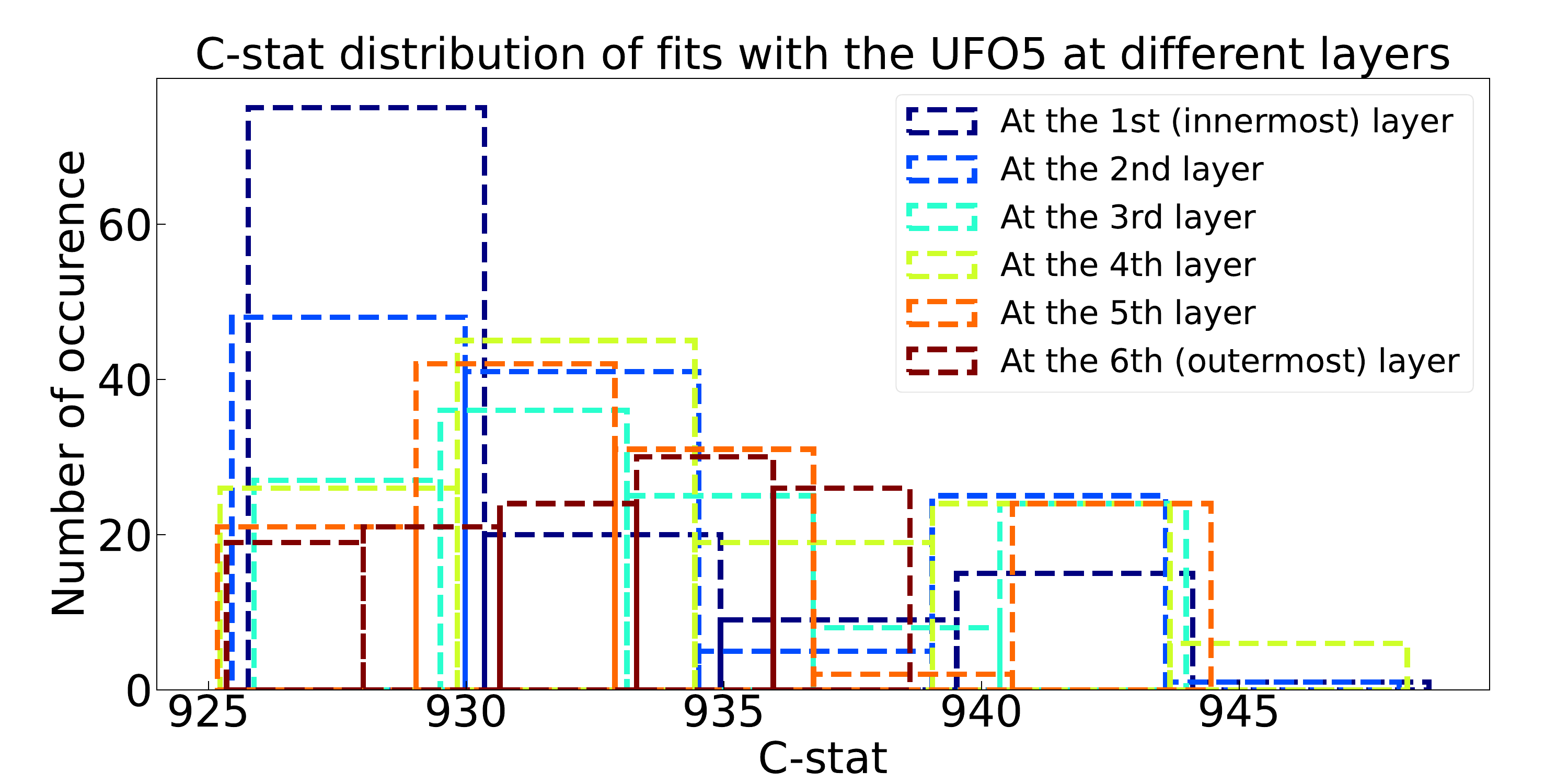} 
 \end{center}
\caption{Similar to Figure \ref{fig:histogram-all}. C-stat distribution of fits with UFO1 (upper left), UFO3 (upper right), UFO4 (lower left), and UFO5 (lower right) placed at different layers. Only UFO5 tentatively prefers the innermost layer based on its distribution, while the other UFOs do not exhibit a preference for their locations.
}
\label{app:fig:histogram-all}
\end{figure*}

\begin{figure*}
 \begin{center}
  \includegraphics[width=0.49\textwidth]{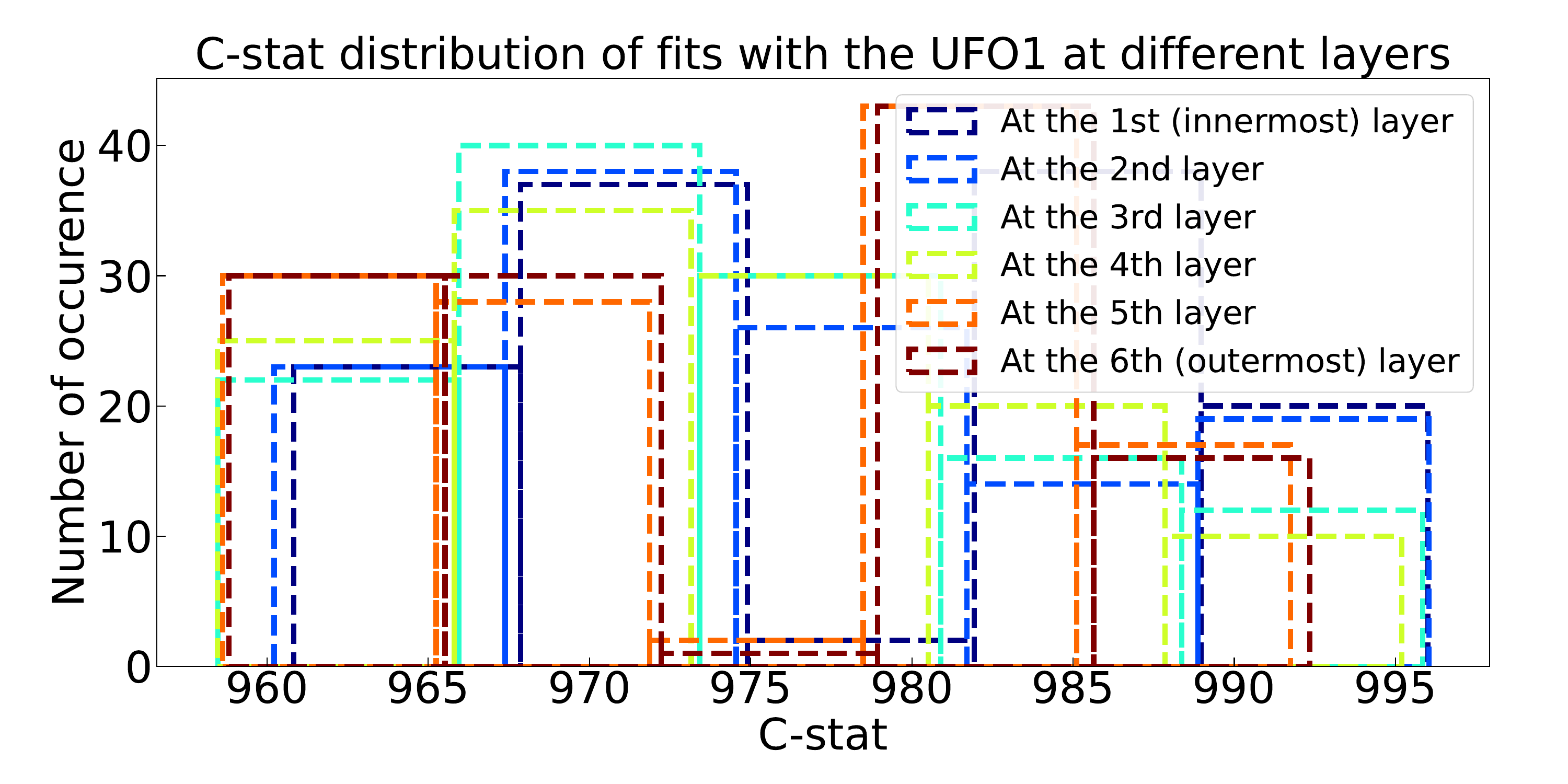} 
  \includegraphics[width=0.49\textwidth]{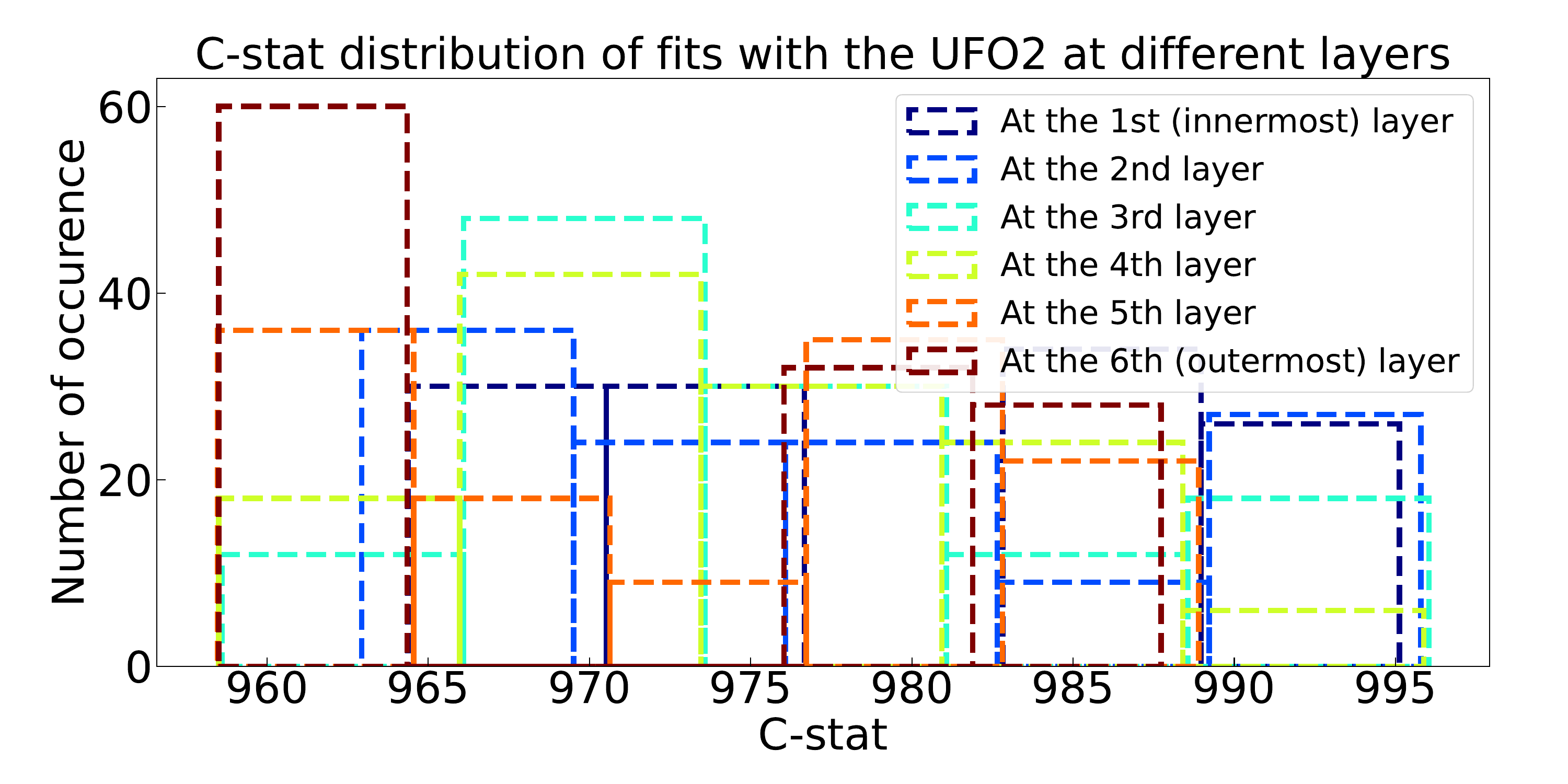} 
  \includegraphics[width=0.49\textwidth]{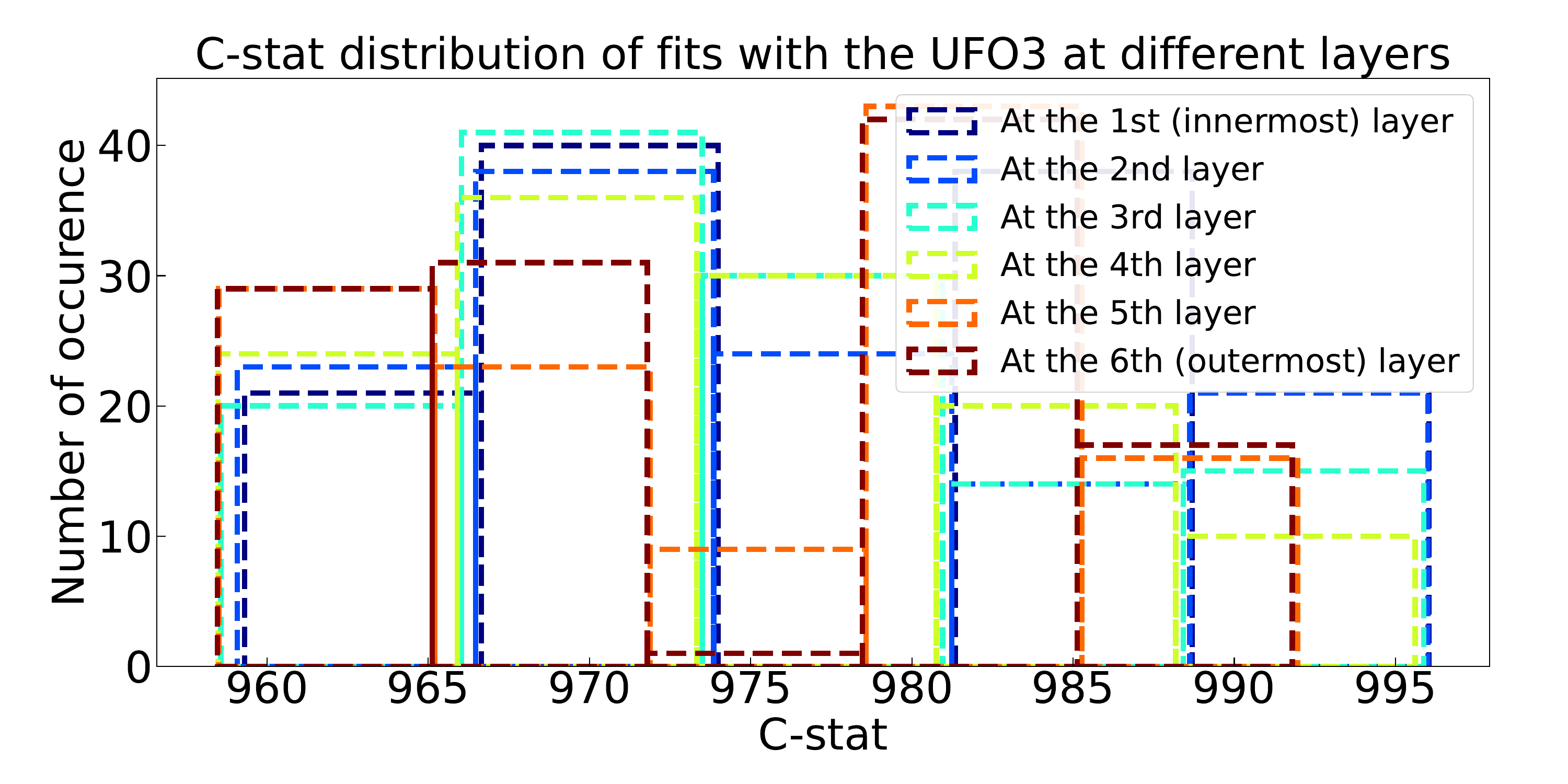} 
  \includegraphics[width=0.49\textwidth]{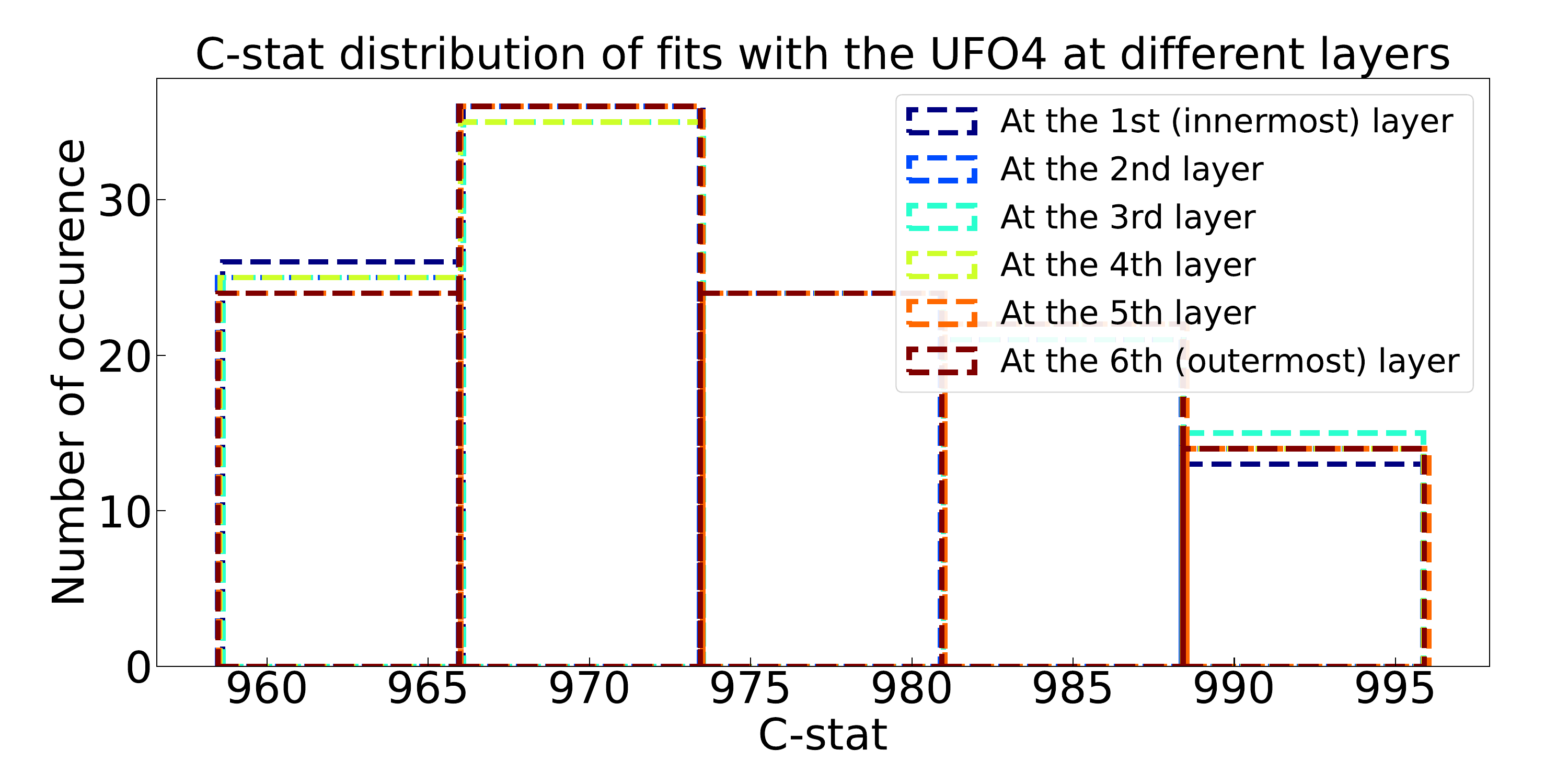} 
  \includegraphics[width=0.49\textwidth]{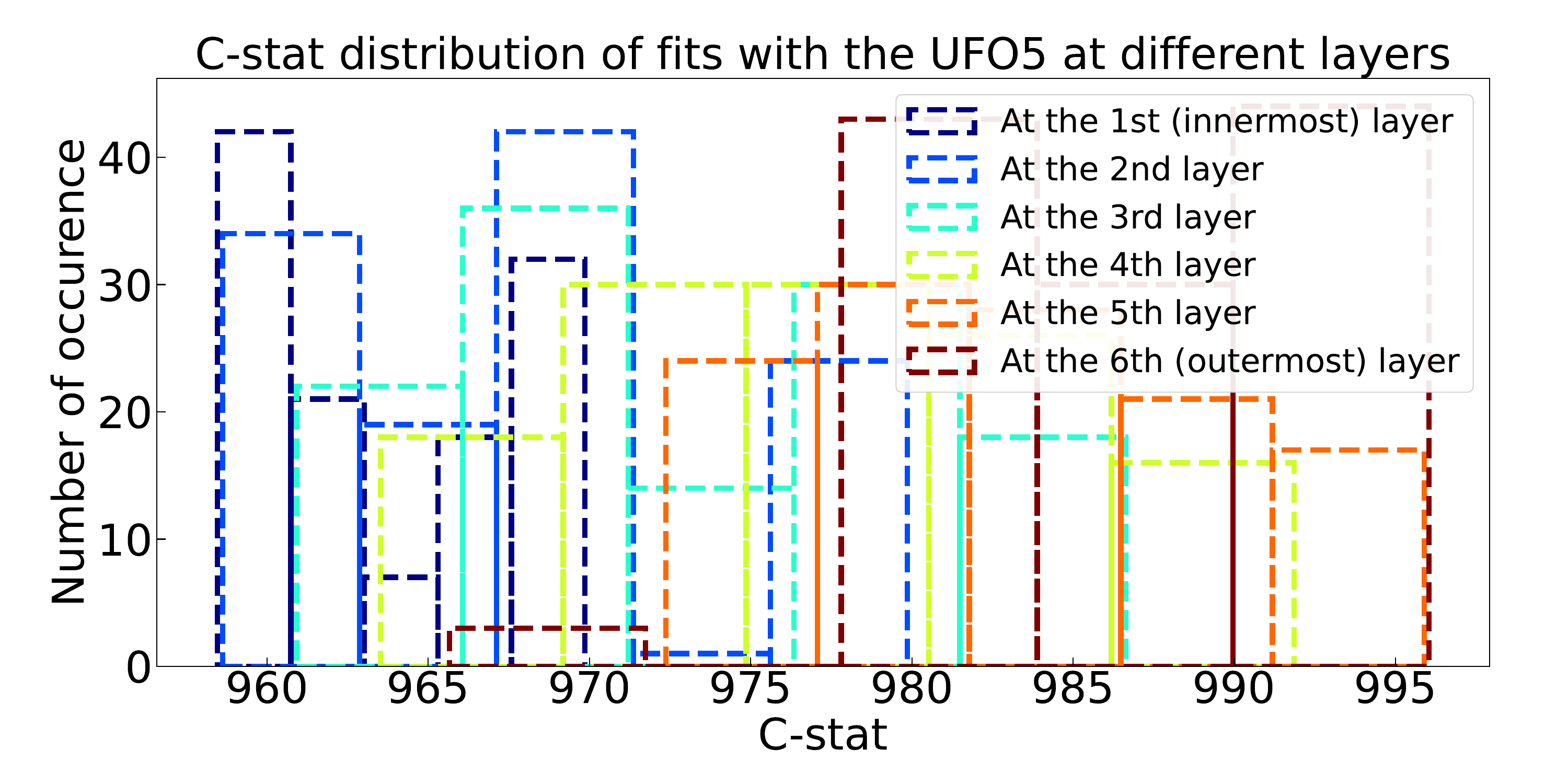} 
  \includegraphics[width=0.49\textwidth]{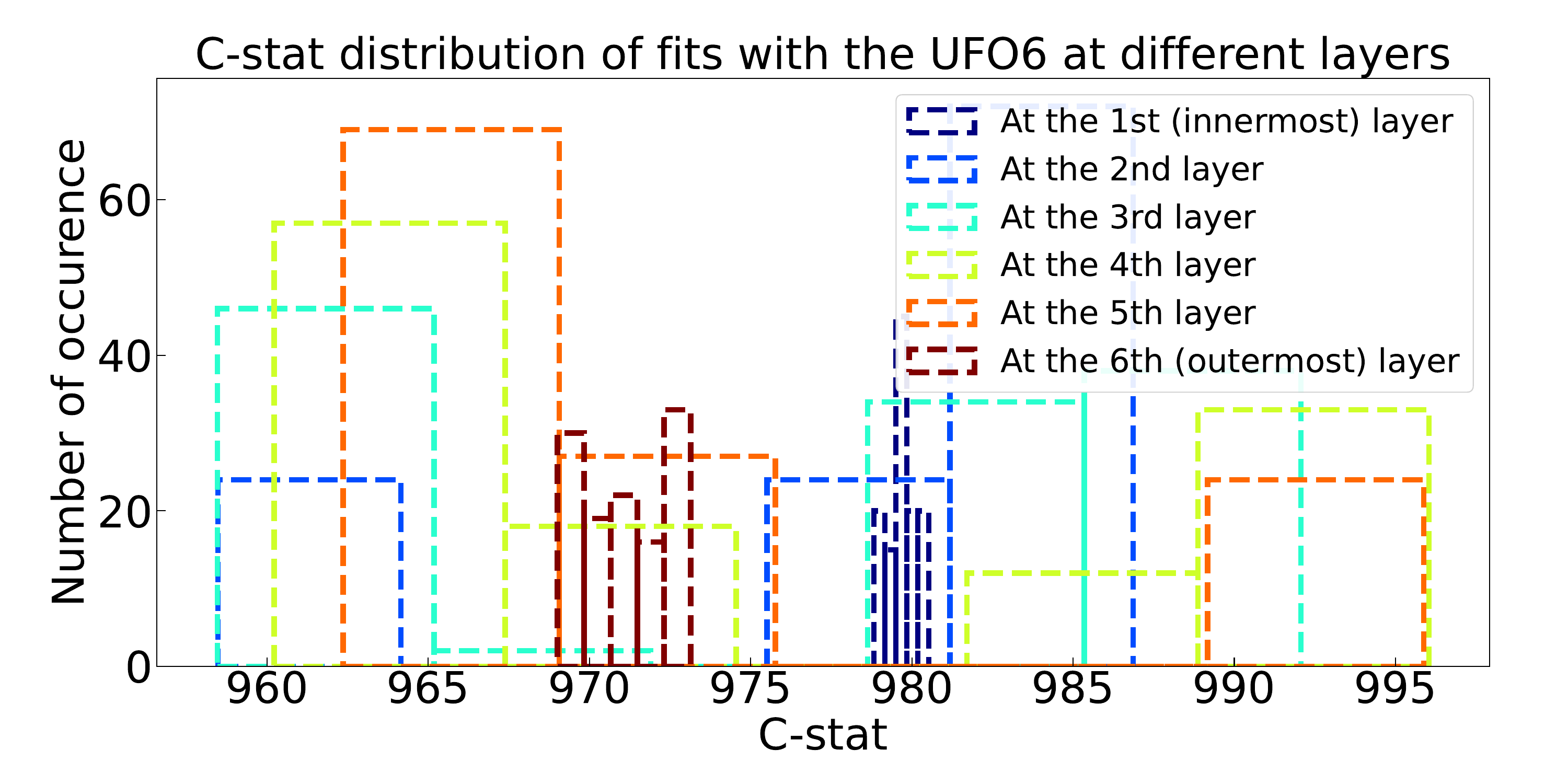} 
 \end{center}
\caption{Similar to Figure \ref{fig:histogram-all} but the analysis was performed on the simulated 100ks XRISM/\textit{Resolve} (GVO) spectrum. C-stat distribution of fits with UFO1-6 (from upper left to lower right panels) placed at different layers. As we discovered from \textit{Xtend} spectrum (Figure \ref{fig:histogram-all} and \ref{app:fig:histogram-all}), UFO2, UFO5, and UFO6 have tentative statistical preferences for the outer, innermost, and third layers, respectively, while the remaining UFOs are insensitive to their sequences. It is clearly shown that the soft X-ray absorber (UFO6) disfavors the innermost or outermost layer ($\Delta\mathrm{C-stat}>10$).}\label{app:fig:histogram-xrism}
\end{figure*}

\begin{figure*}
 \begin{center}
  \includegraphics[width=0.49\textwidth]{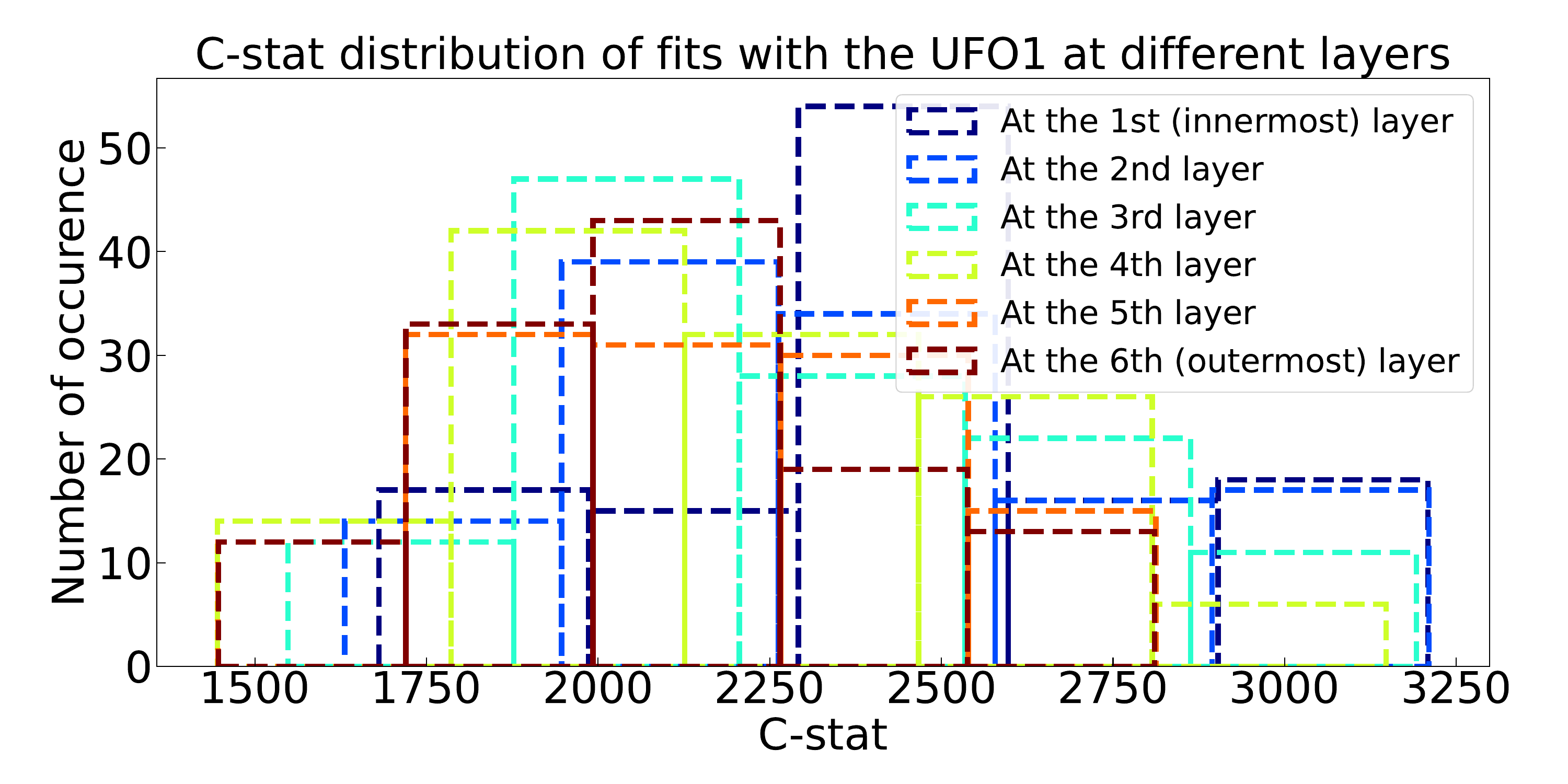} 
  \includegraphics[width=0.49\textwidth]{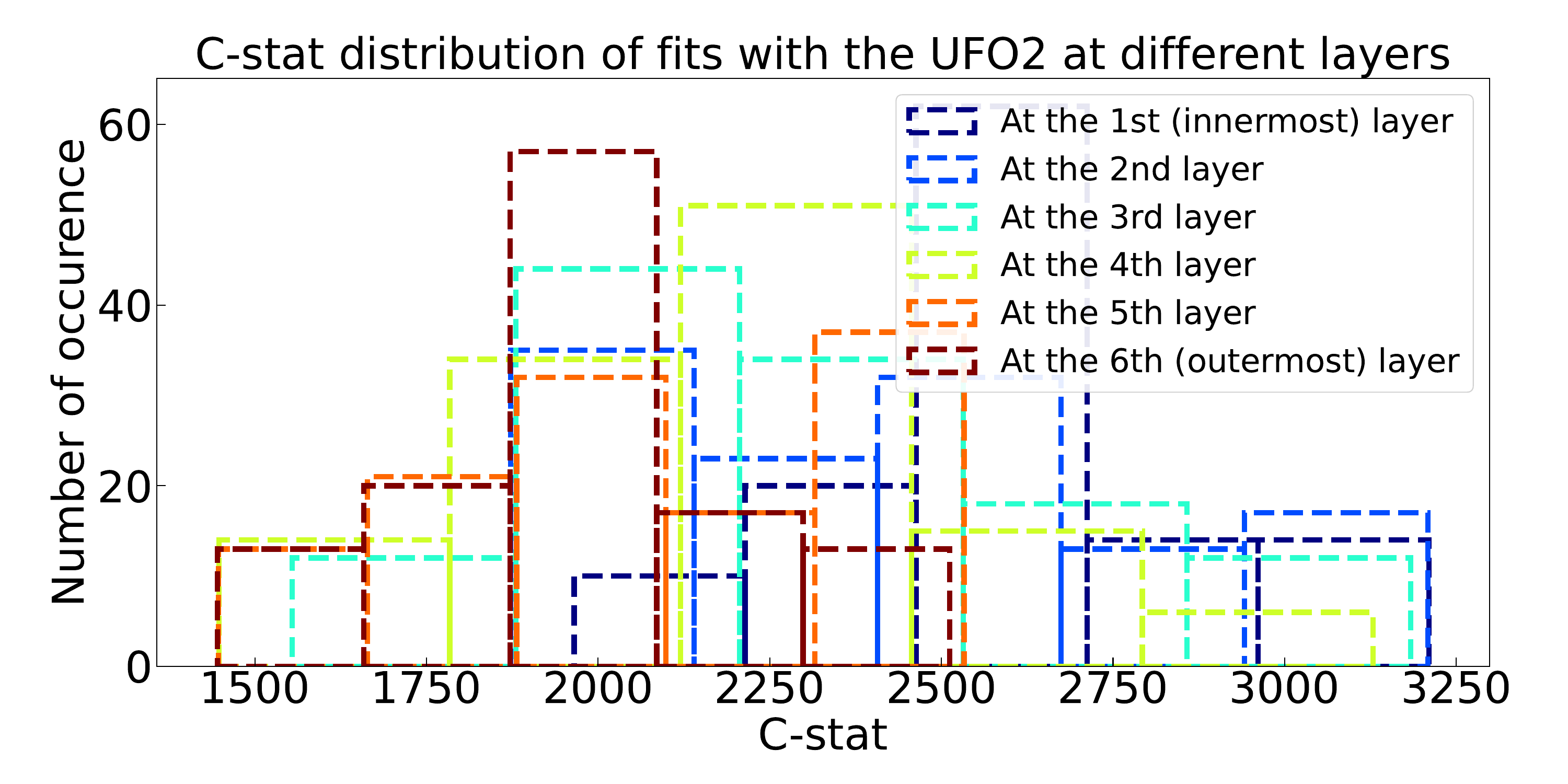} 
  \includegraphics[width=0.49\textwidth]{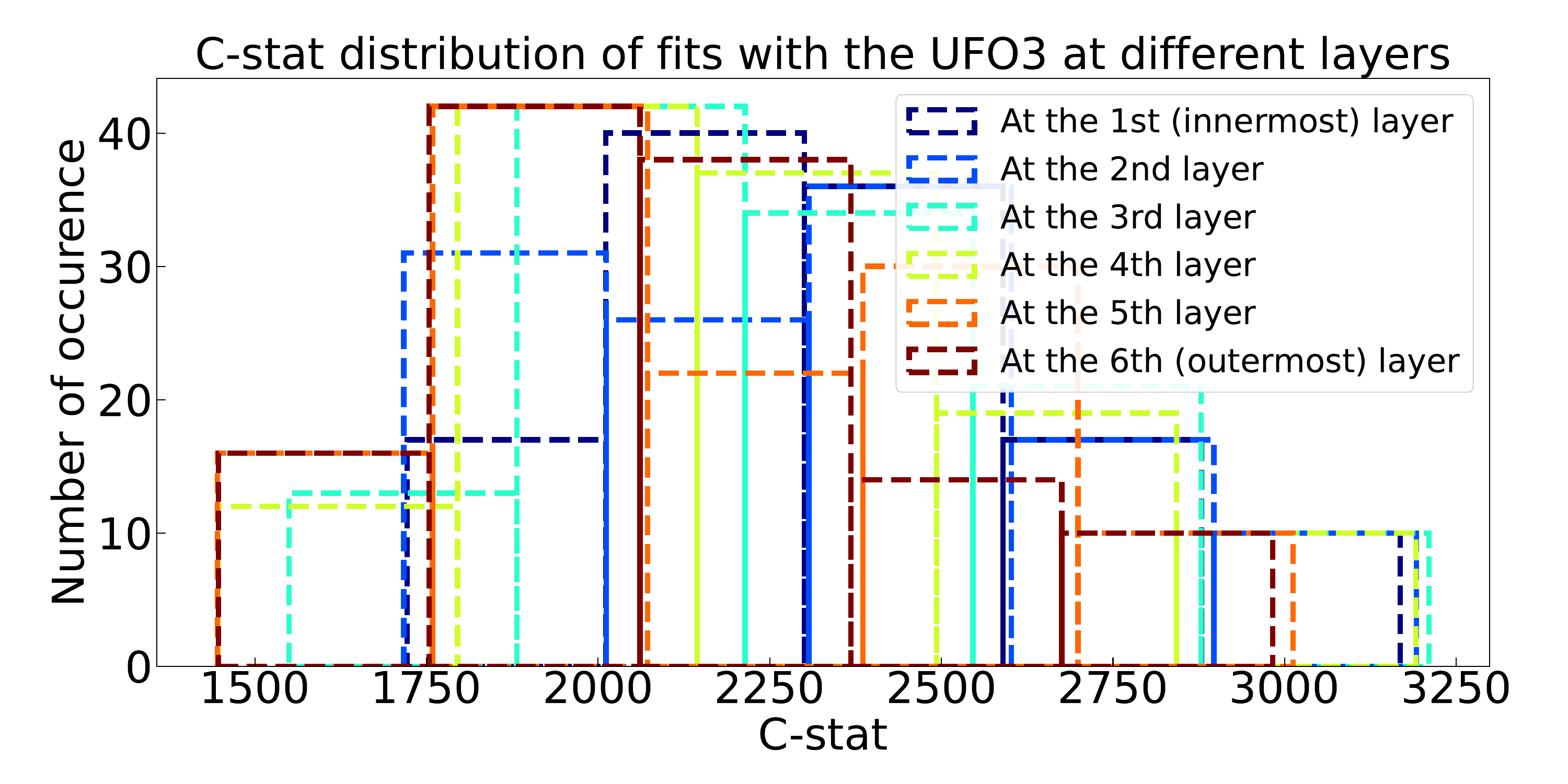} 
  \includegraphics[width=0.49\textwidth]{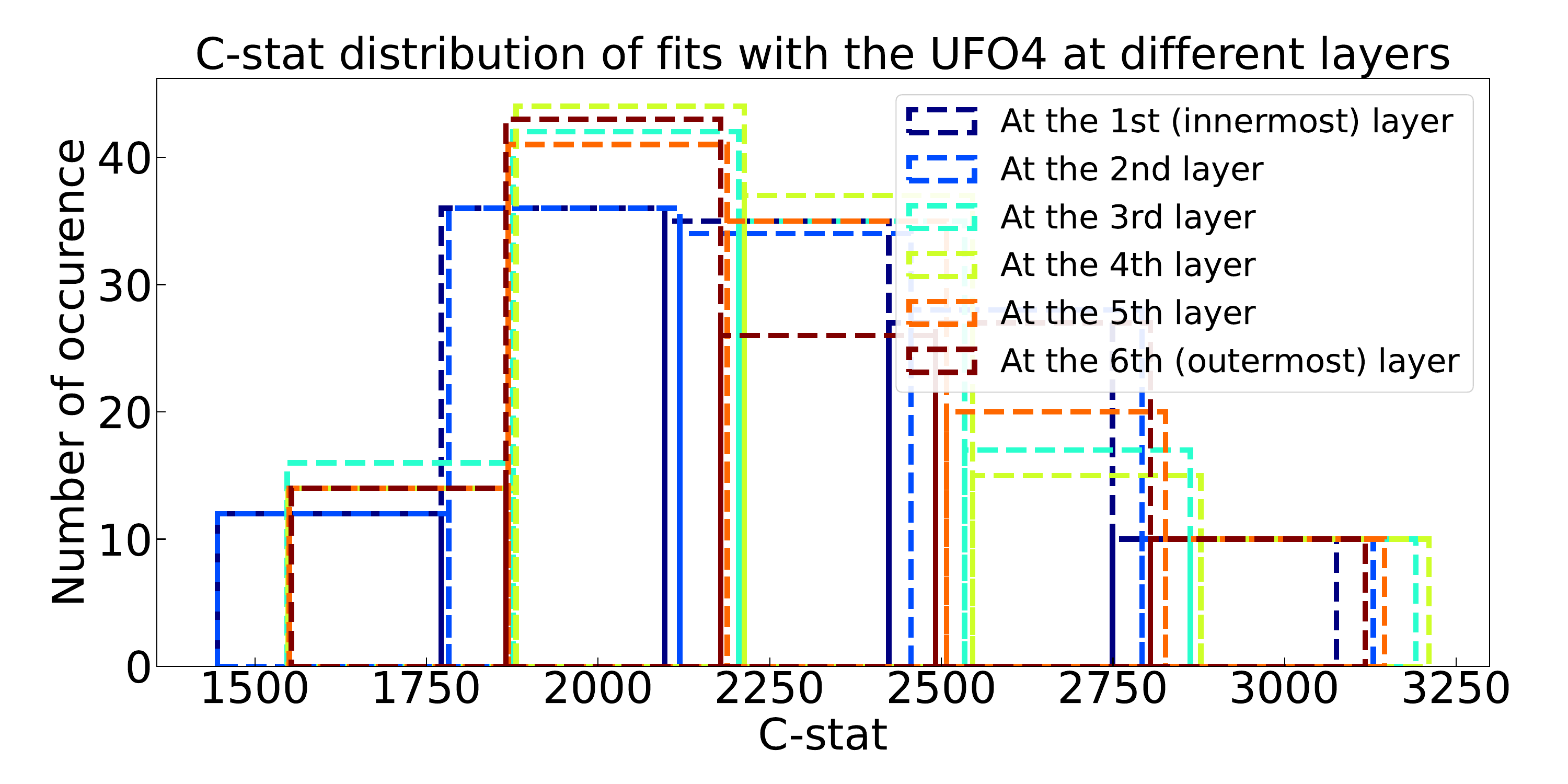} 
  \includegraphics[width=0.49\textwidth]{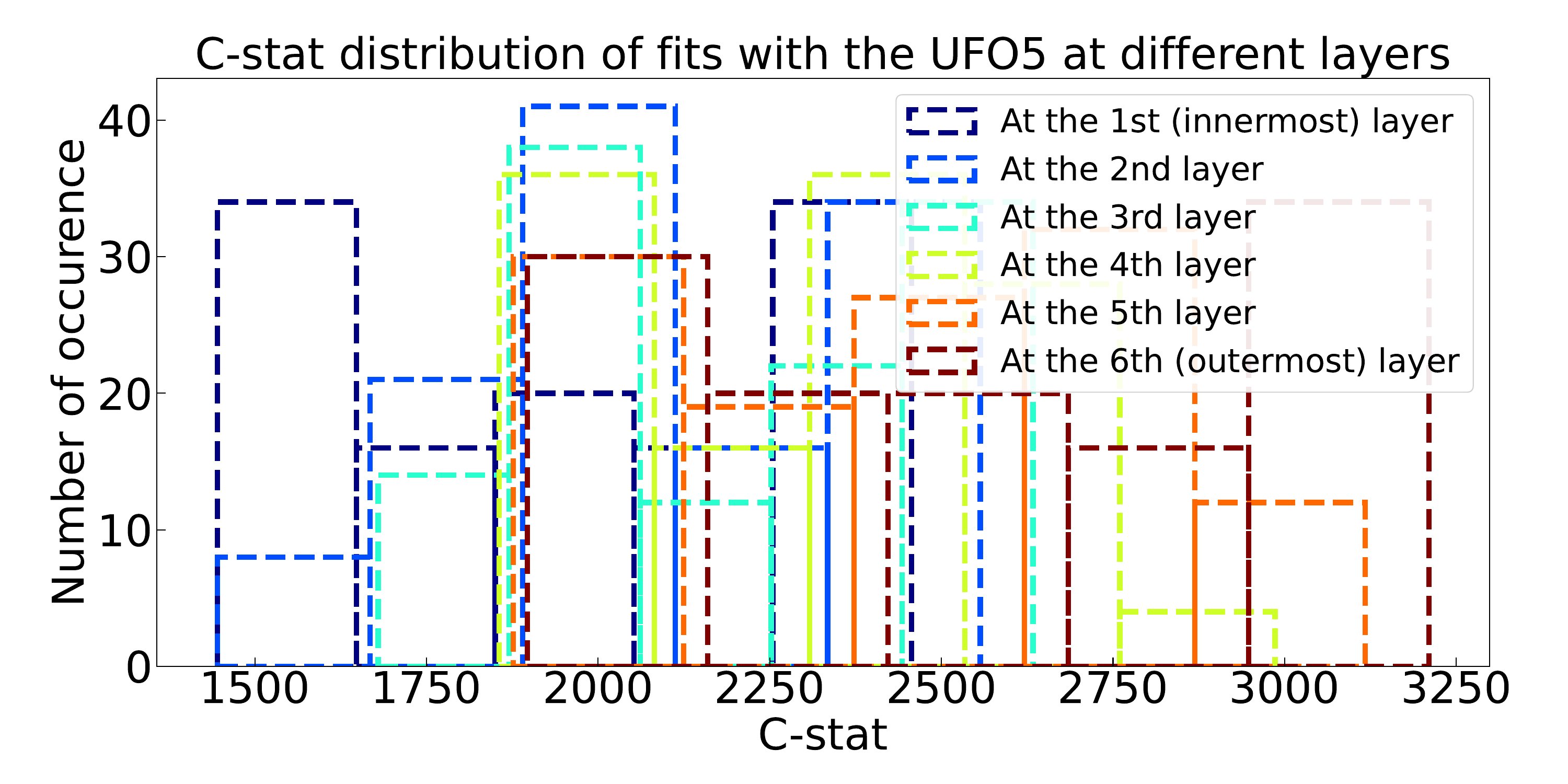} 
  \includegraphics[width=0.49\textwidth]{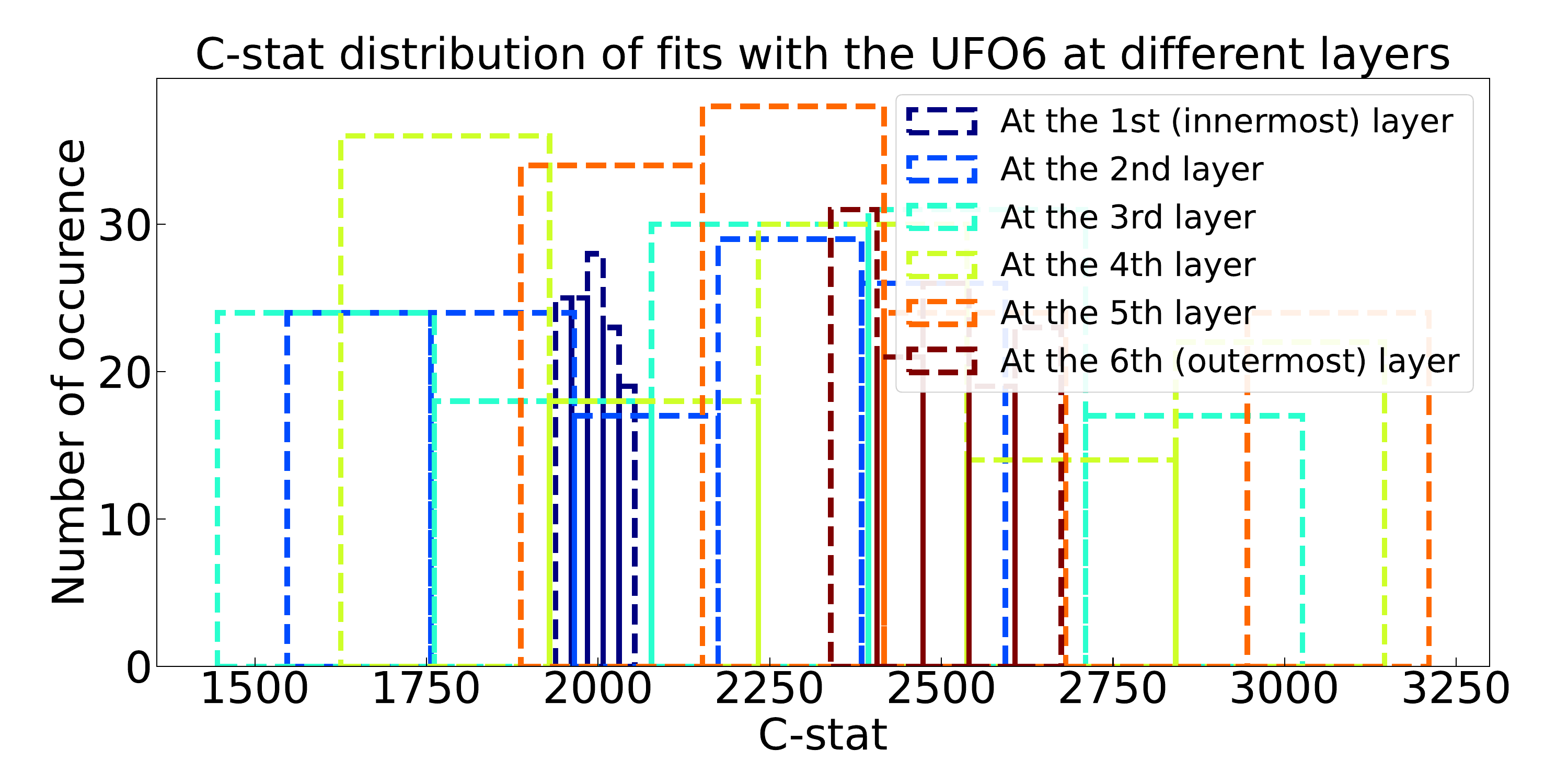} 
 \end{center}
\caption{Similar to Figure \ref{fig:histogram-all} but the analysis was performed on the simulated NewAthena/\textit{X-IFU} spectrum. C-stat distribution of fits with UFO1-6 (from upper left to lower right panels) placed at different layers. UFO6 statistically prefers the third layer, while the positions of the other UFOs still cannot be purely identified by statistics. Despite degeneracies, NewAthena can statistically reveal UFO4-5 at the inner and UFO1-3 at the outer layers.}
\label{app:fig:histogram-athena}
\end{figure*}

The C-stat distributions of spectral fits with multiple UFO components placed at different layers when the soft X-ray absorber was fixed at the outermost layer, are shown in Figure \ref{app:fig:histogram-soft-outermost}. Similar distributions with the position of the soft X-ray absorber taken into consideration are shown in Figure \ref{app:fig:histogram-all}. Figure \ref{app:fig:histogram-xrism} and \ref{app:fig:histogram-athena} illustrate the results when spectral fits were performed on one simulated 100\,ks-exposure spectrum of XRISM with an open gate valve and NewAthena spectra, respectively.

\section{Bayesian posteriors}\label{app:Bayesian}
\renewcommand{\thefigure}{B\arabic{figure}}
\setcounter{figure}{0}

\begin{figure*}
 \begin{center}
  \includegraphics[width=\textwidth]{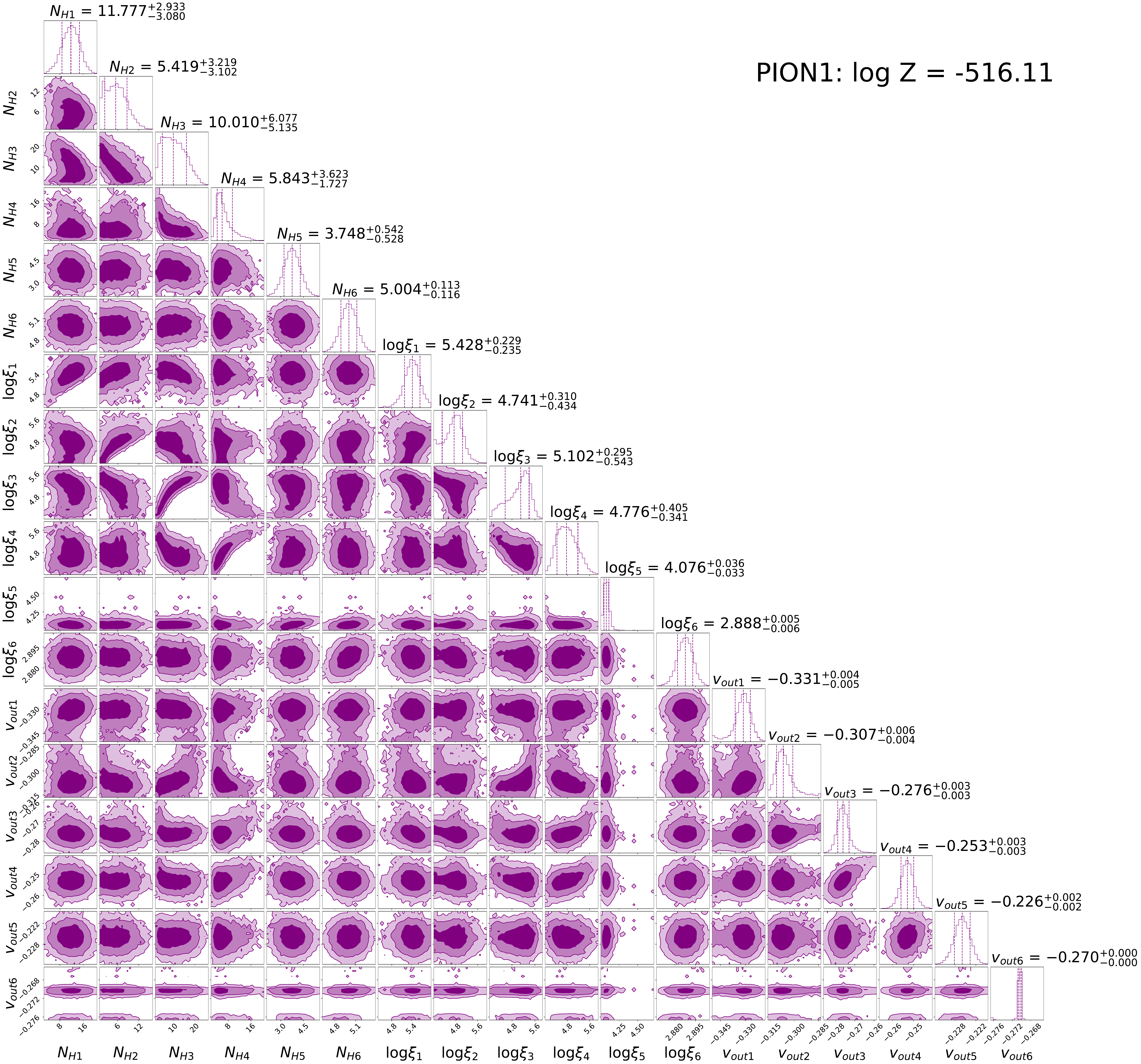} 
 \end{center}
\caption{The normalized probability distributions from \texttt{nautilus} are shown as two-dimensional histograms comparing each pair of free parameters across the six \texttt{PION} components in the PION1 model. Contours represent the 1$\sigma$, 2$\sigma$, and 3$\sigma$ confidence intervals. Column density $N_\mathrm{H}$ is given in units of $10^{22}\,\mathrm{cm}^{-2}$, and outflow velocity is expressed in units of the speed of light $c$.
}\label{app:fig:nautilus-PION1}
\end{figure*}

The Bayesian posterior samples of \texttt{PION} parameters in PION1, PION2, and PION3 models are shown in Figure \ref{app:fig:nautilus-PION1}-\ref{app:fig:nautilus-PION1}, respectively. Bayesian evidences for each model are shown in the upper right corner, showing that PION3 is preferred over PION2 ($\log B_{3,2} = 5$), and PION2 over PION1 ($\log B_{2,1} = 3$), consistent with the results from the C-stat analysis. The measurements of each parameter are compatible with those derived from C-stat analysis. Only the outflow velocity of UFO6 shows degenerated solutions, while those degeneracies are within the instrumental resolution, which has no significant impact on our results.

\begin{figure*}
 \begin{center}
  \includegraphics[width=\textwidth]{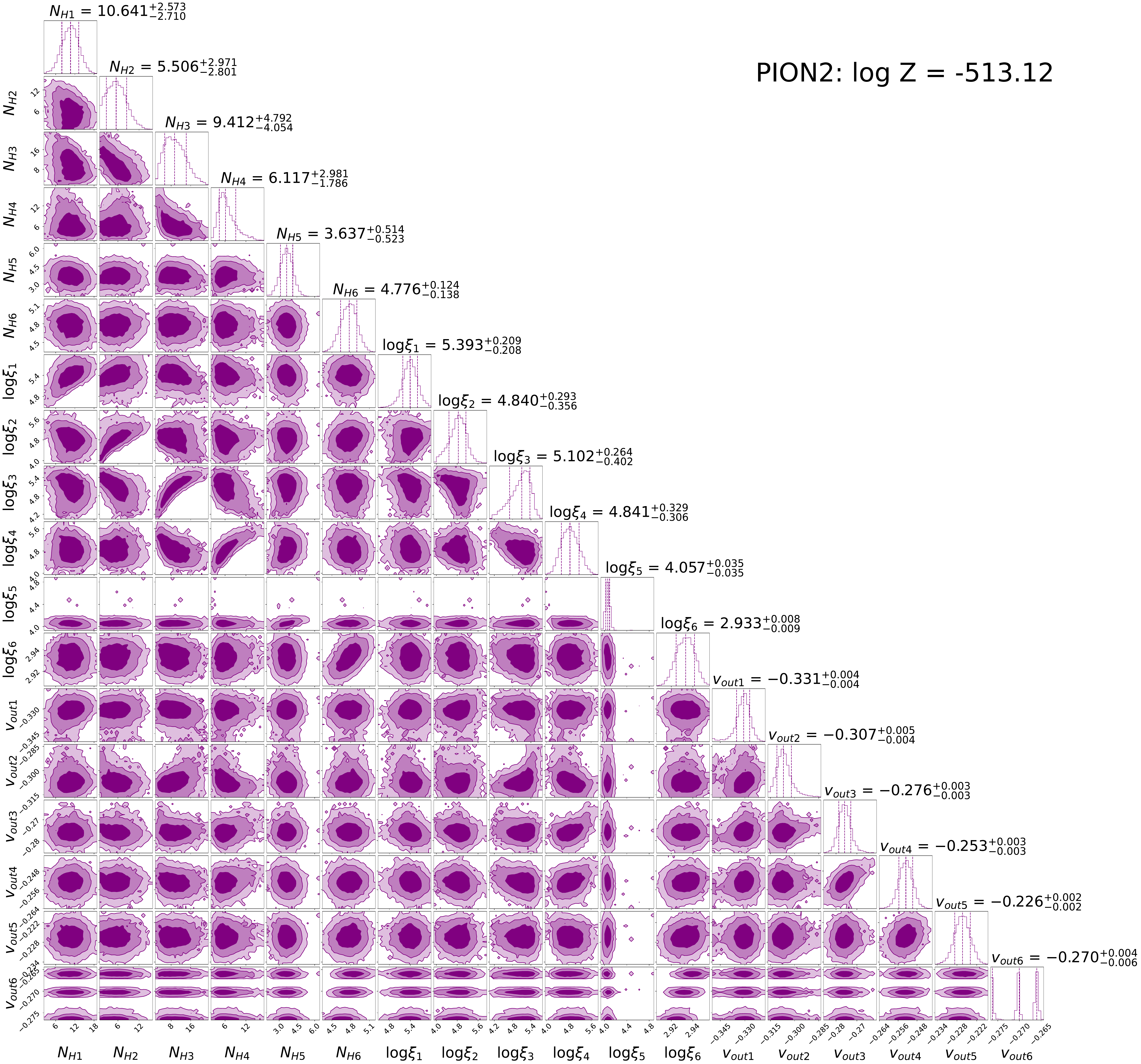} 
 \end{center}
\caption{Simliar to Fig.\ref{app:fig:nautilus-PION1} but for PION2.
}\label{app:fig:nautilus-PION2}
\end{figure*}

\begin{figure*}
 \begin{center}
  \includegraphics[width=\textwidth]{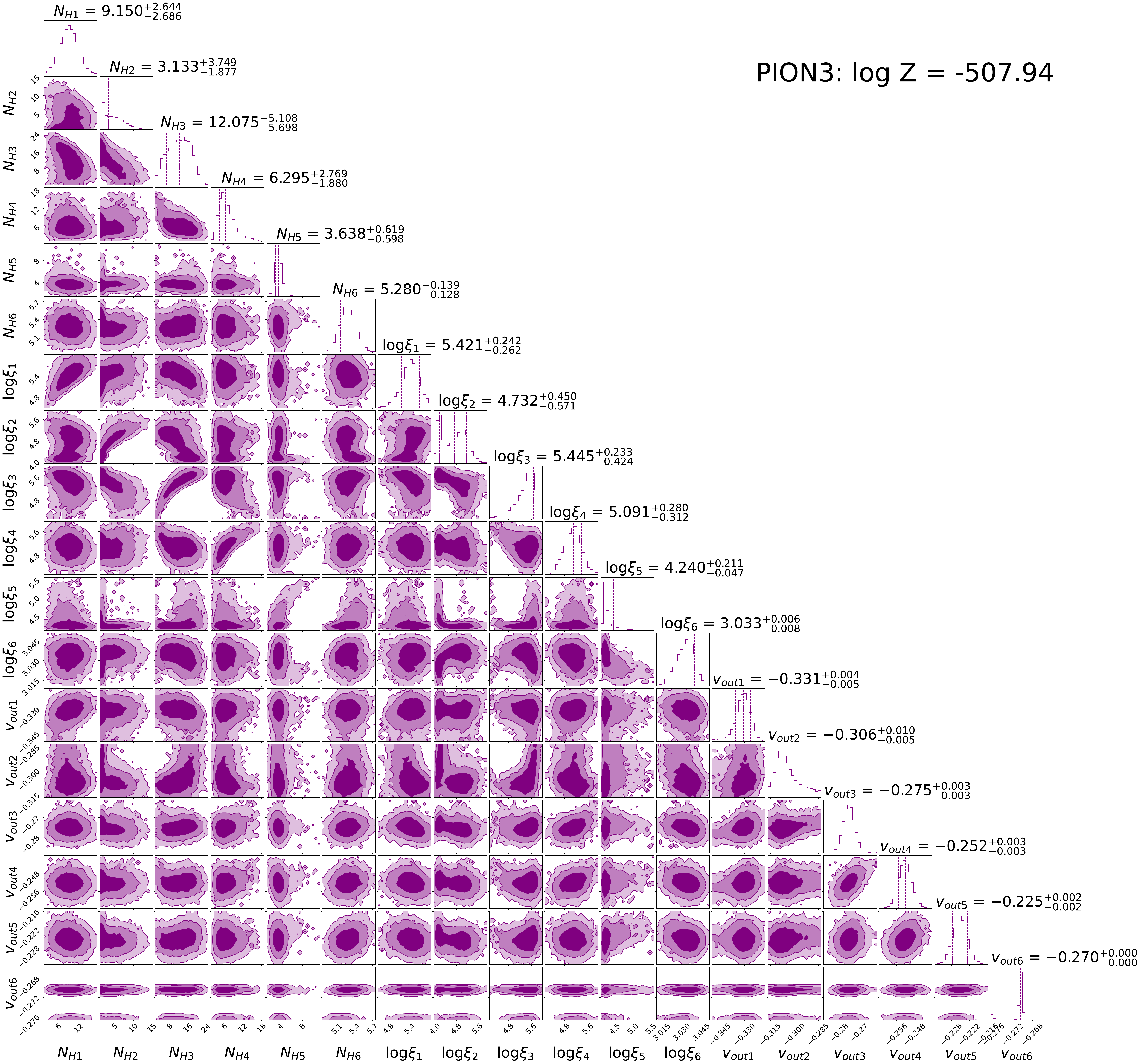} 
 \end{center}
\caption{Simliar to Fig.\ref{app:fig:nautilus-PION1} but for PION3.
}\label{app:fig:nautilus-PION3}
\end{figure*}

\end{appendix}
\bibliographystyle{aa}
\bibliography{ref}




\end{document}